\documentclass[twocolumn,nofootinbib,superscriptaddress,preprintnumbers,floatfix,prd]{revtex4-1}

\usepackage[bookmarks=false,hyperfootnotes=false]{hyperref}
\usepackage{graphicx}
\usepackage{amsmath}
\usepackage{xspace}
\usepackage{xcolor}
\usepackage{float}

\newcommand{\eq}[1]{Eq.~\eqref{eq:#1}}
\newcommand{\eqs}[2]{Eqs.~\eqref{eq:#1} and \eqref{eq:#2}}
\renewcommand{\sec}[1]{Sec.~\ref{sec:#1}}

\newcommand{\subsec}[1]{Sec.~\ref{subsec:#1}}
\newcommand{\fig}[1]{Fig.~\ref{fig:#1}}
\newcommand{\figs}[2]{Figs.~\ref{fig:#1} and \ref{fig:#2}}

\newcommand{\abs}[1]{\lvert#1\rvert}
\newcommand{\ord}[1]{\mathcal{O}(#1)}

\newcommand{\df}{\mathrm{d}}

\newcommand{\as}{\alpha_{\rm s}}
\newcommand{\kt}{{\rm k}_{\rm T}}

\newcommand{\Tau}{\mathcal{T}}

\newcommand{\GeV}{\,\mathrm{GeV}}
\newcommand{\TeV}{\,\mathrm{TeV}}

\newcommand{\nn}{\nonumber}

\newcommand{\cP}{\mathcal{P}}

\newcommand{\cut}{\mathrm{cut}}

\newcommand{\FO}{\mathrm{FO}}

\newcommand{\NLO}{\mathrm{NLO}}

\newcommand{\match}{\mathrm{match}}
\newcommand{\resum}{\mathrm{resum}}
\newcommand{\nons}{\mathrm{nons}}

\newcommand{\one}{{(1)}}

\newcommand{\dsigMC}{\df\sigma^\textsc{mc}}
\newcommand{\obs}{X}

\newcommand{\geneva}{\textsc{Geneva}\xspace}
\newcommand{\mcatnlo}{\textsc{MC@NLO}\xspace}
\newcommand{\powheg}{\textsc{Powheg}\xspace}

\newcommand{\pythia}{\textsc{Pythia}\xspace}

\newcommand{\jetvheto}{\textsc{JetVHeto}\xspace}
\newcommand{\dyqt}{\textsc{DYqT}\xspace}
\newcommand{\dynnlo}{\textsc{Dynnlo}\xspace}
\newcommand{\vrap}{\textsc{Vrap}\xspace}
\newcommand{\hepmc}{\textsc{HepMC}\xspace}
\newcommand{\rivet}{\textsc{Rivet}\xspace}
\newcommand{\fewz}{\textsc{Fewz}\xspace}

\newcommand{\atlas}{ATLAS\xspace}
\newcommand{\cms}{CMS\xspace}
\newcommand{\lhcb}{LHCb\xspace}

\allowdisplaybreaks[2]

\begin{document}


\preprint{\vbox{\hbox{CERN-PH-TH-2015-185}\hbox{DESY 15-136}}}

\title{Drell-Yan Production at NNLL$'+$NNLO Matched to Parton Showers}

\author{Simone Alioli}
\affiliation{CERN Theory Division, CH-1211, Geneva 23, Switzerland\vspace{0.5ex}}

\author{Christian W.~Bauer}
\affiliation{Ernest Orlando Lawrence Berkeley National Laboratory,
University of California, Berkeley, CA 94720\vspace{0.5ex}}

\author{Calvin Berggren}
\affiliation{Texas Lutheran University, TX 78155\vspace{0.5ex}}

\author{Frank J.~Tackmann}
\affiliation{Theory Group, Deutsches Elektronen-Synchrotron (DESY), D-22607 Hamburg, Germany\vspace{0.5ex}}

\author{Jonathan R.~Walsh}
\affiliation{Ernest Orlando Lawrence Berkeley National Laboratory,
University of California, Berkeley, CA 94720\vspace{0.5ex}}

\date{August 06, 2015}

\begin{abstract}

We present results for Drell-Yan production from the \geneva Monte-Carlo framework.
We combine the fully-differential NNLO calculation with higher-order resummation
in the $0$-jettiness resolution variable. The resulting parton-level events
are further combined with parton showering and hadronization provided by \pythia8.
The $0$-jettiness resummation is carried out to NNLL$'$, which consistently incorporates all singular virtual and real NNLO corrections. It thus provides a natural perturbative connection between the NNLO calculation and the parton shower regime, including a systematic assessment of perturbative uncertainties. In this way, inclusive observables are correct to NNLO, up to small power corrections in the resolution cutoff. Furthermore, the perturbative accuracy of $0$-jet-like resummation variables is significantly improved beyond the parton shower approximation.
We provide comparisons with LHC measurements of Drell-Yan production at $7 \TeV$ from \atlas, \cms, and \lhcb.
As already observed in $e^+e^-$ collisions, for resummation-sensitive observables, the agreement with data is noticeably improved by using a lower value of $\alpha_s(M_Z) = 0.1135$.


\end{abstract}

\maketitle


\section{Introduction}
\label{sec:intro}
There are several different ways to obtain theoretical predictions for collider processes involving strongly interacting particles, namely fixed-order (FO) perturbation theory, resummed perturbation theory, and predictions using parton shower algorithms.

In FO perturbation theory, the perturbative expansion is carried out to a given order in the strong coupling constant $\as$. The leading order (LO) corresponds to the lowest order in $\as$ required to obtain the desired process. The next-to-leading order (NLO) then includes all terms of $\ord{\as}$ relative to the LO result, and so on.

When the process or observables in consideration involve large ratios of physical scales, FO perturbation theory can become unreliable or even break down, since for each order in $\as$ one can encounter up to two powers of logarithms of large scale ratios. In this case, the most precise perturbative predictions are obtained by resumming the logarithmic terms to all orders in $\as$. At leading logarithmic (LL) order one includes the most dominant terms of order $\as^n \ln^{2n}$. The next-to-leading logarithmic (NLL) order also includes the next-largest subleading logarithmic terms, and so on. Like at FO, the resummed perturbative expansion happens in a systematic fashion, which is described in more detail later on.

Parton shower (PS) algorithms formally work in the strongly-ordered limit, where each subsequent emission happens at a smaller resolution scale than the previous one. They start from a given LO prediction and effectively multiply it by splitting probabilities for each additional emission. This allows in principle any observable to be predicted at LL (and including some NLL effects) using a probabilistic Markov-Chain algorithm.

Each type of prediction has its advantages as it provides the most accurate description in the limit where other predictions are not applicable. FO predictions are necessary for a precise description of additional hard (i.e. energetic wide-angle) emissions. These are not correctly described even at LO by either resummation or parton showers, since both of these approaches rely on a kinematic expansion in a small resolution variable. For observables sensitive to many soft and collinear emissions generating large logarithms, FO predictions are not suitable and resummation and parton showers are necessary. Resummed perturbation theory allows one to systematically carry out the resummation beyond the LL or strongly-ordered limit and is therefore more accurate than parton showers. On the other hand, the higher-order resummation applies at the level of observables that are still sufficiently inclusive, i.e., that marginalize over many emissions and are only sensitive to a small number of physical scales. To obtain a prediction of the final state that is fully exclusive in all emissions, parton showers are required. In particular, parton shower predictions allow one to further attach a hadronization model to generate fully exclusive hadron-level events. These are an essential requirement for experiments to be able to simulate their data and study detector effects, and allow for the most direct comparison with experimental data.

Clearly, to obtain the best possible predictions, it is desirable to combine the different theoretical descriptions in such a way that one benefits from the advantages of each. In particular, this allows for reliable predictions also in the transition regions in between different parametric limits, which are often important in practice. The (often difficult to answer) question which parametric regime is the most appropriate in a given phase-space region then becomes much less relevant for practical purposes.

Combining resummed and fixed-order perturbation theory is a standard procedure in higher-order resummed calculations, where it is well understood how to systematically match the pure resummed result to the fixed-order result in the appropriate fixed-order limit.

The goal of combining higher-order perturbative calculations with a parton shower is to obtain fully-exclusive hadronized events that also have the perturbative accuracy of the higher-order calculations. So far, there have been many advances to combine FO calculations with parton showers. The combination of LO predictions for several multiplicities with parton showers was pioneered in Refs.~\cite{Catani:2001cc, Lonnblad:2001iq}. The combination of generic NLO calculations with parton showers started with the methods of \mcatnlo~\cite{Frixione:2002ik, Frixione:2003ei} and \powheg~\cite{Nason:2004rx, Frixione:2007vw, Alioli:2010xd}, and is being continuously developed further, see e.g.\ Refs.~\cite{Nagy:2005aa, Kramer:2005hw, Nagy:2007ty, Torrielli:2010aw, Hoche:2010pf, Frixione:2010ra, Frederix:2011ss, Platzer:2011bc, Hartgring:2013jma, Czakon:2015cla, Jadach:2015mza}. By now there are several methods available to combine multiple NLO calculations with parton showers~\cite{Alioli:2011nr, Hoeche:2012yf, Gehrmann:2012yg, Frederix:2012ps, Platzer:2012bs, Alioli:2012fc, Lonnblad:2012ix, Hamilton:2012rf, Luisoni:2013cuh}. More recently, combinations of specific Drell-Yan like NNLO calculations with parton showers have been presented in Refs.~\cite{Hamilton:2013fea, Hoeche:2014aia, Karlberg:2014qua, Hoche:2014dla, Hamilton:2015nsa}. A general method to construct NNLO calculations matched with parton showers has been discussed in Ref.~\cite{Alioli:2013hqa}, and in which the method of Ref.~\cite{Hamilton:2013fea} follows as a special case.

In the above methods, the primary goal is to improve the perturbative accuracy of parton shower Monte Carlo programs for inclusive FO observables, while the description of the resummation region is left to the parton shower. Formally, this combination amounts to matching the FO calculation to the LL resummation in the parton shower's resolution variable. The perturbative accuracy criteria that should be satisfied at NNLO$+$PS have been discussed in detail in Ref.~\cite{Alioli:2013hqa}.

\geneva goes beyond this by also incorporating the higher-order resummation in a suitable resolution variable that separates the FO and parton shower regimes. The perturbative matching essentially happens now from FO to resummation to parton shower. By carrying out the resummation to the appropriate order consistent with the FO accuracy, it effectively mediates between the FO calculation and the parton shower and bridges the perturbative gap between them. This provides several key benefits:
\begin{itemize}
   \item A systematic combination of NNLO calculations with parton showers, which extends to arbitrary processes.
   \item An improved perturbative accuracy of (sufficiently inclusive) resummation observables,
   \item an improved description of the matching/transition between resummation and fixed-order regimes,
   \item a systematic assessment of perturbative uncertainties in the resummation and matching.
\end{itemize}

An implementation of our approach for $e^+ e^-$ collisions was presented in Ref.~\cite{Alioli:2012fc}, where NLO calculations for $e^+ e^- \to 2$ jets and $e^+ e^- \to 3$ jets, and the NNLL$'$ resummation for thrust ($2$-jettiness) as resolution variable are combined with the parton showering and hadronization provided by \pythia8~\cite{Sjostrand:2006za, Sjostrand:2007gs}. For all the tested $2$-jet-like resummation-sensitive observables, the predictions from \geneva closely match the corresponding exact NNLL$^{(\prime)}$ resummed calculations, and after hadronization are in excellent agreement with LEP data. Although not emphasized there, the results in Ref.~\cite{Alioli:2012fc} already include all singular NNLO corrections, and are formally accurate to NNLO up to power corrections in the resolution cutoff.

In this paper, we present results for Drell-Yan production $pp\to\gamma/Z\to \ell^+\ell^-$.
Our implementation combines the general construction of Ref.~\cite{Alioli:2013hqa} with the inclusion of higher-order resummation in the jet resolution variable similar to Ref.~\cite{Alioli:2012fc}. We combine the fully-differential NNLO calculation for Drell-Yan with the NNLL$'$ resummation for $0$-jettiness $\Tau_0$ (aka beam thrust~\cite{Stewart:2009yx}) with \pythia8. The FO accuracy of the generated events is NNLO for $pp \to \ell^+ \ell^-$, NLO for $pp \to \ell^+ \ell^-j$, and LO for $pp \to \ell^+ \ell^-jj$ (which we refer to as NNLO$_0$, NLO$_1$, and LO$_2$, respectively) while the $\Tau_0$ distribution is exactly reproduced at NNLL$'$. We show that by including the higher-order resummation for $\Tau_0$ as the resolution variable, the perturbative accuracy for other 0-jet-like observables in the resummation region is significantly improved, and our predictions agree well with dedicated higher-order resummed calculations.

Comparing our results to Drell-Yan measurements from \atlas, \cms, and \lhcb, obtained during the $7\TeV$ Run at the LHC, we obtain overall good agreement with the data for rapidity distributions, inclusive and exclusive jet cross-sections, as well as the transverse momentum distribution of the hardest jet. For the transverse momentum of the vector boson and the related $\phi^*$ distribution between the leptons, our default settings with $\as(M_Z) = 0.118$ predict a somewhat harder spectrum. We observe a much better agreement with the data when using a lower value of the strong coupling constant $\as(M_Z) = 0.1135$, as was obtained from fits of N$^3$LL$'$ resummed calculations of event shapes to LEP data~\cite{Abbate:2010xh, Abbate:2012jh, Hoang:2014wka, Hoang:2015hka}. The same was also observed for resummation observables in our $e^+e^-$ results in Ref.~\cite{Alioli:2012fc}.

This paper is organized as follows. In \sec{TheoreticalFramework} we discuss the theoretical framework of \geneva. In \subsec{setup} we provide an overview of the general construction and the master formulas for the partonic jet cross sections that serve as the main perturbative input, and in \subsec{details} we discuss specific choices and implementation aspects that are of relevance in more detail. In \sec{shower}, we describe how the partonic events with up to 2 partons are interfaced with a parton shower in a way that avoids double counting and such that the parton shower does not destroy the perturbative accuracy of the partonic calculation. In \sec{ComparisonWithPerturbative}, we compare \geneva with dedicated higher-order perturbative calculations, and in \sec{ComparisonWithData} we compare \geneva's results with measurements from the LHC. We conclude in \sec{conclusions}.

\section{Theoretical Framework}
\label{sec:TheoreticalFramework}

\subsection{General Setup}
\label{subsec:setup}

In this subsection we discuss the general construction we use. Our discussion here follows that of Ref.~\cite{Alioli:2013hqa}, and we refer to that paper for more details.

Drell-Yan production at NNLO receives contributions from partonic processes with up to two final-state partons. In a standard FO calculation, the phase space for these partonic contributions is integrated over in such a way that the cancellation of virtual and real IR divergences happens at the level of observables.

In contrast, an event generator is meant to produce physical events, which means that all IR divergences should cancel on a per-event basis. This implies that an $N$-parton event should correspond to an IR-finite partonic $N$-jet cross section, which is fully-differential in the corresponding partonic $N$-jet phase space. That is, each generated event represents a point in an $N$-jet phase space, rather than an $N$-parton phase space, and each four-momentum in the event provides the energy and direction of a partonic jet.

Therefore, the basis of the \geneva MC framework is the formulation of the perturbative inputs in terms of ``Monte Carlo (MC) cross sections'', which are well-defined partonic jet cross sections according to which the events are distributed, and which can be systematically computed to the desired perturbative accuracy in FO and resummed perturbation theory~\cite{Bauer:2008qh, Bauer:2008qj, Alioli:2012fc}.

The MC cross sections are defined in terms of an $N$-jet resolution variable $\Tau_N$ and formally include the contributions of an arbitrary number of unresolved emissions below a resolution cutoff $\Tau_N < \Tau_N^\cut$. In the present case, we require events with $0$, $1$, and $2$ partons, which are distributed according to
\begin{align} \label{eq:NNLOevents}
\text{$\Phi_0$ events: }
& \qquad \frac{\dsigMC_0}{\df\Phi_0}(\Tau_0^\cut)
\,,\nn \\
\text{$\Phi_1$ events: }
& \qquad
\frac{\dsigMC_{1}}{\df\Phi_{1}}(\Tau_0 > \Tau_0^\cut; \Tau_{1}^\cut)
\,, \nn\\
\text{$\Phi_2$ events: }
& \qquad
\frac{\dsigMC_{\ge 2}}{\df\Phi_{2}}(\Tau_0 > \Tau_0^\cut, \Tau_{1} > \Tau_{1}^\cut)
\,.\end{align}
The exclusive $0$-jet MC cross section is defined by $\Tau_0 < \Tau_0^\cut$, the exclusive $1$-jet MC cross section by $\Tau_0 > \Tau_0^\cut$ and $\Tau_1 < \Tau_1^\cut$, and the inclusive 2-jet MC cross section by $\Tau_0 > \Tau_0^\cut$ and $\Tau_1 > \Tau_1^\cut$. In this way all of partonic phase space is covered. Adding the $1$-jet and $2$-jet events, we can also define the inclusive 1-jet MC cross section as
\begin{align}
\label{eq:inclusive1JetDef}
\frac{\dsigMC_{\ge 1}}{\df\Phi_1}(\Tau_0 > \Tau_0^\cut)
&= \frac{\dsigMC_{1}}{\df\Phi_{1}}(\Tau_0 > \Tau_0^\cut; \Tau_{1}^\cut)
\\ &\quad
+ \!\int \! \frac{\df \Phi_2}{\df \Phi_1} \, \frac{\dsigMC_{\ge 2}}{\df\Phi_{2}}(\Tau_0 \!>\! \Tau_0^\cut, \Tau_{1} \!>\! \Tau_{1}^\cut)
\nn
\,,\end{align}
which is defined by $\Tau_0 > \Tau_0^\cut$ and does not depend anymore on $\Tau_1^\cut$.
In \eq{inclusive1JetDef} we have made use of the shorthand notation 
\begin{align}
\label{eq:dPhiRatio}
\frac{\df \Phi_{M}}{\df \Phi_N} = \df \Phi_{M} \, \delta[ \Phi_N - \Phi_N(\Phi_M) ]
\,.\end{align}

Since the partonic jets are represented by on-shell quarks and gluons in the partonic $\Phi_N$ events, the partonic jet definition used here is quite different from an ordinary jet algorithm. It depends on a phase space map $\Phi_N(\Phi_M)$ that projects the $M$-body phase space of unresolved emissions onto the distributed $\Phi_N$ points, where $N \leq M$. This phase space map must be IR safe, such that the MC cross sections are IR finite. For example, at fixed NNLO the exclusive 0-jet cross section is given by
\begin{align} \label{eq:sig0NNLO}
\frac{\dsigMC_0}{\df\Phi_0}(\Tau_0^\cut) &= (B_0 + V_0 + W_0)(\Phi_0)
\nn \\ & \quad
+ \int \! \frac{\df \Phi_1}{\df \Phi_0}\, (B_1 + V_1)(\Phi_1)\, \theta[\Tau_0(\Phi_1) < \Tau_0^\cut]
\nn \\ & \quad
+ \int \! \frac{\df \Phi_2}{\df \Phi_0}\, B_2 (\Phi_2)\, \theta[\Tau_0(\Phi_2) < \Tau_0^\cut]
\,,\end{align}
where $B_N$ contains the $N$-parton tree-level contributions, $V_N$ the $N$-parton one-loop contributions, and $W_0$ the two-loop contribution. There is considerable freedom in the precise definitions of $\Tau_N$ and the $\Phi_N(\Phi_M)$ map, and we will choose particular definitions based on their theoretical properties and simplicity.

Using the events in \eq{NNLOevents}, the cross section for any observable $X$ is given by
\begin{align} \label{eq:NNLOeventsused}
\sigma(\obs)
&= \int\!\df\Phi_0\, \frac{\dsigMC_0}{\df\Phi_0}(\Tau_0^\cut)\, M_\obs(\Phi_0)
\\ & \quad
+ \int\!\df\Phi_{1}\, \frac{\dsigMC_{1}}{\df\Phi_{1}}(\Tau_0 > \Tau_0^\cut; \Tau_{1}^\cut)\, M_\obs(\Phi_{1})
\nn\\ & \quad
+ \int\!\df\Phi_{2}\, \frac{\dsigMC_{\ge 2}}{\df\Phi_{2}}(\Tau_0 > \Tau_0^\cut, \Tau_{1} > \Tau_{1}^\cut)\, M_\obs(\Phi_{2})
\nn\,,\end{align}
where $M_X(\Phi_N)$ is the measurement function that computes the observable $X$ for the $N$-parton final state $\Phi_N$.
This cross section is not identical to the exact fixed-order result, because for any unresolved emissions the observable is calculated on the projected phase space point $\Phi_N(\Phi_M)$, rather than the exact $\Phi_M$. This is a fundamental limitation inherent to generating IR-finite events beyond LO and for a more detailed discussion we refer to Ref.~\cite{Alioli:2013hqa}. The difference vanishes in the limit $\Tau_N^\cut \to 0$. Hence, we want to choose $\Tau_N^\cut$ as small as possible to have a maximally exclusive description.
For small $\Tau_N^\cut$, however, the MC cross sections contain large logarithms of $\Tau_N$ and $\Tau_N^\cut$, requiring resummation to obtain physically meaningful predictions for them.
This is precisely what a parton shower would do at LL (also including some NLL effects), in which case $\Tau_N^\cut$ would play the role of the shower cutoff. In \geneva we are improving over this by resumming these
logarithms at higher accuracy.

\subsubsection{$0/1$-jet separation}
\label{subsec:01jetsep}

To discuss the construction of the MC cross sections, we first consider the separation between $0$ and $1$ jets which is determined by the $0$-jet resolution variable $\Tau_0$. We write the exclusive 0-jet and inclusive 1-jet MC cross sections as~\cite{Alioli:2012fc, Alioli:2013hqa}
\begin{align} \label{eq:01master}
\frac{\dsigMC_0}{\df\Phi_0}(\Tau_0^\cut)
&= \frac{\df\sigma_{0}^\resum}{\df\Phi_0}(\Tau_0^\cut)
+ \frac{\df\sigma_0^{\rm sing\,match}}{\df\Phi_0}(\Tau_0^\cut)
\nn\\&\quad
+ \frac{\df\sigma_0^{\rm nons}}{\df\Phi_0}(\Tau_0^\cut)
\,,\nn\\
\frac{\dsigMC_{\geq 1}}{\df\Phi_{1}}(\Tau_0 > \Tau_0^\cut)
&= \frac{\df\sigma_{\ge 1}^\resum}{\df\Phi_1}\, \theta(\Tau_0 > \Tau_0^\cut)
\nn\\&\quad
+ \frac{\df\sigma_{\ge 1}^{\rm sing\,match}}{\df\Phi_1}(\Tau_0 > \Tau_0^\cut)
\nn\\&\quad
+ \frac{\df\sigma_{\ge 1}^{\rm nons}}{\df\Phi_1}(\Tau_0 > \Tau_0^\cut)
\,.\end{align}
Here, $\df \sigma_0^\resum$ is the spectrum differential in $\Phi_0$ with the dependence on $\Tau_0^\cut$ resummed to a given logarithmic accuracy, while the remaining two terms give the matching corrections required to reproduce the desired FO accuracy. Their precise form depends on the fixed-order content of the resummed contribution. The singular matching $\df \sigma_0^{\rm sing\,match}$ contains all contributions that do not vanish as $\Tau_0^\cut \to 0$, while the nonsingular matching $\df \sigma_0^{\rm nons}$ only contains contributions that vanish in this limit. The analogous separation is done for $\dsigMC_{\geq1}$, where $\df \sigma_{\ge 1}^\resum$ resums the differential dependence on $\Tau_0$, the singular matching contains contributions that diverge at least like $1/\Tau_0$ for $\Tau_0 \to 0$, and the nonsingular matching contains contributions that contain at most integrable singularities for $\Tau_0\to 0$.

We carry out the resummation of the $\Tau_0$ dependence to NNLL$'$. Thus, the resummed contributions are given by
\begin{align} \label{eq:fulllyDiffRes}
\frac{\df\sigma_0^{\rm resum}}{\df\Phi_0}(\Tau_0^\cut)
&= \frac{\df\sigma^{\rm NNLL'}}{\df\Phi_0}(\Tau_0^\cut)
\,, \nn \\
\frac{\df\sigma_{\ge 1}^{\rm resum}}{\df\Phi_{1}}
&= \frac{\df\sigma^{\rm NNLL'}}{\df\Phi_{0}\df \Tau_0} \, \cP(\Phi_1)
\,.\end{align}
Here, $\df\sigma^{\rm NNLL'}/\df\Tau_0\df\Phi_0$ is the resummed differential $\Tau_0$ spectrum and $\df\sigma^{\rm NNLL'}/\df\Phi_0(\Tau_0^\cut)$ its cumulative integral, see \subsec{SCETResummation} below.
Note that the two-loop virtual corrections, which at FO are proportional to $\delta(\Tau_0)$, are fully incorporated at NNLL$'$. This means they are properly spread to nonzero values of $\Tau_0$ and contribute to the differential spectrum as dictated by the resummation. The $\Tau_0$ resummation is naturally differential in $\Phi_0$, so we can directly use the resummed cumulant in $\df\sigma_0^{\rm resum}$. To make the $\Tau_0$ spectrum fully differential in $\Phi_1$, we have defined a normalized splitting probability $\cP(\Phi_1)$ which satisfies
\begin{align}
\label{eq:cPnorm}
\int \! \frac{\df\Phi_1}{\df \Phi_{0} \df \Tau_0} \, \cP(\Phi_1) = 1
\,.\end{align}
Since $\Tau_0$ can be considered as part of the radiation phase space $\df\Phi_1/\df\Phi_0$, the integration here is effectively over two remaining radiation variables, e.g. an energy splitting and an azimuthal angle. Thus, the integral of the splitting probability over all $\Phi_1$ points restricted to any fixed values of $\Phi_0(\Phi_1)$ and $\Tau_0(\Phi_1)$ is equal to 1. This will be discussed in more detail in \subsec{normalizedSplitting}.

At NNLL$'$, the resummation fully incorporates all singular contributions to $\ord{\as^2}$, implying that the singular matching vanishes,
\begin{align} \label{eq:01match}
\frac{\df\sigma_0^{\rm sing\,match}}{\df\Phi_0}(\Tau_0^\cut)
&= 0
\,, \nn \\
\frac{\df\sigma_{\ge 1}^{\rm sing\,match}}{\df\Phi_1}(\Tau_0 > \Tau_0^\cut)
&= 0
\,.\end{align}

At our desired NNLO accuracy, $\dsigMC_0$ and $\dsigMC_{\geq 1}$ must be correct to NNLO$_0$ and NLO$_1$, respectively, which determines the nonsingular matching corrections to be
\begin{align} \label{eq:01nons}
&\frac{\df\sigma_0^\nons}{\df\Phi_{0}}(\Tau_0^\cut)
\nn \\
& \quad
= \frac{\df\sigma_0^{{\rm NNLO_0}}}{\df\Phi_{0}}(\Tau_0^\cut)
- \biggl[\frac{\df\sigma_0^\resum}{\df\Phi_{0}}(\Tau_0^\cut) \biggr]_{\rm NNLO_0}
\,, \nn \\
&\frac{\df\sigma_{\ge 1}^\nons}{\df\Phi_{1}}(\Tau_0 > \Tau_0^\cut)
\\ \nn & \quad
= \frac{\df\sigma_{\ge 1}^{{\rm NLO_1}}}{\df\Phi_{1}}(\Tau_0 > \Tau_0^\cut)
-  \biggl[\frac{\df\sigma_{\geq 1}^{\resum}}{\df\Phi_1} \biggr]_{\rm NLO_1} \!\!\!\theta(\Tau_0 > \Tau_0^\cut)
\,.\end{align}
The terms in square brackets are the FO expansions to $\as^2$ of the resummed cumulant and spectrum in \eq{fulllyDiffRes}. The NNLO$_0$ result for the cumulant is given in \eq{sig0NNLO}. The NLO$_1$ result for the fully-differential spectrum is given by
\begin{align} \label{eq:NLO1}
\frac{\df\sigma_{\geq 1}^{\NLO_1}}{\df\Phi_1}(\Tau_0 > \Tau_0^\cut)
&= (B_1 + V_1)(\Phi_1)\,\theta[\Tau_0(\Phi_1) > \Tau_0^\cut]
\nn \\ & \quad
+ \int\!\!\frac{\df\Phi_2}{\df\Phi_1^\Tau}\, B_2(\Phi_2)\,\theta[\Tau_0(\Phi_2) \!>\! \Tau_0^\cut]
.\end{align}
It depends on the projection
\begin{equation} \label{eq:Phi1TauProj}
\frac{\df\Phi_2}{\df\Phi_1^\Tau} \equiv \df\Phi_2\,\delta[\Phi_1 - \Phi_1^\Tau(\Phi_2)]\,\Theta^\Tau(\Phi_2)
\,,\end{equation}
where $\Theta^\Tau(\Phi_2)$ defines the region of $\Phi_2$ that can be projected onto massless $\Phi_1$ via the IR-safe phase space map $\Phi_1^\Tau(\Phi_2)$. Only this projectable region of $\Phi_2$ is included in $\df\sigma_{\geq 1}^\nons/\df\Phi_1$, while the remainder will be included in the $\Phi_2$ events below.

The fact that the NNLL$'$ resummation reproduces all singular terms of the full FO result to ${\cal O}(\as^2)$ implies that the resummed expanded result in $\df\sigma_{\geq 1}^\nons/\df\Phi_1$ acts precisely as a differential NNLO $\Tau_0$-subtraction~\cite{Gaunt:2015pea} (see also Ref.~\cite{Boughezal:2015dva}). The cancellation of the singular terms between the FO and resummed expanded results is not point-by-point in $\Phi_1$ but nonlocal, which means it only holds ``on average'' upon integrating over $\df\Phi_1/\df\Phi_0\df\Tau_0$. As discussed in detail in Ref.~\cite{Gaunt:2015pea}, the cancellation can be made more local by considering more-differential resummations (e.g.\ utilizing the results of Refs.~\cite{Jain:2011iu, Procura:2014cba, Gaunt:2014xxa}). For the cancellation to be point-by-point in $\Tau_0$, $\df\sigma_{\geq 1}^{\NLO_1}/\df\Phi_1$ has to reproduce the right singular $\Tau_0$-dependence when projected onto $\df\Tau_0\df\Phi_0$, which means the map $\Phi_1^\Tau(\Phi_2)$ has to preserve the value of $\Tau_0$,
\begin{equation} \label{eq:Tau0map}
\Tau_0(\Phi_1^\Tau(\Phi_2)) = \Tau_0(\Phi_2)
\,.\end{equation}
In other words, if this was not the case, we would not obtain the correct nonsingular corrections for the $\Tau_0$ spectrum, and this would destroy its NNLL$'$ accuracy. The underlying reason is that we perform the FO calculation in an event generator form as in \eq{NNLOeventsused}, and so to compute the FO $\Tau_0$ spectrum we always project some $\Phi_2$ points, namely those below $\Tau_1 < \Tau_1^\cut$, onto $\Phi_1$ and only then onto $\Tau_0$. More details on this map are given in~\subsec{PSMaps}.

\subsubsection{$1/2$-jet separation}

Next, we consider the separation of the inclusive 1-jet cross section into an exclusive 1-jet MC cross section and an inclusive 2-jet MC cross section, using the $1$-jet resolution variable $\Tau_1$. We write them as
\begin{widetext}
\begin{align} \label{eq:12master}
\frac{\dsigMC_{1}}{\df\Phi_{1}} (\Tau_0 > \Tau_0^\cut; \Tau_{1}^\cut)
&= \frac{\df\sigma_1^\resum}{\df\Phi_{1}}(\Tau_0 > \Tau_0^\cut; \Tau_{1}^\cut)
+ \frac{\df\sigma_1^\match}{\df \Phi_{1}}(\Tau_0 > \Tau_0^\cut; \Tau_{1}^\cut)
\,,\nn\\
\frac{\dsigMC_{\geq 2}}{\df\Phi_{2}} (\Tau_0 > \Tau_0^\cut, \Tau_{1}>\Tau_{1}^\cut)
&= \frac{\df\sigma^\resum_{\geq 2}}{\df\Phi_2}(\Tau_0 > \Tau_0^\cut)\,\theta(\Tau_{1} > \Tau_{1}^\cut)
+ \frac{\df \sigma^\match_{\geq 2}}{\df \Phi_2}(\Tau_0 > \Tau_0^\cut, \Tau_1 > \Tau_1^\cut)
\,.\end{align}
Here, $\df\sigma_1^\resum$ and $\df\sigma_{\geq 2}^\resum$ contain the resummation of the $\Tau_1^\cut$ and differential $\Tau_1$ dependence. The $\df\sigma_1^\match$ and $\df\sigma_{\geq 2}^\match$ contain the matching corrections required to achieve the desired FO accuracy.

For the $\Tau_1$ resummation we limit ourselves to LL, which is sufficient for matching to the parton shower.
At this level, we can write the resummed contributions as
\begin{align} \label{eq:Tau1resum}
\frac{\df\sigma_1^\resum}{\df\Phi_{1}}(\Tau_0 > \Tau_0^\cut; \Tau_{1}^\cut)
&= \frac{\df\sigma_{\geq 1}^C}{\df\Phi_1} \, U_1(\Phi_1, \Tau_1^\cut)\, \theta(\Tau_0 > \Tau_0^\cut)
\,,\nn\\
\frac{\df\sigma^\resum_{\geq 2}}{\df\Phi_2}(\Tau_0 > \Tau_0^\cut)
&=
\frac{\df\sigma_{\geq 1}^C}{\df\Phi_1}\, U_1'(\Phi_1, \Tau_1)\, \theta(\Tau_0 > \Tau_0^\cut) \Big\vert_{\Phi_1 = \Phi_1^\Tau(\Phi_2)} \!\! \cP(\Phi_2) \, \theta(\Tau_1 > \Tau_1^\cut)
\,.\end{align}
Here, $U_1(\Phi_1, \Tau_1^\cut)$ denotes the Sudakov factor that resums the dependence on $\Tau_1^\cut$ to at least LL accuracy, and $U_1'(\Phi_1, \Tau_1)$ denotes its derivative with respect to $\Tau_1^\cut$ and resums the differential $\Tau_1$ dependence. Its details are given in \subsec{U1Sudakov}.

The differential $\Tau_1$ resummation in $\df\sigma_{\geq 2}$ is evaluated at the projected $\Phi_1$ point $\Phi_1^\Tau(\Phi_2)$. As we will see below, this is required so the $\Tau_0$ resummation is evaluated at the correct $\Tau_0$ value. The normalized splitting probability $\cP(\Phi_2)$ is defined analogous to $\cP(\Phi_1)$ in \eq{cPnorm} but for $\Phi_1 \to \Phi_2$ and using the $\Phi_1^\Tau(\Phi_2)$ map, with the details given in \subsec{normalizedSplitting}.

The $\Tau_1$ resummation acts on top of the overall $\df\sigma_{\geq 1}^C / \df\Phi_1$ in \eq{Tau1resum}, which is the inclusive $1$-jet cross section in the $\Tau_1$-singular limit. Its NLO$_1$ expansion therefore has to be
\begin{align} \label{eq:NLO1singular}
\biggl[\frac{\df\sigma^{C}_{\geq 1}}{\df\Phi_{1}}\biggr]_{\NLO_1}
= (B_1 + V_1)(\Phi_1) + \int\!\frac{\df\Phi_2}{\df \Phi_1^C}\, C_2(\Phi_2)
\equiv (B_1 + V_1^C)(\Phi_1)
\,.\end{align}
(Its full form will include the $\Tau_0$ resummation and is derived below.)
Here, $C_2(\Phi_2)$ denotes a standard NLO subtraction~\cite{Catani:1996jh, Frixione:1995ms} that reproduces the pointwise singular behavior of $B_2(\Phi_2)$. We use the FKS subtractions~\cite{Frixione:1995ms} in our implementation. Note that the subtraction term here uses its own projection $\df\Phi_2/\df\Phi_1^C \equiv \df\Phi_2\,\delta[\Phi_1 - \Phi_1^C(\Phi_2)]$.
Using \eqs{Tau1resum}{NLO1singular} and requiring that $\dsigMC_1$ and $\dsigMC_{\geq 2}$ are correct to NLO$_1$ and LO$_2$, respectively, then determines the matching corrections to be
\begin{align} \label{eq:Phi1Nons}
\frac{\df\sigma_1^\match}{\df\Phi_1}(\Tau_0 > \Tau_0^\cut; \Tau_1^\cut)
&= \int\!\biggl[\frac{\df\Phi_{2}}{\df\Phi_1^\Tau}\,B_{2}(\Phi_2)\, \theta[\Tau_0(\Phi_2) > \Tau_0^\cut]\,\theta(\Tau_{1} < \Tau_1^\cut)
- \frac{\df\Phi_2}{\df \Phi_1^C}\, C_{2}(\Phi_{2})\, \theta(\Tau_0 > \Tau_0^\cut) \biggr]
\nn \\ & \quad
- B_1(\Phi_1)\, U_1^\one(\Phi_1, \Tau_1^\cut)\, \theta(\Tau_0 > \Tau_0^\cut)
\,,\\
\label{eq:Phi2Nons}
\frac{\df\sigma_{\geq 2}^\match}{\df\Phi_2}(\Tau_0 > \Tau_0^\cut, \Tau_1 > \Tau_1^\cut)
&= \bigl\{ B_2(\Phi_2)\,[1 - \Theta^\Tau(\Phi_2)\,\theta(\Tau_1 < \Tau_1^\cut)]
\nn \\ & \quad
- B_1(\Phi_1^\Tau)\,U_1^{\one\prime}(\Phi_1^\Tau, \Tau_1)\,\cP(\Phi_2)\, \theta(\Tau_1 > \Tau_1^\cut)
\bigr\}\, \theta[\Tau_0(\Phi_2) > \Tau_0^\cut]
\,.\end{align}
Here, $U_1^{\one(\prime)}(\Phi_1, \Tau_1^\cut)$ denotes the $\ord{\as}$ term in the expansion of the $U_1^{(\prime)}(\Phi_1, \Tau_1^\cut)$ Sudakov factor (or its derivative). These terms cancel the leading double-logarithmic terms in $\Tau_1$ in the FO pieces (the $B_2$ contributions). As long as we only include the LL $\Tau_1$ resummation, there will be some subleading single-logarithmic terms present in the matching corrections.

So far, we have discussed the construction of an additive NLO$_1+$LL$_1$ matching. The important point is now that this should also include the $\Tau_0$ resummation which is important at small $\Tau_0$. The condition for this is that the integral of NLO$_1+$LL$_1$ result with the $\Phi_1^\Tau(\Phi_2)$ map must reproduce the $\Tau_0$-resummed result for the inclusive $1$-jet MC cross section in \eq{01master}. That is,
\begin{align} \label{eq:MC1plusMC2cond}
\frac{\dsigMC_{\geq 1}}{\df\Phi_1}(\Tau_0 > \Tau_0^\cut)
&= \frac{\dsigMC_{1}}{\df\Phi_{1}} (\Tau_0 > \Tau_0^\cut; \Tau_{1}^\cut)
+ \int\!\frac{\df\Phi_{2}}{\df\Phi_{1}^\Tau}\, \frac{\dsigMC_{\ge 2}}{\df\Phi_{2}} (\Tau_0 > \Tau_0^\cut, \Tau_{1} > \Tau_{1}^\cut)
\nn \\
&= \frac{\df\sigma_{\geq 1}^C}{\df\Phi_1}\, \theta(\Tau_0 > \Tau_0^\cut)
+ \int \! \biggl[
\frac{\df \Phi_2}{\df \Phi_1^\Tau}\, B_2(\Phi_2)\, \theta[\Tau_0(\Phi_2) > \Tau_0^\cut]
- \frac{\df\Phi_2}{\df\Phi_1^C}\, C_2(\Phi_2)\, \theta(\Tau_0 > \Tau_0^\cut) \biggr]
\,,\end{align}
\end{widetext}
where we have used the identity
\begin{align}
\label{eq:unitarity}
U_1(\Phi_1, \Tau_1^\cut) + \int \! \frac{\df \Phi_2}{\df \Phi_1^\Tau}\, U_1'(\Phi_1, \Tau_1)\, \cP(\Phi_2)\,
\theta(\Tau_1 > \Tau_1^\cut) = 1
\,.\end{align}
Inserting the expression for $\dsigMC_{\geq 1}$ in terms of $\df\sigma_{\geq 1}^\resum$ and $\df\sigma_{\geq 1}^\nons$ in \eqs{fulllyDiffRes}{01nons}, we obtain the result for $\df\sigma_{\geq 1}^C/\df\Phi_1$ beyond NLO$_1$,
\begin{align}
\frac{\df\sigma^{C}_{\geq 1}}{\df\Phi_{1}}
&= \frac{\df\sigma_{\geq 1}^\resum}{\df\Phi_1}
+ (B_1 + V_1^C)(\Phi_1)
- \biggl[\frac{\df\sigma_{\geq 1}^\resum}{\df\Phi_1}\,\biggr]_{\NLO_1}
\,.\end{align}
It contains the full differential $\Tau_0$ resummation via $\df\sigma_{\geq 1}^\resum$. The difference to the full inclusive $1$-jet cross section is that the $\Tau_0$-nonsingular terms are now evaluated in the $\Tau_1$-singular approximation, and this mismatch is compensated by the $\Tau_1$ matching contributions.

The above expressions completely define the fully-differential jet cross sections. In the next subsection we provide some additional details on the specific implementation of the various pieces in \geneva.

\subsection{Implementation Details}
\label{subsec:details}

\subsubsection{Choice of the jet resolution variables}
\label{subsec:TauNDefs}

We choose $N$-jettiness~\cite{Stewart:2010tn} as our $N$-jet resolution variables. It is defined as
\begin{align} \label{eq:TauNdef}
\Tau_N = \sum_k \min \Bigl\{ \hat q_a \cdot p_k, \hat q_b \cdot p_k, \hat q_1 \cdot p_k, \ldots , \hat q_N \cdot p_k \Bigr\}
\,,\end{align}
where the sum over $k$ runs over all final-state particles, excluding the vector boson and all its decay products. We use a canonical geometric measure, where $\hat q_i = n_i = (1, \vec n_i)$ are lightlike reference vectors along the jet and beam directions.

$N$-jettiness is a global event shape that is explicitly designed as an $N$-jet resolution variable, as it measures the degree to which the final state is $N$-jet-like for a given $N$ and automatically clusters the final state into $N$ jet and beam regions in an IR-safe way and without any dependence on an additional jet clustering algorithm. Furthermore, it is theoretically well understood and its all-orders singular structure and resummation are known.

For Drell-Yan, we need $0$-jettiness $\Tau_0$ and $1$-jettiness $\Tau_1$, and we always evaluate \eq{TauNdef} in the longitudinally boosted frame where the vector boson has zero rapidity. The reference vectors for the beam directions are always along the beam directions, so $\vec n_a = \hat z$ and $\vec n_b = -\hat z$. For $\Tau_1$, we also need to define a jet direction $\vec n_J$. For our calculation we only need to define it for a $\Phi_2$ event, in which case we can choose $\vec n_J$ by minimizing $\Tau_1$.
We define a clustering metric
\begin{align}
\label{eq:NJettinessMetric}
d(p) &= \abs{\vec p} - \abs{p^z}
\,, \nn\\
d(p_1, p_2) & = \abs{\vec p_1} + \abs{\vec p_2} - \abs{\vec p_1 + \vec p_2}
\,,\end{align}
and find the smallest of $d(p_1)$, $d(p_2)$, and $d(p_1, p_2)$, where $p_1$ and $p_2$ are the two parton momenta in the $\Phi_2$ event. If $d(p_i)$ is smallest, we can think of $p_i$ being clustered with one of the beams, and then $\vec n_J = \vec p_j / \abs{\vec p_j}$ with $j\neq i$. If $d(p_1, p_2)$ is the smallest, we can think of $p_1$ and $p_2$ being clustered together, and then $\vec n_J = (\vec p_1 + \vec p_2)/\abs{\vec{p}_1 + \vec{p}_2}$. It is not hard to show that this minimizes $\Tau_1$ and its final value is determined by the minimum metric as
\begin{align}
\Tau_1(\Phi_2) = \sum_{k=1}^2 (E_k - \abs{p_k^z}) + \min\bigl\{d(p_1), d(p_2), d(p_1, p_2) \bigr\}
\,.\end{align}

\subsubsection{The $\Tau_0$ spectrum at NNLL$'$ from SCET}
\label{subsec:SCETResummation}

The all-orders parton-level factorization theorem for $0$-jettiness (aka beam thrust) is given by~\cite{Stewart:2009yx, Stewart:2010pd}
\begin{align}
\label{eq:Tau0Factorization}
\frac{\df \sigma^{\rm SCET}}{\df \Phi_0 \df \Tau_0}
&=
\sum_{ij} \frac{\df\sigma_{ij}^B}{\df\Phi_0} H_{ij} (Q^2, \mu) \int\!\df t_a\, \df t_b \, B_i (t_a, x_a, \mu)
\nn \\ & \quad \times
B_j (t_b, x_b, \mu) \, S\Bigl(\Tau_0 - \frac{t_a + t_b}{Q}, \mu \Bigr)
\,.\end{align}
Here, $\df\sigma_{ij}^B/\df\Phi_0$ is the Born cross section for the $ij \to Z/\gamma^* \to \ell^+ \ell^-$ hard scattering. The hard function $H_{ij}(Q)$ contains the corresponding Born and virtual squared matrix elements, and the sum runs over all possible $q\bar q$ pairs $ij = \{u\bar u, \bar u u, d\bar d, \bar d d, \ldots\}$. The $B_i(t, x)$ are inclusive (anti)quark beam functions~\cite{Stewart:2009yx}, with the momentum fractions $x_{a,b}$  given in terms of the total rapidity $Y$ and invariant mass $Q \equiv m_{\ell^+\ell^-}$ of the vector-boson final state by
\begin{equation}
x_a = \frac{Q}{E_{\rm cm}}\, e^{Y}
\,, \qquad x_b = \frac{Q}{E_{\rm cm}}\, e^{-Y}
\,.\end{equation}
The beam functions are computed perturbatively in terms of standard PDFs $f_j$, schematically $B_i = \sum_j \mathcal{I}_{ij}\otimes f_j$. Finally, $S(k)$ is the quark hemisphere soft function for beam thrust.

Equation~\eqref{eq:Tau0Factorization} is derived using Soft Collinear Effective Theory (SCET)~\cite{Bauer:2000ew, Bauer:2000yr, Bauer:2001ct, Bauer:2001yt, Bauer:2002nz}. Its key feature is that each of $H$, $B$, $S$ only depend on a single characteristic scale. Therefore, there are no large logarithms in their perturbative expansion when each is evaluated at its own characteristic scale, which is given by
\begin{align}
\label{eq:canonicalScales}
\mu_H = Q\,, \quad \mu_B = \sqrt{Q \Tau_0}\,, \quad \mu_S = \Tau_0
\,.\end{align}
In \eq{Tau0Factorization}, all ingredients have to be evaluated at a common scale $\mu$, which is arbitrary and whose dependence exactly cancels between the different functions. The renormalization group evolution in the effective theory is then used to evolve each function from its own scale to $\mu$. This gives the resummed $\Tau_0$ spectrum used in \eq{fulllyDiffRes} as
\begin{align} \label{eq:resummedspectrum}
\frac{\df \sigma^{\rm NNLL'}}{\df \Phi_0 \df \Tau_0}
&= \sum_{ij} \frac{\df \sigma_{ij}^B}{\df \Phi_0} H_{ij}(Q^2, \mu_H)\, U_H(\mu_H, \mu)
\nn\\
& \quad 
\times \bigl[ B_i (x_a, \mu_B) \otimes U_B(\mu_B, \mu) \bigr]
\nn\\
& \quad
\times \bigl[ B_j (x_b, \mu_B) \otimes U_B(\mu_B, \mu) \bigr]
\nn\\
& \qquad
\otimes \bigl[ S(\mu_S) \otimes U_S(\mu_S, \mu) \bigr]
\,,\end{align} 
where we have now written the convolutions between the different functions in schematic form. 

The RGE factors $U_X(\mu_X, \mu)$ resum the large logarithmic terms, and the functions evaluated at their own scale provide the boundary conditions for the evolution. At NNLL$'$ accuracy, we need the boundary conditions at 2-loop order, and the evolution to 3(2)-loops in the cusp (noncusp) anomalous dimensions. All required expressions are known in the literature~\cite{Idilbi:2006dg, Becher:2006mr, Stewart:2010qs, Monni:2011gb, Kelley:2011ng, Hornig:2011iu, Gaunt:2014xga, Gaunt:2014cfa, Kang:2015moa, Gaunt:2015pea} and we do not reproduce them here.

\begin{figure}[t!]
\includegraphics[width=\columnwidth]{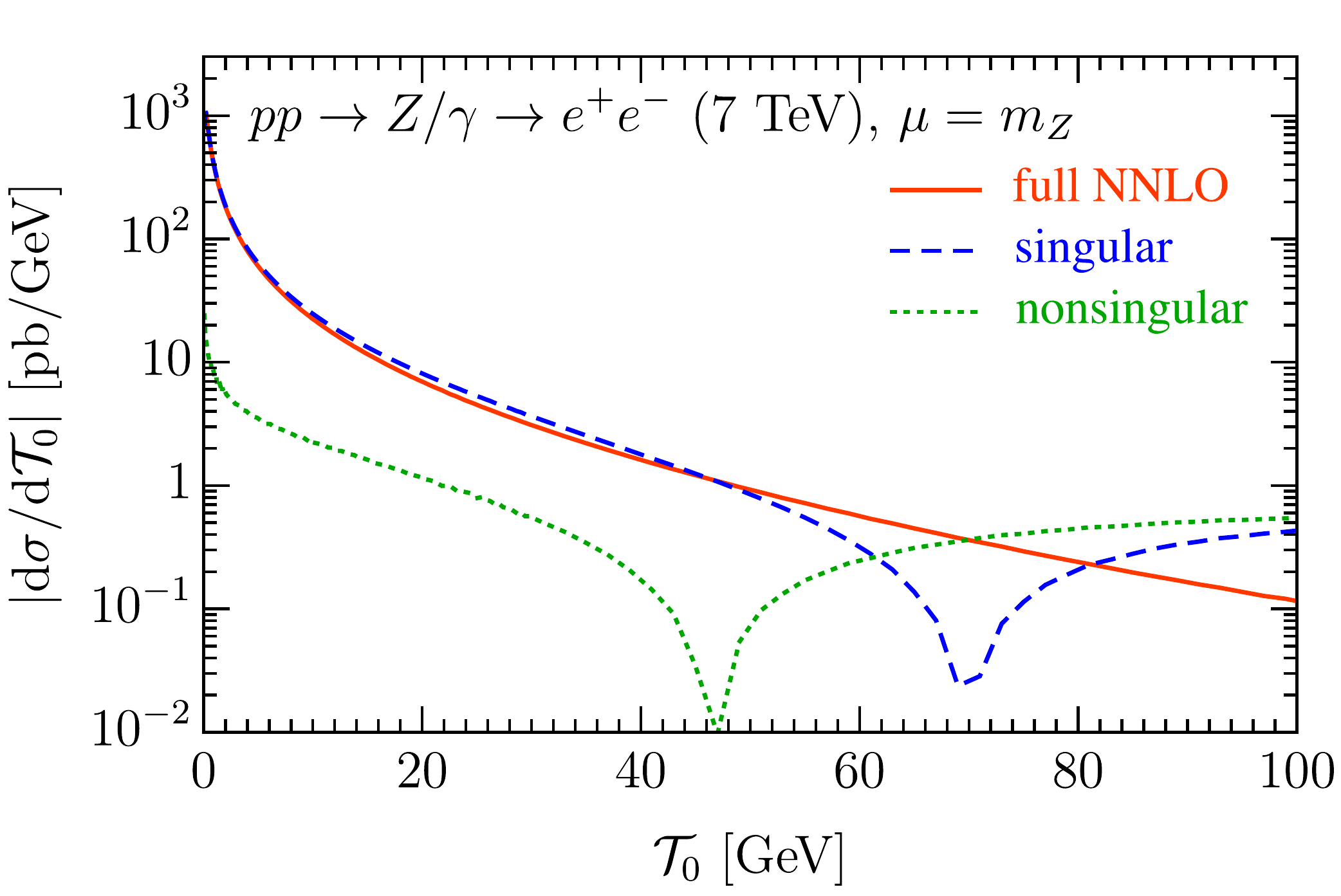}%
\vspace{-2ex}
\caption{Comparison of the full, singular, and nonsingular $\Tau_0$ distribution at NNLO$_0$.}
\label{fig:Tau0pieces}
\end{figure}

In practice, the canonical scales in \eq{canonicalScales} are appropriate in the resummation region, where the singular corrections dominate. In \fig{Tau0pieces}, we compare the singular, nonsingular, and full results for the $\Tau_0$ distribution at NNLO$_0$. We can see that up to $\Tau_0 \lesssim 50\GeV$ the singular dominate and the nonsingular are suppressed by an order of magnitude or more. On the other hand, in the FO region for larger values of $\Tau_0\sim Q$, the resummation must be turned off since here the singular terms being resummed become meaningless and there are large cancellations between the singular and nonsingular terms, which must be preserved to reproduce the correct FO result. This is clearly visible in \fig{Tau0pieces} for $\Tau_0 \gtrsim 80\GeV$, where the magnitude of the singular and nonsingular are larger than the full result (the singular turn negative at the dip at $\Tau_0\simeq 70\GeV$). For this reason the resummation must be turned off in this region. This is done by taking all scales to be equal to the common FO scale $\mu_S = \mu_B = \mu_H = \mu_\FO$. A smooth transition between the canonical and FO limits is achieved by using profile scales~\cite{Ligeti:2008ac, Abbate:2010xh}, where $\mu_B$ and $\mu_S$ are smooth functions of $\Tau_0$, interpolating between the exact canonical scaling in \eq{canonicalScales} and the FO scale. The uncertainties from the resummation and the transition between resummation and FO can be estimated by choosing different sets of profile scales, which provide a sensible variation around the central scale choice. In other words, for each event we perform the calculation of its MC cross section with a different set of profile scales, producing different weights for the event, and the variation in these define our event-by-event uncertainties as described next.

Profile scales and their variations are used in many SCET calculations and have been shown to provide reliable perturbative uncertainty estimates for resummed and matched predictions for observables in a variety of contexts, see e.g.\ Refs.~\cite{Abbate:2010xh, Berger:2010xi, Stewart:2013faa, Jouttenus:2013hs, Gangal:2014qda, Neill:2015roa}. For our central profiles we take
\begin{align} \label{eq:centralscale}
\mu_H &= \mu_\FO
\,, \nn \\
\mu_S(\Tau_0) & = \mu_\FO f_{\rm run}(\Tau_0/Q)
\,, \nn \\
\mu_B(\Tau_0) &=  \mu_\FO \sqrt{f_{\rm run}(\Tau_0/Q)}
\,,\end{align}
where the common profile function $f_{\rm run}(x)$ is as in Ref.~\cite{Stewart:2013faa},
\begin{align}
f_{\rm run}(x) &=
\begin{cases} x_0 \bigl[1+ (x/x_0)^2/4 \bigr] & x \le 2x_0\,,
 \\ x & 2x_0 \le x \le x_1\,,
 \\ x + \frac{(2-x_2-x_3)(x-x_1)^2}{2(x_2-x_1)(x_3-x_1)} & x_1 \le x \le x_2\,,
 \\  1 - \frac{(2-x_1-x_2)(x-x_3)^2}{2(x_3-x_1)(x_3-x_2)} & x_2 \le x \le x_3\,,
 \\ 1 & x_3 \le x\,.
\end{cases}
\label{eq:frun}
\,,\end{align}
which has strict canonical scaling below $x_1$ and the resummation is turned off above $x_3$. From \fig{Tau0pieces} we choose for our central scale the parameters
\begin{align} \label{eq:TauBprofile}
\mu_\FO &= Q
\,, \qquad
x_0 = 2.5\GeV/Q
\,, \nn \\
\{x_1,x_2, x_3\} &= \{0.2, 0.5, 0.8\}
\,.\end{align}
For the FO variations we vary $\mu_\FO$ up and down to $2Q$ and $Q/2$, taking the maximal absolute deviation in the results from the central value as the FO uncertainty. For our resummation uncertainties, we use the profile variations designed for $\Tau$-like resummation variables from Ref.~\cite{Gangal:2014qda}, which give us four independent up and down variations for $\mu_B$ and $\mu_S$ that probe deviations from the canonical scaling but never violate the parametric scaling $\mu_B^2 \sim \mu_S\mu_H$ and turn off beyond $x_3$. In addition, we include two more profiles where we vary all $x_i$ transition points by $\pm0.05$, giving us a total of 6 resummation profile variations. We then take the maximal absolute deviation in the result from the central value among all 6 profiles as the resummation uncertainty. The total perturbative uncertainty is obtained by adding the FO and resummation uncertainties in quadrature. Note that the profile scales depend on the underlying $\Phi_0$ point through the value of $Q$, but we choose them to be independent of $Y$.

The same resummation formalism can be used to obtain an expression for the resummed cumulant of the $\Tau_0$ spectrum in \eq{fulllyDiffRes}
\begin{align}
\label{eq:derivativeRelation}
\frac{\df \sigma^{\rm NNLL'}}{\df \Phi_0}(\Tau_0^\cut) = \int_0^{\Tau_0^\cut} \!\!\! \df \Tau_0 \, \frac{\df \sigma^{\rm NNLL'}}{\df \Phi_0 \df \Tau_0}
\,.\end{align}
Here, we first integrate the resummed expression for the $\Tau_0$ distribution in \eq{resummedspectrum}, and then choose the scales using the same profile scales but as a function of $\Tau_0^\cut$. In particular, in the resummation region, the canonical values are
\begin{align}
\mu_H = Q\,, \quad \mu_B = \sqrt{Q \Tau_0^\cut}\,, \quad \mu_S = \Tau_0^\cut
\,.\end{align}
Since the operations of integrating the factorization theorem and choosing the scales do not commute with one another, the resulting expression for the cumulant is not exactly equal to the integral of the $\Tau_0$ spectrum, where the scales would be chosen before performing the integration.  The resulting differences are formally always of higher order, beyond NNLL$'$ in our case (see e.g.\ Ref.~\cite{Almeida:2014uva} for a detailed discussion). However, they can be numerically relevant, especially if one aims at preserving the NNLO${_0}$ total cross section as accurately as possible.

In principle, one could enforce the equality between the cumulant and the integral of the spectrum in \eq{derivativeRelation} exactly, by simply defining the $\Tau_0$ spectrum as the derivative of the $\Tau_0^\cut$ cumulant. However, this can give rise to unreliable uncertainty estimates in the spectrum as well as a poor description of the region of $\Tau_0$ where the resummation is turned off and FO predictions are valid~\cite{Abbate:2010xh}.  In \geneva we circumvent these problems by adding to the $\Tau_0$ spectrum the quantity
\begin{align} 
\label{eq:higherorder}
\kappa(\Tau_0) \left[ \frac{\df}{\df \Tau_0} \frac{\df \sigma^{\rm NNLL'}}{\df \Phi_0}(\Tau_0, \mu_h)
 - \frac{\df \sigma^{\rm NNLL'}}{\df \Phi_0 \df \Tau_0} (\mu_h) \right]
\,,\end{align}
which is explicitly beyond NNLL$'$ order.  The profile scales $\mu_h$ are chosen to turn off the resummation somewhat earlier than the profile scales for the rest of the resummed calculation, in order to preserve the accurate spectrum prediction in the tail of the distribution.  Furthermore, $\mu_h$ and its derivative are smooth functions of $\Tau_0$, which is needed because the difference of resummed terms in \eq{higherorder} is proportional to $\df\mu_h / \df\Tau_0$~\cite{Almeida:2014uva}.  The function $\kappa(\Tau_0)$ is a function that smoothly interpolates from a constant value for $\Tau_0 \ll Q$ to zero at the point where $\mu_h (\Tau_0) = Q$, and further reduces the effects of the higher order correction terms in the tail of the distribution.  By adjusting the constant value at small $\Tau_0$ within $\ord{1}$ factors, the overall size of these terms can be adjusted (in practice this constant value is approximately 2).

These higher-order terms may be used to capture the correct FO scale uncertainties in the inclusive cross section upon integrating over the full spectrum.  We use the same profile for each of the scale variation described above, but scaled by the value of $\mu_\FO/Q$ (so 1 for the central scale and resummation scale variations, 2 for the up FO scale variation, and 1/2 for the down FO scale variation).  By adjusting the constant value of $\kappa(\Tau_0 \to 0)$ separately for each $\mu_\FO / Q$ value, we can correctly obtain the total integrated FO cross section for each FO scale variation.

Technically, each of the resummation scale variations should yield strictly the same total cross section, such that the resummation uncertainties vanish for inclusive observables. However, in practice, the profile scale variations are primarily meant and designed to capture the resummation uncertainties in the peak and transition regions. Implementing a constraint that they all reproduce exactly the same numerical integral of the resummed spectrum as the central profile is very challenging, especially since the resummation uncertainties in the peak region can be sizeable, while the inclusive cross section uncertainties are very small. (Alternatively, we could separately finetune the $\kappa(\Tau_0)$ function for each of the different resummation profiles.) Instead, what we will do is to simply drop the resummation uncertainties whenever we consider fully inclusive observables.

\subsubsection{NNLO nonsingular corrections}
\label{subsec:GetNNLO}

As discussed in \subsec{01jetsep}, the fact that we are resumming the $\Tau_0$ spectrum and cumulant to NNLL$'$ accuracy implies that there are no singular terms in the matching to the NNLO calculation. Therefore, \geneva directly computes the fully-differential inclusive 0-jet cross section at NNLO, up to power corrections in $\Tau_0^\cut$. The power-suppressed missing terms are precisely the $0$-jet nonsingular contributions in \eq{01nons}, which we can write as
\begin{align} \label{eq:Sigmanons}
\frac{\df\sigma_0^\nons}{\df\Phi_{0}}(\Tau_0^\cut)
&= \bigl[ \as f_1(\Tau_0^\cut, \Phi_0) + \as^2 f_2(\Tau_0^\cut, \Phi_0) \bigr]\Tau_0^\cut
\,, \nn \\
\Sigma_\nons(\Tau_0^\cut) &= \int\!\df\Phi_0\,\frac{\df\sigma_0^\nons}{\df\Phi_{0}}(\Tau_0^\cut)
\,.\end{align}
The functions $f_k(\Tau_0^\cut, \Phi_0)$ can still have divergences as $\Tau_0^\cut\to 0$, but these divergences are at most logarithmic. Thus, these corrections vanish in the limit $\Tau_0^\cut\to 0$, and for small enough $\Tau_0^\cut$ they can be neglected, allowing to obtain NNLO accuracy without doing the full NNLO calculation (which would determine the function $f_2(\Phi_0)$).

\begin{figure}[t!]
\includegraphics[width=\columnwidth]{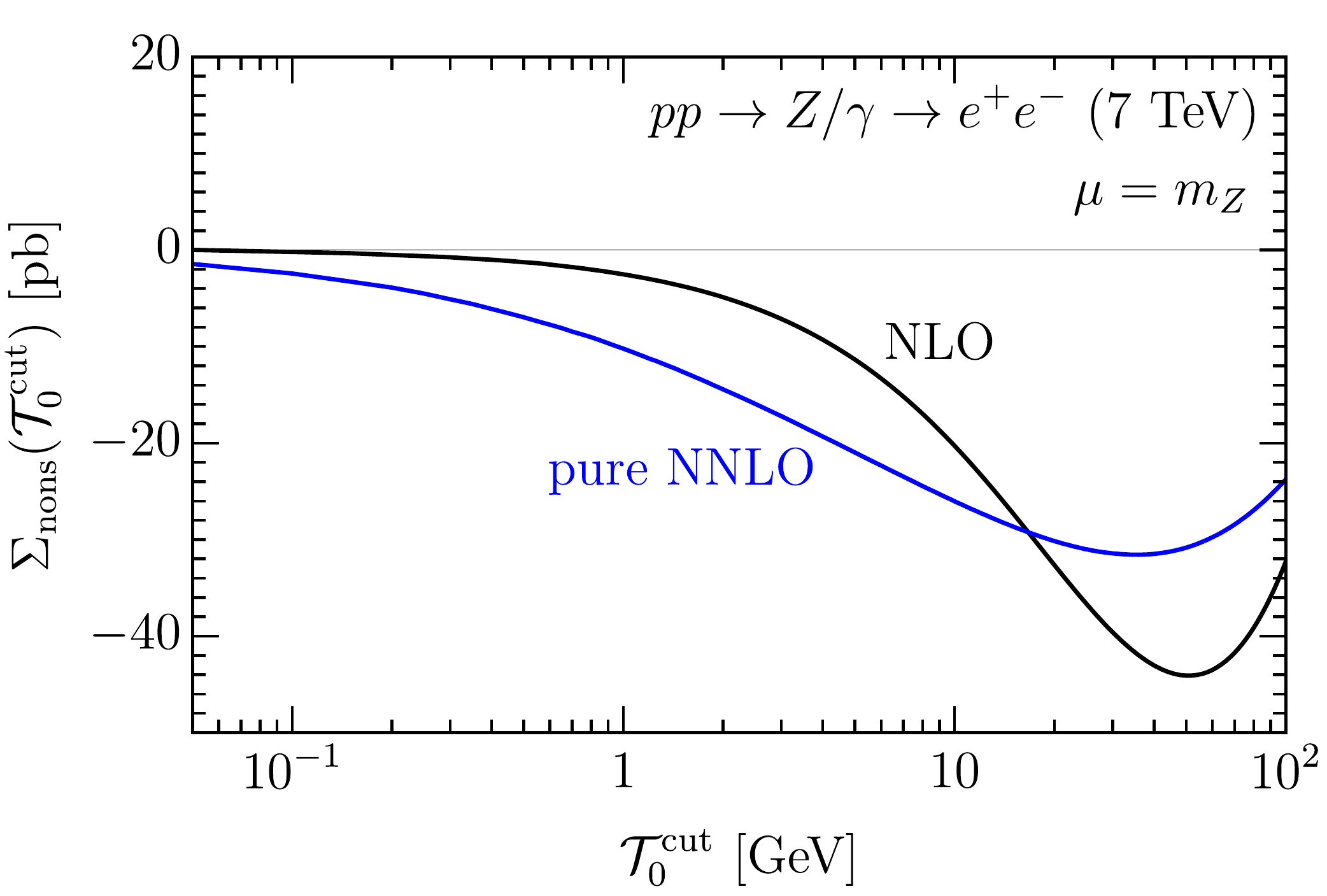}%
\vspace{-2ex}
\caption{The nonsingular cumulant for $0$-jettiness, as a function of $\Tau_0^\cut$. The black (blue) curve shows the NLO (pure NNLO) nonsingular cumulant, coming from the $f_1$ and $f_2$ terms in \eq{Sigmanons}.}
\label{fig:Tau0nonscumulant}
\end{figure}

We show the size of the total correction terms $\Sigma_\nons(\Tau_0^\cut)$ in \fig{Tau0nonscumulant} for the NLO and pure NNLO contribution. At our default value of $\Tau_0^\cut = 1\GeV$ these terms are very small but also not completely negligible. In the current implementation we include the full NLO term, i.e., the $f_1(\Phi_0, \Tau_0^\cut)$ piece in $\df\sigma_0^\nons(\Tau_0^\cut)$ by performing the corresponding $0$-jet NLO calculation on-the-fly. For the NNLO term, we neglect the $\Phi_0$ dependence and only include its total integral. Technically, this is done by obtaining the pure NNLO contribution to $\Sigma_\nons(\Tau_0^\cut)$ from the known dependence of the total cross section on $\Tau_0^\cut$ (i.e.\ analogous to how we obtain \fig{Tau0nonscumulant}). This correction is then included through a simple rescaling of $\dsigMC_0/\df\Phi_0(\Tau_0^\cut)$.

\subsubsection{The normalized splitting probabilities}
\label{subsec:normalizedSplitting}

In \eq{cPnorm} we introduced a normalized splitting probability $\cP(\Phi_1)$, which is used to make the resummed $\Tau_0$ spectrum fully-differential in $\Phi_1$ in \eq{fulllyDiffRes}. A similar splitting probability $\cP(\Phi_2)$ is used in \eq{Tau1resum} to make the $\Tau_1$ resummed term fully-differential in $\Phi_2$. As mentioned before, the splitting probabilities are normalized,
\begin{align}
\int \! \frac{\df \Phi_{N+1}}{\df \Phi_N \df\Tau_N} \, \cP(\Phi_{N+1}) = 1
\,,\end{align}
and provide the dependence on the two extra emission variables required besides $\Tau_N$ to define a full splitting from $\Phi_N \to \Phi_{N+1}$. The two extra variables are chosen as usual as an energy ratio $z$ and an azimuthal angle $\phi$. In the collinear and soft limit, the $z$ variable denotes the ratio $z = E_M / E_S$ for a splitting $M \to D S$, where a mother particle $M$ splits into a daughter particle $D$ and a sister particle $S$. For ISR splittings, the daughter particle is the final-state particle, while for FSR splittings it is always the gluon for $q \to qg$ splittings, the quark for  $g \to q\bar{q}$  ones and the softer gluon for $g\to gg$ splittings.

We first need to decide which two particles in $\Phi_{N+1}$ are interpreted as arising from the splitting and which are then labeled with $D$ and $S$. Here we have to choose the two particles which set the value of $\Tau_N$, which are the two particles that are closest according to the $N$-jettiness clustering metric, as described in \eq{NJettinessMetric}.

The normalized splitting probability is then given by
\begin{align}
\cP(\Phi_{N+1}) = \frac{p_{\rm sp}(z, \phi)}{\sum_{\rm sp} \int_{z_{\rm min}(\Tau_N)}^{z_{\rm max}(\Tau_N)} \!\! \df z \df \phi \, p_{\rm sp}(z, \phi)} \frac{\df \Phi_{N} \df \Tau_N \df z \df \phi}{\df \Phi_{N+1}}
\,,\end{align}
where the splitting function $p_{\rm sp}(z, \phi)$ depends on the chosen splitting $\rm sp$ which as explained above depends on the phase-space point ${\rm sp} \equiv {\rm sp}(\Phi_{N+1})$. With this definition, the splitting probability is indeed normalized:
\begin{align}
\label{eq:Pnorm}
&\int \! \frac{\df\Phi_{N+1}}{\df \Phi_{N} \df \Tau_N}\cP(\Phi_{N+1})
\nn\\
& \qquad = \sum_{\rm sp} \int \! \df z \df \phi \, \frac{p_{\rm sp}(z, \phi)}{\sum_{\rm sp} \int_{z_{\rm min}}^{z_{\rm max}} \!\! \df z \, p_{\rm sp}(z, \phi)}
\nn\\
& \qquad= 1
\,.\end{align}

The splitting functions for FSR and ISR are chosen to be 
\begin{align}
p_{{\rm sp}\in \rm FSR}(z, \phi) &= {\rm AP}_{\rm sp}(z, \phi)
\,,\nn\\
p_{{\rm sp}\in \rm ISR}(z, \phi) &= \frac{f(x_M / z, \mu)}{f(x_S, \mu)}{\rm AP}_{\rm sp}(z, \phi)
\,,\end{align}
with ${\rm AP}_{\rm sp}(z, \phi)$ denoting the unregularized Altarelli-Parisi splitting function, and $f(x, \mu)$ the standard parton distributions function evaluated at the momentum fraction $x$ and the factorization scale $\mu$. The variables $x_M$ and $x_S$ denote the momentum fractions of the mother and sister particles in the ISR splitting $M \to D S$.

\subsubsection{$\Tau_1$ resummation}
\label{subsec:U1Sudakov}

The Sudakov factor $U_1(\Tau_1^{\max}, \Tau_1)$ in \eq{Tau1resum} resums the $\Tau_1$ dependence to LL accuracy. We use the expression that is obtained from the $\Tau_1$ factorization theorem~\cite{Stewart:2010tn} analogous to the one given for $\Tau_0$ in \eq{Tau0Factorization}, where all ingredients are calculated at tree level, and the running is performed at LL only. The $\ord{\as}$ expansion of this Sudakov factor has the simple expression
\begin{align}
U_1^{(1)}(\Tau_1^{\rm max}, \Tau_1) = - \frac{\as(\Tau_1^{\rm max})( 2 C_F + C_A)}{2\pi}  \, \ln^2\frac{\Tau_1}{\Tau_1^{\rm max}}
\,.\end{align}

The above minimal LL resummation is sufficient for our current purposes, where we focus on $0$-jet resummation variables and inclusive observables and where we primarily need it to have a proper matching to the parton shower. A more precise description of $1$-jet resummation variables will require an improved $\Tau_1$ resummation, which we leave for future work. Extending the $\Tau_1$-resummation beyond LL accuracy is of course possible. One can also consider including the resummation through a \powheg-like Sudakov factor, where the ratio of the real matrix element $B_2$ over the born $B_1$ would be exponentiated. We have verified that our underlying NLO$_1+$LL$_1$ matched calculation reasonably agrees with a \powheg implementation of $V+1$ jet~\cite{Alioli:2010qp} in regions where the $0$-jet resummation is not relevant.

\subsubsection{NLO$_1$ calculation and phase-space map}
\label{subsec:PSMaps}

The last point addressed in this section concerns the phase-space map $\Phi_1^\Tau(\Phi_2)$ that is used to project a partonic phase-space point $\Phi_2$ onto the $\Phi_1$ point in the NLO$_1$ calculation, see \eqs{NLO1}{Phi1Nons}. It is also used in the $\Tau_1$ resummation in \eq{Tau1resum}. Just requiring to have some IR-safe map as usual for fixed-order calculations (like for example the FKS map $\Phi_1^C(\Phi_2)$ used in the $C_2$ subtraction term in \eq{Phi1Nons}) is not enough anymore.

As discussed around \eq{Tau0map}, we have the additional constraint that the $\Phi_1^\Tau$ map must preserve the value of $\Tau_0$ in order for the NLO$_1$ calculation to be consistent with the $\Tau_0$ resummation. This property was also important for the $1/2$-jet separation to preserve the $\Tau_0$ resummation.

Technically, the only real-emission integral that we have to perform as part of the NLO$_1$ calculation is the integral in \eq{Phi1Nons}. We perform this integral numerically by Monte-Carlo integration over the radiation phase space for a given $\Phi_1$ event. The $B_2$ and $C_2$ terms in \eq{Phi1Nons} are allowed to use different maps, since all IR-safe maps agree in the IR-singular limit. However, we have to be able to invert both maps, i.e., we have to reconstruct all $\Phi_2$ points that would project to the given $\Phi_1$ point. We then parametrize the radiation phase space for both maps in terms of a common set of radiation variables, which are then Monte-Carlo sampled. While the usual FKS maps are designed to be invertible, the invertibility turns out to be a very nontrivial requirement on the $\Phi_1^\Tau(\Phi_2)$ map.

Formally, we only need to preserve $\Tau_0$ for the singular terms, i.e., we can use any map which preserves a variant of $\Tau_0$ with the same singular structure to the order we want to carry out the resummation, while any nonsingular difference is captured by the nonsingular corrections. Therefore, to enable us to invert the map, we hold a recursive definition of $\Tau_0$ fixed, which is effectively defined by the action of the map,
\begin{equation}
\Tau_0^{\rm FR}(\Phi_2) = \Tau_0[\Phi_1^\Tau(\Phi_2)]
\,.\end{equation}

The map itself is constructed as follows: We first cluster (by simply adding/subtracting four-momenta) the two partons in $\Phi_2$ that set the $\Tau_1$ value, i.e., that are closest according to the $N$-jettiness metric. If the two final-state partons are clustered (FSR clustering), then this yields a $\Phi_1$ point with an off-shell final-state momentum. If one of the final-state partons in $\Phi_2$ is closest to a beam direction (ISR clustering), then this yields a $\Phi_1$ point with one of the incoming momenta off shell (and off axis). Next, we compute $\Tau_0$ on this off-shell $\Phi_1$ point. Finally, we construct an on-shell $\Phi_1$ point that has the same $\Tau_0$ value and the same $q^\mu$ of the vector boson. With $q^\mu$ and $\Tau_0$ given, the resulting $\Phi_1$ point is fully determined by four-momentum conservation and on-shell conditions.
Hence, by construction, the map also preserves the momentum of the vector boson $q^\mu(\Phi_2) = q^\mu(\Phi_1^\Tau(\Phi_2))$.
It is easy to see that this map is IR safe, and it is sufficiently simple that it can be inverted.

For all ISR clusterings as well as FSR clusterings with both partons in the same hemisphere (defined by the vector-boson rapidity) this definition of $\Tau_0^{\rm FR}$ is identical to the exact value of $\Tau_0(\Phi_2)$. The only case where it differs is for FSR clusterings where the two final-state partons are in opposite hemispheres.
In the $\Tau_0$-singular limit, this can only happen for soft emissions, and hence will in principle affect the constant terms in the two-loop soft function. (Collinear emissions in the beam functions always correspond to ISR clusterings.) However, this can only happen in a small region of phase space where the soft emitter was already accidentally close in rapidity to the vector boson. We therefore expect this to be a small effect, which we have verified numerically. We can therefore safely ignore it for our resummation and simply take it into account as part of the nonsingular matching corrections.

Finally, we need to comment on the $\Theta^\Tau(\Phi_2)$ appearing in \eq{Phi1TauProj}. It is easy to see that an on-shell massless $\Phi_1$ point must satisfy $\Tau_0 \leq \abs{\vec{q}_T}$. However, there can be $\Phi_2$ points that can have a larger $\Tau_0$ value, and hence such $\Phi_2$ points cannot be projected onto a massless $\Phi_1$ point. The $\Theta^\Tau(\Phi_2)$ is defined to be $1$ for points that can be projected and $0$ otherwise. Equivalently, these points also cannot be reached by the inverse map from splitting a $\Phi_1$ point, and so are never part of the integral in \eq{Phi1Nons}. Since the map is IR safe, the nonprojectable $\Phi_2$ region is nonsingular and is included in \eq{Phi2Nons} by the $1 - \Theta^\Tau(\Phi_2)$ constraint.

\section{Interfacing with a Parton Shower}
\label{sec:shower}

In the previous section we have given all required formulas for the jet cross sections $\dsigMC_0$, $\dsigMC_1$ and $\dsigMC_{\ge 2}$. As discussed in detail  there, these jet cross sections include the contributions of higher multiplicity phase space points, as long as the jet resolution variable for these points in phase space is below the $\Tau_k^\cut$ value. In Table~\ref{tab:ShowerConstraints} we summarize how the phase-space points of different multiplicities contribute to the given jet cross section. Only the first three columns were necessary for the partonic calculation of the previous section. The last column shows instead the the constraints to be imposed on events with higher multiplicities. We will detail how these affect the development of the parton shower below. 
  
\begin{table*}[t]
\begin{tabular}{l || c | c | c || c }

\hline\hline
 & $\Phi_0$ & $\Phi_1$ & $\Phi_2$ & $\Phi_N$
\\ \hline
$\dsigMC_0/\df\Phi_0$
& 
{\rm All} & 
$\theta_{\Tau_0}(\Phi_1)\, {\rm and} \, \theta_{\rm map}(\Phi_0; \Phi_1)$ &
$\theta_{\Tau_0}(\Phi_2)$ &
$\theta_{\Tau_0}(\Phi_N)$
\\ \hline
$\dsigMC_{1}/\df\Phi_{1}$
& -- &
$\overline\theta_{\Tau_0}(\Phi_1)\, {\rm or} \, \overline\theta_{\rm map}(\Phi_1)$ &
$\overline\theta_{\Tau_0}(\Phi_2)\, {\rm and} \, \theta_{\Tau_1}(\Phi_2) \, {\rm and} \, \theta_{\rm map}(\Phi_1; \Phi_2)$ &
$\overline\theta_{\Tau_0}(\Phi_N)\, {\rm and} \, \theta_{\Tau_1}(\Phi_N)$
\\ \hline
$\dsigMC_{\geq 2}/\df\Phi_{2}$
& 
-- & 
-- &
$\overline\theta_{\Tau_0}(\Phi_2)\, {\rm and} \, \left[\overline\theta_{\Tau_1}(\Phi_2) \, {\rm or} \, \overline\theta_{\rm map}(\Phi_2)\right]$ & 
$\overline\theta_{\Tau_0}(\Phi_N)\, {\rm and} \, \overline\theta_{\Tau_1}(\Phi_N)$
\\ \hline\hline
\end{tabular}
\caption{Table showing how the different partonic phase-space points are contributing to the different jet multiplicities. Here we have defined $\theta_{\Tau_N}(\Phi_M)  \equiv \theta[ \Tau_N(\Phi_M) < \Tau_N^\cut ]$,
$\overline\theta_{\Tau_N}(\Phi_M)  \equiv \theta[ \Tau_N(\Phi_M) > \Tau_N^\cut ]$, 
$\theta_{\rm map}(\Phi_0; \Phi_1) \equiv [ \Phi_1 \, \mbox{projects onto} \, \Phi_0 ]$, 
$\theta_{\rm map}(\Phi_1; \Phi_2) \equiv [ \Phi_2 \, \mbox{projects onto} \, \Phi_1 ]$, $\overline\theta_{\rm map}(\Phi_1) \equiv [ \Phi_1 \, \mbox{does not project onto any} \, \Phi_0 ]$, 
$\overline\theta_{\rm map}(\Phi_2) \equiv [ \Phi_2 \, \mbox{does not project onto any} \, \Phi_1 ]$.
}
\label{tab:ShowerConstraints}
\end{table*}

The purpose of the parton shower is to make the calculation fully differential in the higher multiplicities, which can be viewed as filling the exclusive 0- and 1-jet bins with radiation, as well as adding extra jets to the inclusive 2-jet multiplicity. A parton shower acts on a partonic event by adding extra radiation in a recursive and unitary fashion. If the evolution variable of the parton shower were chosen to be $N$-jettiness~\footnote{Here and in the following we treat $\Tau_N$ as the jet resolution scale of a \emph{single} emission. Since $\Tau_N$ is a global variable that depends on all  emissions, this warrants some explanation. What we intend here proceeds from the condition that  for massless particles  $\Tau_N(\Phi_{N+1})$  is always the total plus component of the closest pair of particles, using  e.g. the metric defined in \eq{NJettinessMetric}, relative to the direction of the sum of their momenta.  Moreover,  $\Tau_N \to 0$ when  $\Phi_{N+1} \to \Phi_{N}$. Near the singular limit, where the shower is important and a good approximation of the underlying physics, one can thus assume that the direction  of the axes entering the definition of $\Tau_N$  is aligned to the direction of the $N$ hard partons. In this limit, a singular emission does not change these directions. Therefore, the value of $\Tau_N$ does in fact represent the hardness of the emission and can be used as evolution variable.}, a single emission would be given by
\begin{align}
\label{eq:ShowerOneEmission}
S^{(1)}\left[\df \sigma_N\right]
&\equiv
\df \sigma_N \biggl [ U_N(\Tau_{N}^{\max}, \Lambda) + U_N'(\Tau_N^{\max}, \Tau_N)
\\\nn & \quad \times
\frac{\df \Phi_{N+1}}{\df \Phi_N}\, \cP(\Phi_{N+1})\, \theta(\Tau_{N}^{\max} > \Tau_N > \Lambda) \biggr]
\,,\end{align}
where $\Tau_N^{\max} \sim {\cal O}(\Tau_{N-1})$ is the maximum value of $\Tau_N$ that can be reached given the value $\Tau_{N-1}$  that characterizes the hardness of the configuration before the emission and $\Lambda$ is the shower cutoff. This means that the parton shower either keeps the $N$-body event unchanged with the no-emission probability determined by the Sudakov factor $U_N(\Tau_N^{\max}, \Lambda)$, or it adds an extra emission according to the probability
\begin{align}
U_N'(\Tau_{N}^{\max}, \Tau_N) \, \cP(\Phi_{N+1})
\,,\end{align}
which we define here in terms of the derivative of the Sudakov form factor and the normalized splitting probability given in \subsec{normalizedSplitting}.

From \eq{ShowerOneEmission} and using the analog of \eq{unitarity} it is clear that the parton shower is unitary.
Defining
\begin{align}
\label{eq:dfSigmaNp1Def}
\df \sigma_{N+1}^S
&\equiv \df \sigma_N \, U_N'(\Tau_{N}^{\max}, \Tau_N) \, \cP(\Phi_{N+1}) \,
\nn\\ & \quad \times
\frac{\df \Phi_{N+1}}{\df \Phi_{N}} \,  \theta(\Tau_{N}^{\max} > \Tau_N > \Lambda)
\,,\end{align}
the recursive shower can now be written as
\begin{align}
\label{eq:RecursiveShower}
S\bigl[\df \sigma_N  \bigr]
\equiv \df \sigma_N \, U_N(\Tau_{N}^{\max}, \Lambda) + S\bigl[ \df \sigma_{N+1}^S \bigr]
\,.\end{align}
Note that the definition of the parton shower requires a phase-space map. This is clear from \eq{dfSigmaNp1Def}, which depends on $\df \Phi_{N+1} / \df \Phi_N$.

The above equation assumes that the evolution variable of the parton shower is $N$-jettiness. \pythia8, which we interface to for showering, as well as any other parton showers currently existing, use a different evolution variables, e.g. the transverse momentum or the angular separation of the emission.  
However, one can imagine taking the output of any of these showers and recluster the partons according to the $N$-jettiness metric defined in~\eq{NJettinessMetric}. The resulting history of splittings is equal at LL with the recursive expression given in~\eq{RecursiveShower}. In the remaining discussion, we will work with \eqs{ShowerOneEmission}{RecursiveShower}, and it is understood that this is valid only after the aforementioned reclustering. 
Note that this construction implies a recursive definition of $N$-jettiness, which in general is not identical to the standard definition given in \subsec{TauNDefs}. It does agree, however, with the $\Tau_0^{\rm FR}$ definition discussed in \subsec{PSMaps}, which is used in \geneva.

The parton shower is not allowed to affect results of the jet cross sections at the accuracy they were calculated in the previous section. Starting from this condition, let us first consider the constraints we would have to impose on a parton shower that is strongly ordered in $N$-jettiness, as discussed above. Strong ordering in $\Tau_N$ means that the $\Tau_N$ value at each emission is much smaller than the $\Tau_{N}^{\max}$ upper limit determined by the $\Tau_{N-1}$ value of the previous emission. In this case, we would need to ensure that at each stage in the shower each partonic phase space point satisfies the constraints given in Table~\ref{tab:ShowerConstraints}.
For example, when showering a 0-jet event defined by $\Phi_0$, the phase space point $\Phi_1$ reached after the first emission must have $\Tau_0(\Phi_1) < \Tau_0^\cut$ and $\Phi_1$ must be projectable onto $\Phi_0$. Each subsequent emission only needs to satisfy $\Tau_0(\Phi_N) < \Tau_0^\cut$. When showering a 1-jet event defined by $\Phi_1$, the $\Phi_2$ point obtained after the first emission would need to have $\Tau_1(\Phi_2) < \Tau_1^\cut$, and $\Phi_2$ would have to be projectable onto $\Phi_1$ with the $\Tau_0$-preserving map discussed in \subsec{PSMaps}. This implies that the $\Phi_2$ point obtained after one emission has to satisfy $\Tau_0(\Phi_2) = \Tau_0(\Phi_1)$ as well as $q^\mu(\Phi_2) = q^\mu(\Phi_1)$, $q^\mu$ being the vector boson four-momentum. Each subsequent emission then only needs to satisfy $\Tau_1(\Phi_N) < \Tau_1^\cut$.

For a shower ordered in a different evolution variable, very similar restrictions hold, however the special conditions regarding the phase space maps are not required on the first emission, but rather on the emission with the largest value of the jet resolution scale, which can happen at a much later stage in the shower. Failing to properly account for this could affect the accuracy at which observables are predicted and ultimately spoil the leading-logarithmic terms and the color-coherence effects that are built into the shower Sudakov factors.

A solution to this problem is to ensure that the first emission has
indeed the largest value of the jet resolution scale. This can be
achieved by performing the first emission in \geneva, with a
probability obtained from a simple LL Sudakov factor and using a
splitting that preserves the map that was so carefully constructed. One
would then only need to ensure that subsequent emissions from the parton
shower have a lower jet resolution scale than the first, which can be
achieved by properly setting the starting scale for the showering and
applying standard veto techniques. 

The general idea of this approach is to split an $N$-jet differential cross section into two pieces
\begin{align}
&\frac{\dsigMC_{N\to N}}{\df\Phi_{N}} (\Tau_{N}^\cut; \Lambda_N)
= \frac{\dsigMC_{N}}{\df\Phi_{N}} (\Tau_{N}^\cut) \, U_N(\Tau_N^\cut, \Lambda_N)
\,, \\
&\frac{\dsigMC_{N \to N+1}}{\df\Phi_{N+1}} (\Tau_N > \Lambda_N, \Tau_N^\cut)
\nn\\ &\qquad
= \frac{\df}{\df \Tau_N} \left[\frac{\dsigMC_{N\to N}}{\df\Phi_{N}} (\Tau_{N}^\cut; \Tau_{N}) \right] \, \cP(\Phi_{N+1})
\nn\\ & \qquad \quad \times
\theta(\Tau_N^\cut > \Tau_N > \Lambda_N)
\,,\end{align}
using a Sudakov factor $U_N$ and splitting probability $\cP$ to define a branching and no-branching probability as in a unitary parton shower. The first term is now an exclusive $N$-jet cross-section with a jet resolution at $\Lambda_N$, which can be much smaller than the $\Tau_N^\cut$ value used in \geneva. This process can also be iterated further.

Applying this to our $\Phi_0$ and $\Phi_1$ events, and adding all emissions up to two partons, we find
\begin{widetext}
\begin{align}
\label{eq:analyticaShower0}
\frac{\dsigMC_{0}}{\df\Phi_{0}} (\Tau_{0}^\cut; \Lambda_0)
&= \frac{\dsigMC_{0}}{\df\Phi_{0}} (\Tau_{0}^\cut) \, U_0(\Tau_0^\cut, \Lambda_0)
\,,\\
\frac{\dsigMC_{1}}{\df\Phi_{1}} (\Tau_0 > \Lambda_0, \Tau_0^\cut, \Tau_1^\cut, \Lambda_1)
&= \frac{\df\sigma_{1}}{\df\Phi_{1}} (\Tau_0 > \Tau_0^\cut, \Tau_1^\cut) \, U_1(\Tau_1^\cut, \Lambda_1)
\nn\\
\label{eq:analyticaShower1}
& \quad
+ \frac{\df}{\df \Tau_0} \, \frac{\dsigMC_{0}}{\df\Phi_{0}} (\Tau_{0}^\cut; \Tau_{0}) \, \cP(\Phi_1) \, \theta(\Tau_0^\cut > \Tau_0 > \Lambda_0) \, U_1(\Tau_1^{\rm max}, \Lambda_1)
\,,\\
\frac{\dsigMC_{\geq 2}}{\df\Phi_{2}} (\Tau_0 > \Lambda_0, \Tau_{1}>\Lambda_1, \Tau_0^\cut, \Tau_1^\cut)
&= \frac{\dsigMC_{\geq 2}}{\df\Phi_{2}} (\Tau_0 > \Tau_0^\cut, \Tau_{1}>\Tau_{1}^\cut) 
\nn\\
\label{eq:analyticaShower2}
& \quad
+ \frac{\df}{\df \Tau_1} \, \frac{\dsigMC_{1}}{\df\Phi_{1}} (\Tau_0 > \Lambda_0, \Tau_0^\cut, \Tau_1^\cut, \Tau_1) \, \cP(\Phi_2) \,  \theta(\widetilde \Tau_1^{\max} > \Tau_1 > \Lambda_1)
\,.\end{align}
\end{widetext}
The $\widetilde \Tau_1^{\max}$ in the theta-function in the last line above is either $\Tau_1^\cut$ or $\Tau_1^{\rm max}$, depending on whether the derivative is on the first or the second term of~\eq{analyticaShower1}. Once the first emissions have been performed as described above, the only remaining constraint on the parton shower is what is given in the last column of Table~\ref{tab:ShowerConstraints}, namely the constraint on the jet resolution variable. In practice, we will only apply this to the $\Phi_1$ events, since as discussed below, the showering of the $\Phi_0$ events is not affecting the accuracy of the $\Tau_0$ distribution.

In the remainder of this section we give an argument why the parton shower acting on the jet events as described above does not affect the NNLL$'+$NNLO accuracy.
The $\Phi_0$ events include all phase-space points for which $\Tau_0 < \Tau_0^\cut$, and by definition \geneva is only predicting the normalization, and not the distribution in $\Tau_0$. Therefore, as long as we constrain the parton shower to only shower these events up to $\Tau_0^\cut$, the shower will exactly provide the missing $\Tau_0$ shape below $\Tau_0^\cut$ which is integrated over by \geneva. On the other hand, due to the unitarity constraint, the shower will not change the normalization of the events below  $\Tau_0^\cut$, which is thus still the NNLL$'+$NNLO$_0$ one predicted by \geneva.

By performing the first emission of our $\Phi_1$ events as described above, and taking $\Lambda_1$ as small as possible (we use a value of $0.1$~GeV), the remaining 1-jet cross section is only about $0.1\%$ of the total cross section.
The theoretical problem of how the shower affects the accuracy of this small contribution to the total cross section, which requires imposing all the constraints detailed above, can then be ignored for practical purposes.

What is therefore left to show is that the shower does not affect the $\Tau_0$ distribution when showering $\Phi_2$ events.
The $\Tau_0$ distribution of the original $2$-body phase space point is given by
\begin{align}
\frac{\df \sigma}{\df \Tau_0}
= \int\! \df \Phi_2\, \frac{\df \sigma_2}{\df \Phi_2} \, \delta[\Tau_0 - \Tau_0(\Phi_2)]
\,,\end{align}
while the expression after the first emission done by the shower is given by
\begin{align}
\frac{\df \sigma^S}{\df \Tau_0}
&= \int\! \df \Phi_2\, \frac{\df \sigma_2}{\df \Phi_2}\, U_2(\Tau_2^{\rm max}, \Lambda_2) \,
\delta[\Tau_0 - \Tau_0(\Phi_2)]
\nn \\ &\quad
+  \int\! \df \Phi_{3}\, \frac{\df \sigma^S_{3}}{\df \Phi_{3}} \, \delta[\Tau_0 - \Tau_0(\Phi_{3})]
\,.\end{align}
Clearly, thanks to the infrared safety of $\Tau_0$, in the soft and collinear limit of $\Phi_{3}$ one has
\begin{align}
\label{eq:Tau0Preserv}
\Tau_0(\Phi_3) = \Tau_0(\Phi_{2})
\,.
\end{align}
If the phase-space map of the parton shower were to preserve the value of $\Tau_0$, the relation~\eq{Tau0Preserv} would remain exact even away from the collinear and soft limits. The first emission done by the shower would then not change the $\Tau_0$ spectrum
\begin{align}
\frac{\df \sigma^S}{\df \Tau_0} = \frac{\df \sigma}{\df \Tau_0}
\,.\end{align}

If the phase-space map of the shower does not preserve $\Tau_0$, instead, a difference between these two values exists. We can assume this difference to be proportional to $\Tau_2(\Phi_{3})$, which measures the distance from the soft and collinear limits
\begin{align}
\Tau_0(\Phi_{2}) - \Tau_0(\Phi_{3}) &= a(\Phi_{3}) \, \Tau_2(\Phi_{3})
\,.\end{align}
Here $a(\Phi_3)$ is well behaved in the singular limit
of $\Phi_3$, because all the singular behavior is
incorporated in $\Tau_2(\Phi_{3}) \to 0$.  For the sake of simplicity
and to avoid rendering the notation too heavy we will drop the
$(\Phi_3)$ dependence of $a$ and treat it as constant in the remainder
of this section. The reader can easily convince themselves that the
argument given below will not be affected by this approximation.
 
This allows us to write
\begin{align}
\frac{\df \sigma^S}{\df \Tau_0}
&= \int \!\df \Phi_2\, \frac{\df \sigma_2}{\df \Phi_2} \,U_2(\Tau_1^{\rm max}, \Lambda_2) \, \delta[\Tau_0(\Phi_2) - \Tau_0]
\nn\\
&\quad
+  \int \!\df \Phi_3\, \frac{\df \sigma^S_3}{\df \Phi_3} \, \delta[\Tau_0(\Phi_2) - \Tau_0 + a \, \Tau_2(\Phi_3)]
\,.\end{align}
Thanks to the normalization of the splitting probabilities,  one can easily perform the integrals over the  two radiation variables other than $\Tau_2$  to obtain
\begin{align}
\frac{\df \sigma^S}{\df \Tau_0}
&= \int \!\df \Phi_2\, \frac{\df \sigma_2}{\df \Phi_2}U_2(\Tau_2^{\rm max}, \Lambda_2) \, \delta[\Tau_0(\Phi_2) - \Tau_0]
\nn\\
&\quad
+  \int \!\df \Phi_{2}\, \frac{\df \sigma_{2}}{\df \Phi_{2}}  \, \int_{\Lambda_2}^{\Tau_2^{\max}} \!\!\!\!\! \df \Tau_2 \, U_2'(\Tau_2^{\rm max}, \Tau_2)
\nn\\
& \qquad \times
\delta[\Tau_0(\Phi_{2}) - \Tau_0 + a \, \Tau_2]
\,.\end{align}
Next, the delta function is Taylor expanded to obtain
\begin{align}
\label{eq:difffirstemi}
\frac{\df \sigma}{\df \Tau_0} - \frac{\df \sigma^S}{\df \Tau_0} 
&= a
\int \!\df \Phi_{2}\, \frac{\df \sigma_{2}}{\df \Phi_{2}}  \, \delta'[\Tau_0 - \Tau_0(\Phi_{2})] \,
\nn\\
& \qquad 
\int_{\Lambda_2}^{\Tau_2^{\max}} \!\!\!\!\! \df \Tau_2 \, \Tau_2 \, U_2'(\Tau_{2}^{\rm max}, \Tau_2)\,\nn\\
&= a
\int \!\df \Tau_0'\, \frac{\df \sigma_{2}}{\df \Tau_0'} \, \left\langle \Tau_{2} \right\rangle(\Tau_0') \, \delta'[\Tau_0 - \Tau_0'] 
\nn\\
&= -a \frac{\df}{\df \Tau_0}  \left[ \frac{\df \sigma}{\df \Tau_0} \, \left\langle \Tau_{2} \right \rangle (\Tau_0) \right]
\,,\end{align}
where we have defined the average $\Tau_2$ value
\begin{align}
\left\langle \Tau_{2} \right \rangle \equiv \int_{\Lambda_2}^{\Tau_2^{\max}} \!\!\! \df \Tau_2 \, \Tau_2 \, U_2'(\Tau_{2}^{\rm max}, \Tau_2)
\,.\end{align}
In the last lines of \eq{difffirstemi}, we have written $\left\langle \Tau_{2} \right \rangle (\Tau_0) $  to make the dependence of the average value of  $2$-jettiness on $\Tau_0$ explicit.
In fact, one can derive this dependence using a  simple  LL  Sudakov factor for the parton shower. Let us first consider a single emission governed by
\begin{align}
U(\Tau^{\max}, \Tau) \sim \exp\left[ - C \, \frac{\as}{\pi} \ln^2 \frac{\Tau}{\Tau^{\max}}\right]
\,,\end{align}
with $C$ the appropriate color factor. We can now easily show that
\begin{align}
\langle \Tau \rangle
&\equiv \int_0^{\Tau^{\max}} \!\!\!\!\!\!\! \df \Tau \, \Tau \, U'(\Tau^{\max}, \Tau)
\nn\\
&\sim \, \Tau^{\max} \left[ 1 - \frac{e^{\pi/(4 \as C)} \pi \, {\rm Erfc}\left(\frac{\sqrt{\pi}}{2\sqrt{ \as C}}\right)}{2\sqrt{\as C}}\right]
\nn\\
&\sim  2 \, \frac{C \,\as}{\pi} \, \Tau^{\max} + {\cal O}(\as^2)
\,.\end{align}
Iterating this over two emissions, we find
\begin{align}
\langle \Tau_2 \rangle = & \, \int_0^{\Tau_0} \!\!\!\! \df \Tau_1 \, U_1'(\Tau_0, \Tau_1)   \int_0^{\Tau_1} \!\!\! \df \Tau_2 \, \Tau_2 \,  U_2'(\Tau_1, \Tau_2)
\nn\\
\sim & \, 2 \, \frac{C_2 \as}{\pi}  \int_0^{\Tau_0} \!\!\!\! \df \Tau_1 \, \Tau_1 \, U'(\Tau_0, \Tau_1)   
\nn\\
\sim & \, 4 \, \frac{C_1 C_2 \, \as^2}{\pi^2} \, \Tau_0  + {\cal O}(\as^3) 
\,,\end{align}
where we used the scaling $\Tau_N^{\max} \sim \Tau_{N-1}$. 

We can also numerically extract the dependence  of $\langle \Tau_2 \rangle$  on $\Tau_0$  from an actual shower in the following way: Starting from the showered expression  $\df \sigma_3^S / \df \Phi_3$ defined as in \eq{dfSigmaNp1Def}, we have
\begin{align}
& \int \! \df \Phi_3 \, \frac{\df \sigma_3^S}{\df \Phi_3} \, \delta(\Tau_0(\Phi_2) - \Tau_0) \, \Tau_2
\nn\\
&\quad=
\int \!\df \Phi_{2}\, \frac{\df \sigma_{2}}{\df \Phi_{2}}  \, \delta[\Tau_0 - \Tau_0(\Phi_{2})] \,
\int_{\Lambda_2}^{\Tau_2^{\max}} \!\!\!\!\!\! \df \Tau_2 \, \Tau_2 \, U_2'(\Tau_{2}^{\rm max}, \Tau_2)\,\nn\\
&\quad=
\int \!\df \Tau_0'\, \frac{\df \sigma_{2}}{\Tau_0'}  \, \delta[\Tau_0 - \Tau_0'] \,
\langle \Tau_2 \rangle(\Tau_0') 
\nn\\
&\quad =
\frac{\df \sigma_2}{\df \Tau_0} \langle \Tau_2 \rangle (\Tau_0) 
\,.\end{align}
This gives
\begin{align}
\left\langle \Tau_{2} \right \rangle = \left[ \frac{\df \sigma_2}{\df \Tau_0}\right]^{-1} \, \int \! \df \Phi_3 \, \frac{\df \sigma_3^S}{\df \Phi_3} \, \delta(\Tau_0(\Phi_2) - \Tau_0) \, \Tau_2
\,.
\end{align}
Taking the $\Phi_2$ events from \geneva, running them through \pythia8, and calculating this ratio for various values of $\Tau_0$, we obtain a very good fit to a straight line with
\begin{align}
\left\langle \Tau_{2} \right \rangle \approx 0.06 \, \Tau_0
\,.\end{align} 

Taking advantage of the linear scaling of $\langle \Tau_{2} \rangle$ with $\Tau_0$, and dropping the explicit dependence in the notation from here forward, we can now rewrite \eq{difffirstemi} as
\begin{align}
\frac{\df \sigma}{\df \Tau_0} - \frac{\df \sigma^S}{\df \Tau_0} 
&= 
-a  \left[ \frac{\df}{\df \Tau_0}  \, \Tau_0 \, \frac{\df \sigma}{\df \Tau_0} \right] \, \frac{\left\langle \Tau_{2} \right \rangle}{\Tau_0}
\,.
\end{align}
This leads to
\begin{align}
\frac{\frac{\df \sigma}{\df \Tau_0} - \frac{\df \sigma^S}{\df \Tau_0}}{\frac{\df \sigma}{\df \Tau_0}} 
&= 
f(\Tau_0) \frac{\left\langle \Tau_{2} \right \rangle}{\Tau_0}
\,,
\end{align}
where
\begin{align}
f(\Tau_0) &\equiv -a \, \frac{\df}{\df \ln \Tau_0} \ln \left[ \frac{\df \sigma}{\df \ln \Tau_0}\right]
\,.
\end{align}
Since the dominant contribution to $\df \sigma / \df \ln \Tau_0$ is given by
\begin{align}
\frac{\df \sigma}{\df \ln \Tau_0} \sim - \as \ln \Tau_0 \, e^{-\as \ln^2 \Tau_0}
\,,
\end{align}
we find
\begin{align}
f(\Tau_0) \sim \frac{1}{\ln \Tau_0}
\,.\end{align}
This gives our final result for the change of the $\Tau_0$ spectrum after the first emission done by the parton shower
\begin{align}
\label{eq:diff1stshower}
\frac{\frac{\df \sigma}{\df \Tau_0} - \frac{\df \sigma^S}{\df \Tau_0}}{\frac{\df \sigma}{\df \Tau_0}} 
\sim 
\frac{1}{\ln \Tau_0} \frac{\left\langle \Tau_{2} \right \rangle}{\Tau_0}
\sim 
\frac{\alpha_s^2}{\ln \Tau_0}
\,.\end{align}
Since the parton shower is strongly ordered in $\Tau_N$, in the sense discussed at the beginning of this section, all the subsequent emissions will not alter this result to the order we are working.

This result can be compared to the relative uncertainty in our NNLL$'$ calculation of the $\Tau_0$ spectrum, which can be estimated considering the dominant term beyond the NNLL$'$ order. This term scales as $\as^3/\Tau_0$. Comparing this to the first dominant one that we include in the  NNLL$'$ prediction, we find the relative uncertainty to be
\begin{align}
\label{eq:diffpastNNLLp}
\frac{\Delta \frac{\df \sigma^{\rm NNLL'}}{\df \Tau_0}}{\frac{\df \sigma^{\rm NNLL'}}{\df \Tau_0}} \sim \frac{\as^3 / \Tau_0}{\as \ln \Tau_0 / \Tau_0} \sim \frac{\as^2}{\ln \Tau_0}
\,.\end{align}
Upon inspection of  \eq{diff1stshower} and \eq{diffpastNNLLp}, we can see that the shift in the spectrum due to \pythia's showering is of the same size as the missing higher-order terms in the perturbative calculation. Therefore, the showering only affects our prediction beyond the claimed accuracy.

\section{Comparison with Dedicated Perturbative Calculations}
\label{sec:ComparisonWithPerturbative}

\begin{figure*}[t!]
\begin{center}
\includegraphics[width=0.33\textwidth]{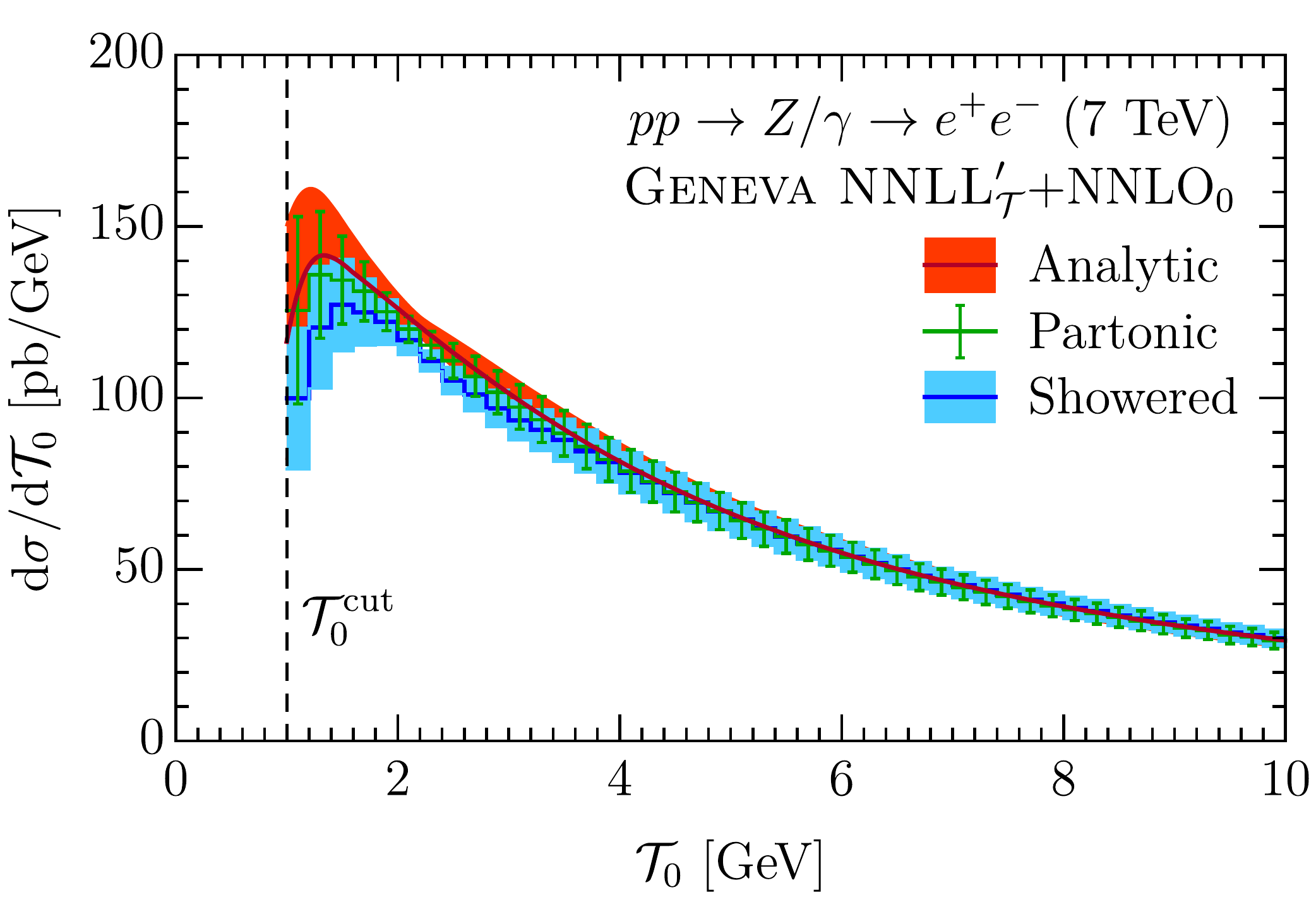}%
\hfill%
\includegraphics[width=0.323\textwidth]{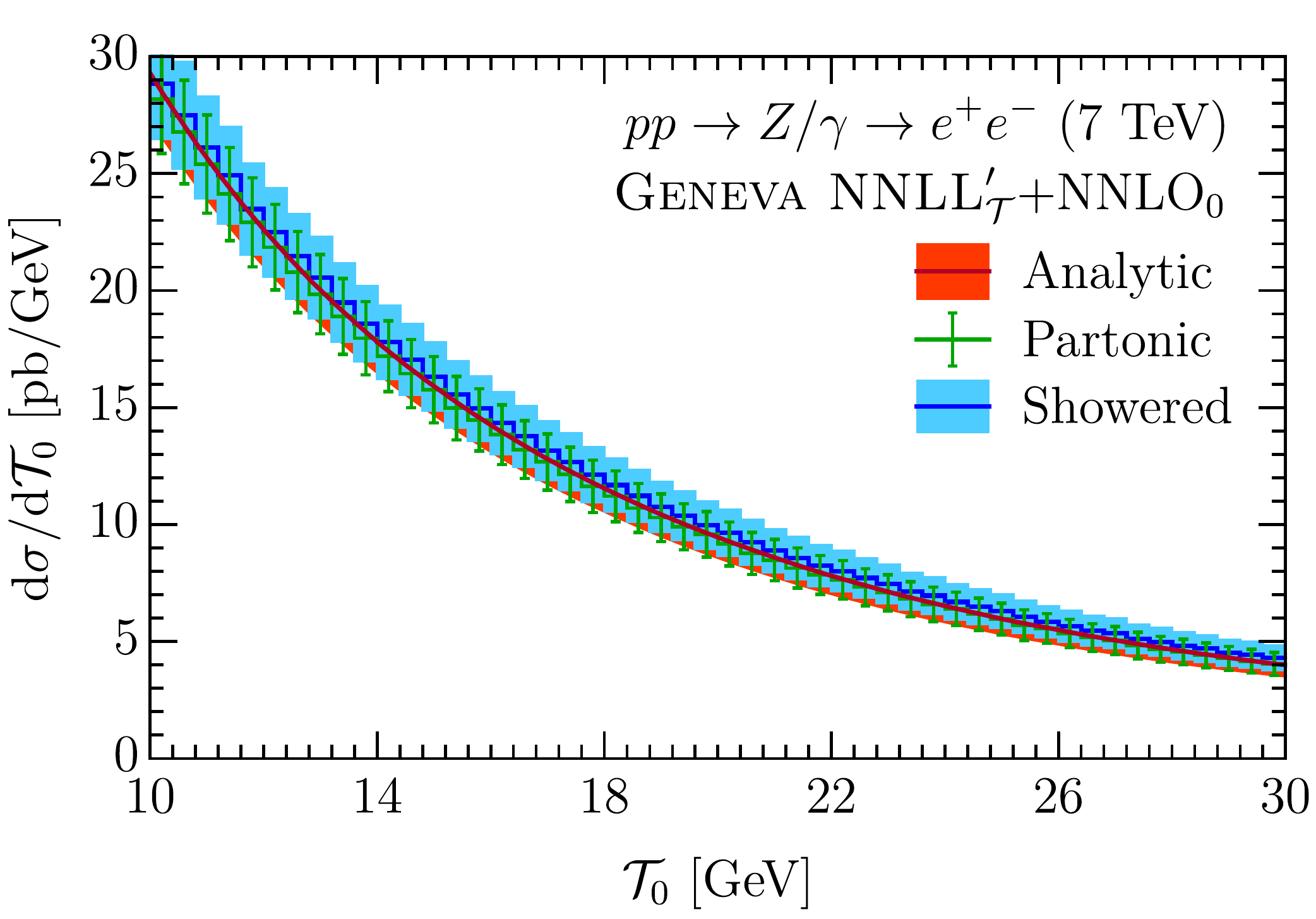}%
\hfill%
\includegraphics[width=0.34\textwidth]{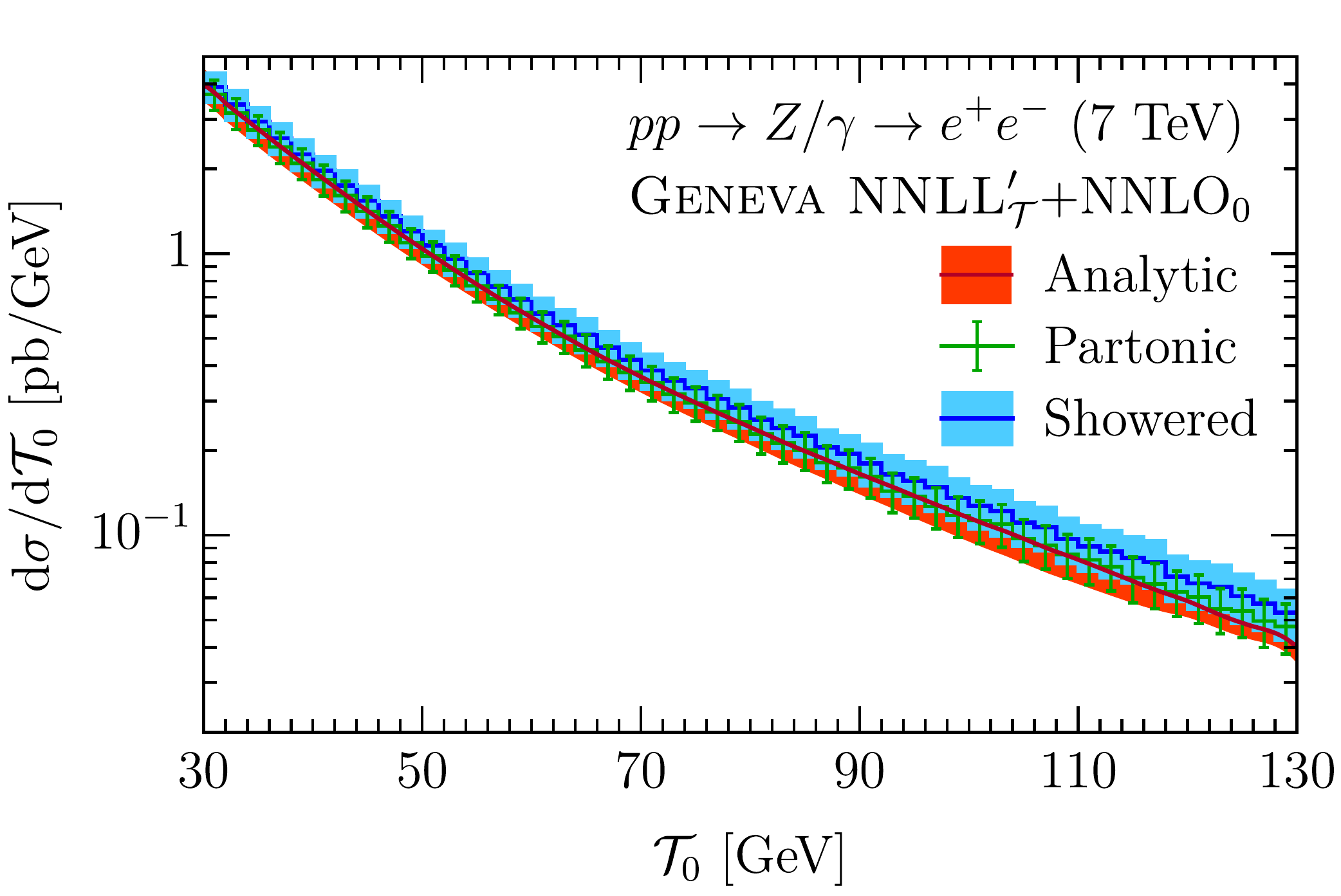}%
\caption{Validation of the $\Tau_0$ spectrum in \geneva.  The analytic NNLL$'$+NNLO $\Tau_0$ resummation is compared to the partonic and showered results of \geneva, but before addition of nonperturbative effects.}
\label{fig:Tau0validation}
\end{center}
\end{figure*}

\begin{figure*}[t!]
\begin{center}
\includegraphics[width=0.33\textwidth]{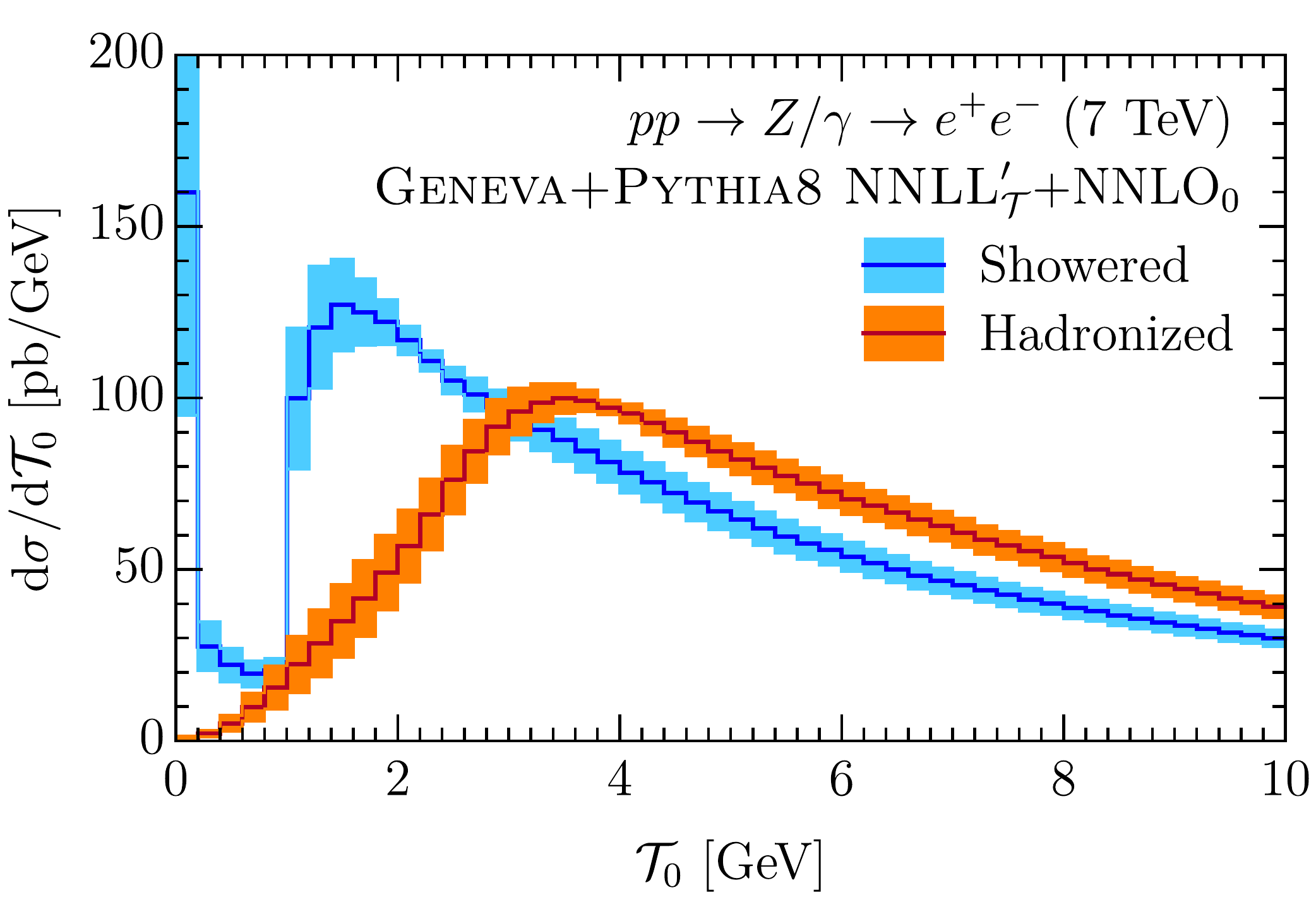}%
\hfill%
\includegraphics[width=0.323\textwidth]{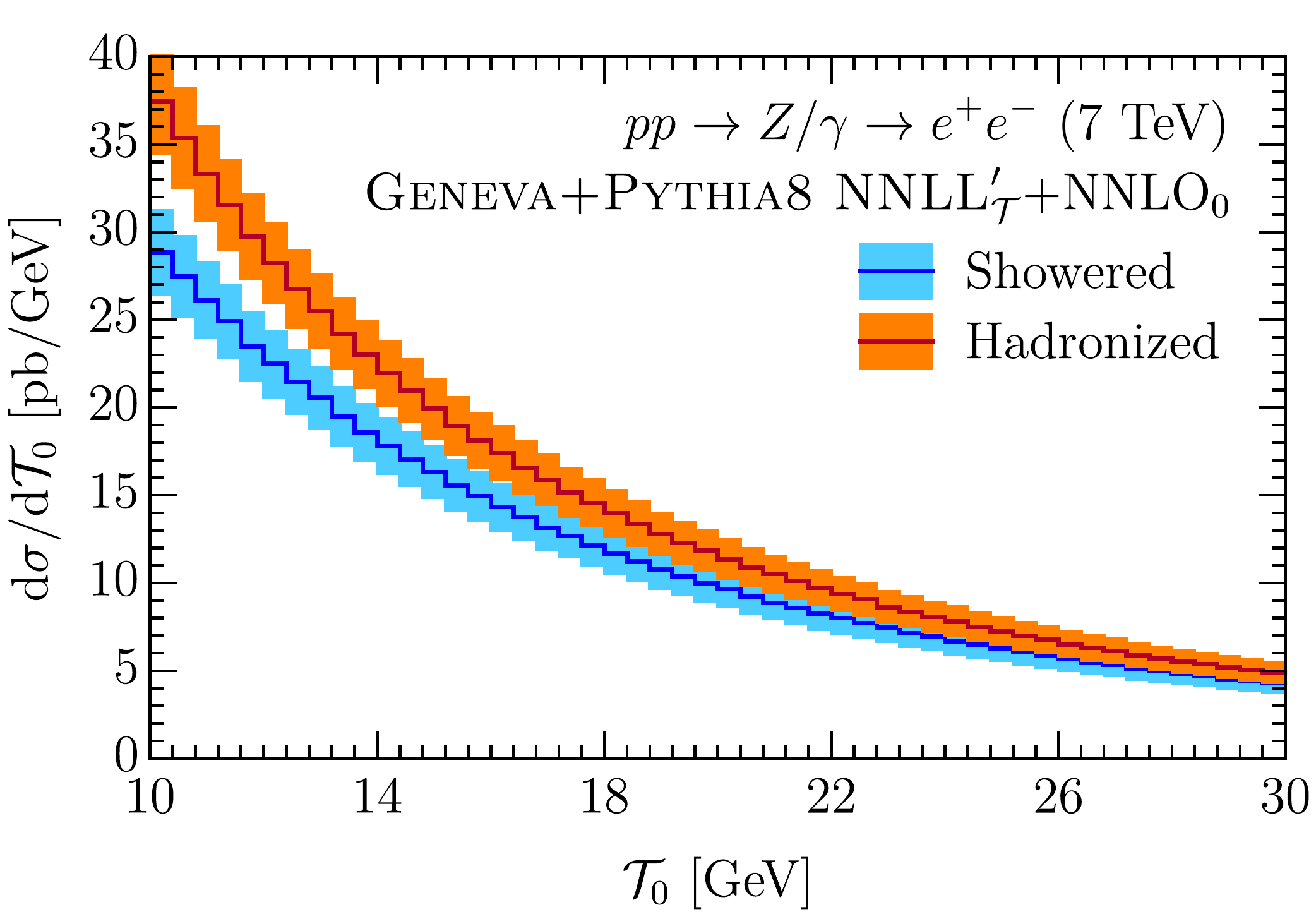}%
\hfill%
\includegraphics[width=0.34\textwidth]{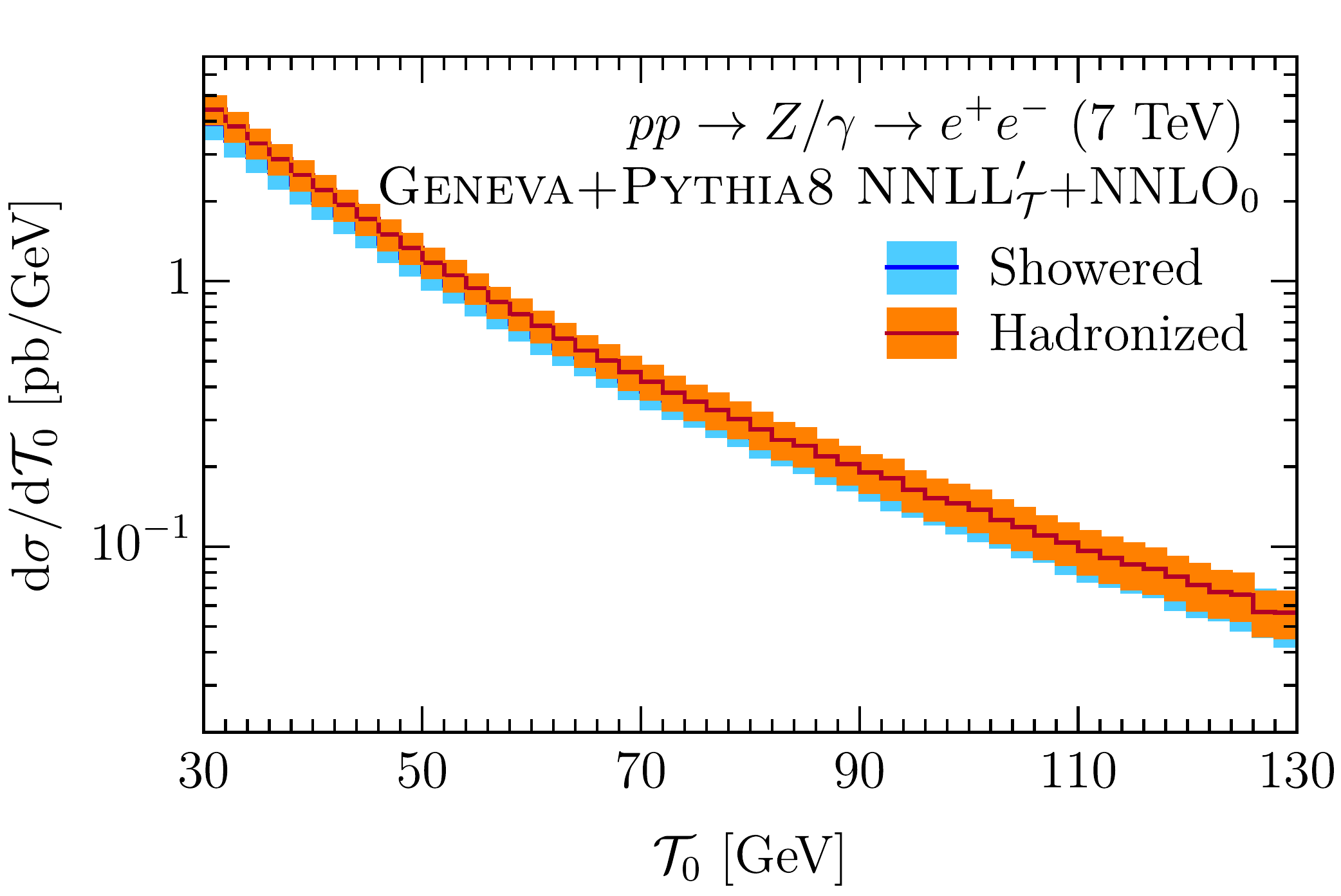}%
\caption{Comparison of the showered and hadronized $\Tau_0$ spectra in \geneva.}
\label{fig:Tau0hadronized}
\end{center}
\end{figure*}

In this section, we present the results of \geneva and the comparisons with dedicated FO and resummed predictions.
All calculations are performed for $pp$ collisions with $7 \TeV$ center-of-mass energy. We use the \textsc{CT10nnlo}~\cite{Gao:2013xoa} set from \textsc{Lhapdf6}~\cite{Buckley:2014ana} as our default PDFs, together
with its default value of $\as(M_Z) = 0.118$. We include the full interference effects between $Z/\gamma^*$ and restrict the invariant mass for the dilepton pair to the range $60 < m_{\ell^+\ell-} < 120$ GeV during the generation of events.
The other parameters relevant for our calculation are
\begin{eqnarray}
&&M_Z =  91.1876~{ \rm GeV},\quad \Gamma_Z =  2.4952~{\rm GeV}, \\&& \sin^2 \theta_W^{\rm eff} =  0.2226459 ,\quad   \alpha_{\rm em }^{-1}(M_Z) = 132.338\,. \nn
\end{eqnarray}
The $Z$ couplings are given by
\begin{equation}
g_{\ell/q}=\frac{e}{\sin\theta_W^{\rm eff}\cos\theta_W^{\rm eff}} 
\left[T_3^{(\ell/q)}-q_{\ell/q}\sin^2\theta_W^{\rm eff}\right]\,,
\end{equation}
where $\ell/q$ denotes the given left or right component of a lepton or a
quark and $ e=\sqrt{4\pi \alpha_{\rm em}(M_Z)}$. For the parton shower, we use \pythia 8.205~\cite{Sjostrand:2006za, Sjostrand:2007gs}, with a slightly modified version of Tune 4C, namely with $\alpha_s = 0.118$ for both the ISR and FSR shower, $p_{T_0}^{\rm ref} = 4$ GeV, and the hard primordial $k_T = 2.5$ GeV. Unless otherwise stated,  the  \geneva predictions include showering and hadronization effects, but presently do not include multiple parton interactions.

We first present the $\Tau_0$ spectrum, and show that we exactly reproduce the input resummed spectrum, and that the parton shower does not modify this distribution as required.

To validate the FO accuracy of \geneva, we next consider the rapidity distribution of the $Z$-boson, as well as that of the negatively charged lepton from the $Z$ decay. These distributions are known to NNLO accuracy. As discussed in \sec{TheoreticalFramework}, \geneva reproduces all $\ord{\as^2}$ contributions up to small power-suppressed effects, and these observables allow us to test the size of these effects. In addition, it allows us to validate that the procedure described in \subsec{SCETResummation} to turn off the resummation and numerically reproduce the total cross section from the integrated resummed spectrum works.

Finally, to study the accuracy of our resummation, we show the results for various observables that are sensitive to the higher logarithms included in \geneva. The first is the 0-jet cross section, which only contains events with jets below a given $p_T$ cut and is known analytically to NNLL. The corresponding observable in Higgs production is an important ingredient in Higgs analyses. The other two resummation sensitive observables we consider are the transverse momentum distribution of the $Z$ boson, as well as the $\phi^*$ distribution~\cite{Banfi:2011dm}. For all these observables we compare our results to the analytic resummed results at NNLL matched to NNLO$_0$ or NLO$_1$.

\subsection{The $\Tau_0$ Spectrum}
\label{subsec:Tau0}

\begin{figure}[t]
\begin{center}
\includegraphics[width=\columnwidth]{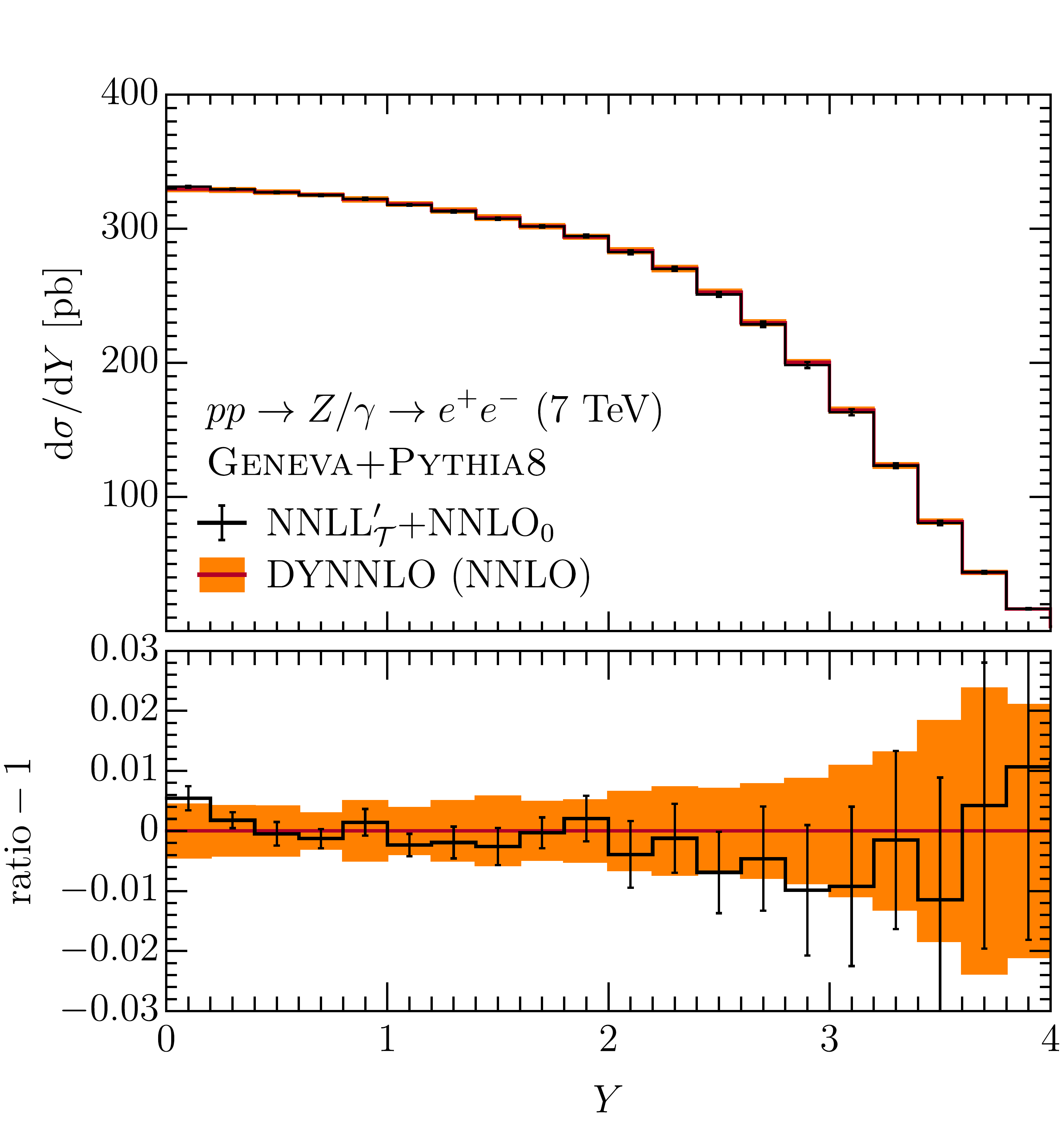}%
\vspace{-1ex}
\caption{Comparison of \geneva with the NNLO rapidity distribution of the vector boson. The orange curve shows the results from \dynnlo, while the black histogram shows the \geneva result.  For \geneva, the uncertainties shown are the FO uncertainties as described in the text.}
\label{fig:YVNNLO}
\end{center}
\end{figure}
\begin{figure}[t]
\includegraphics[width=\columnwidth]{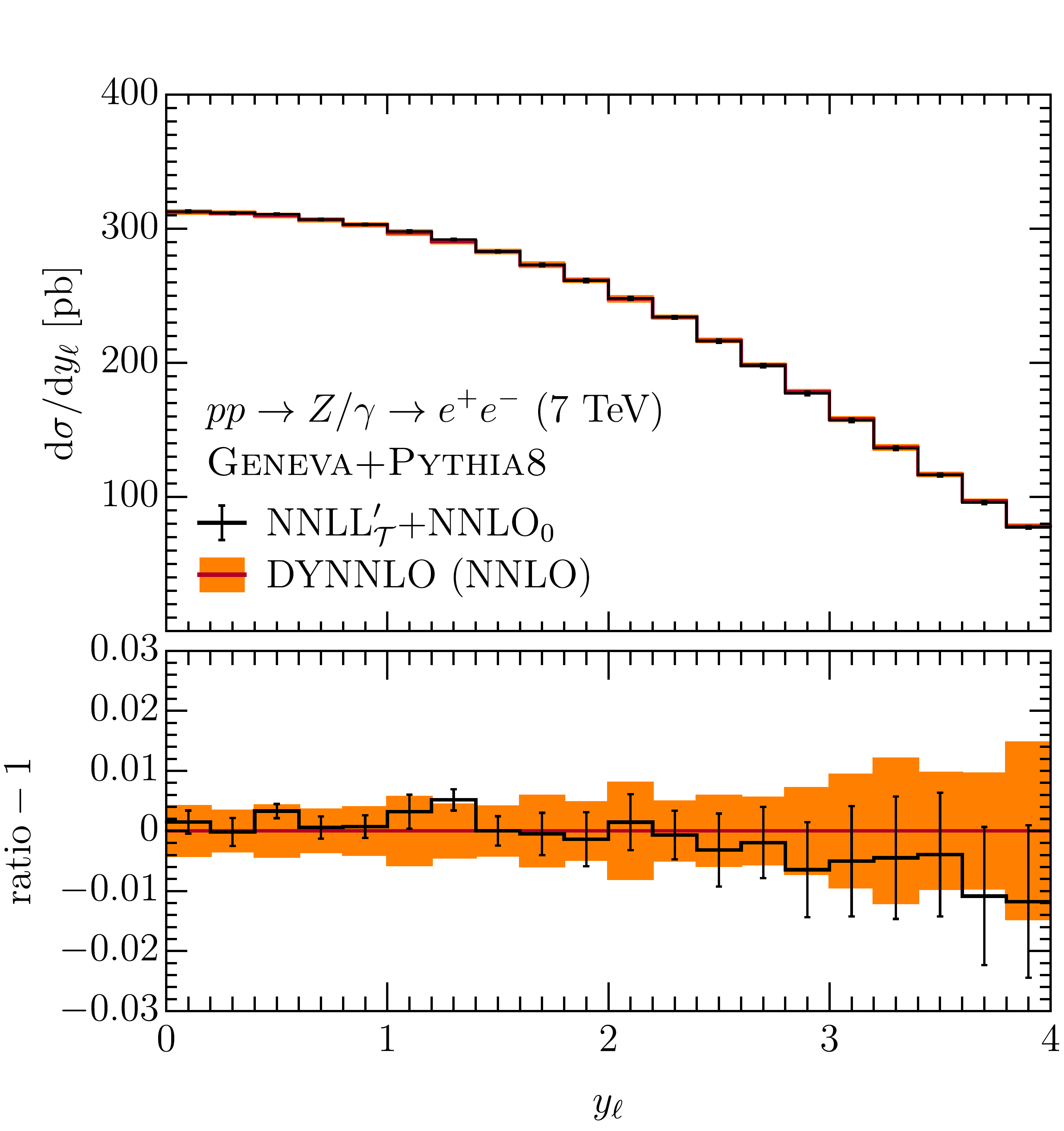}%
\vspace{-1ex}
\caption{Comparison of \geneva with the NNLO rapidity distribution of the negatively charged lepton. The orange curve shows the results from \dynnlo, while the black histogram shows the \geneva result.  For \geneva, the uncertainties shown are the FO uncertainties as described in the text.}
\label{fig:YlepNNLO}
\end{figure}
\begin{figure}[t!]
\includegraphics[width=\columnwidth]{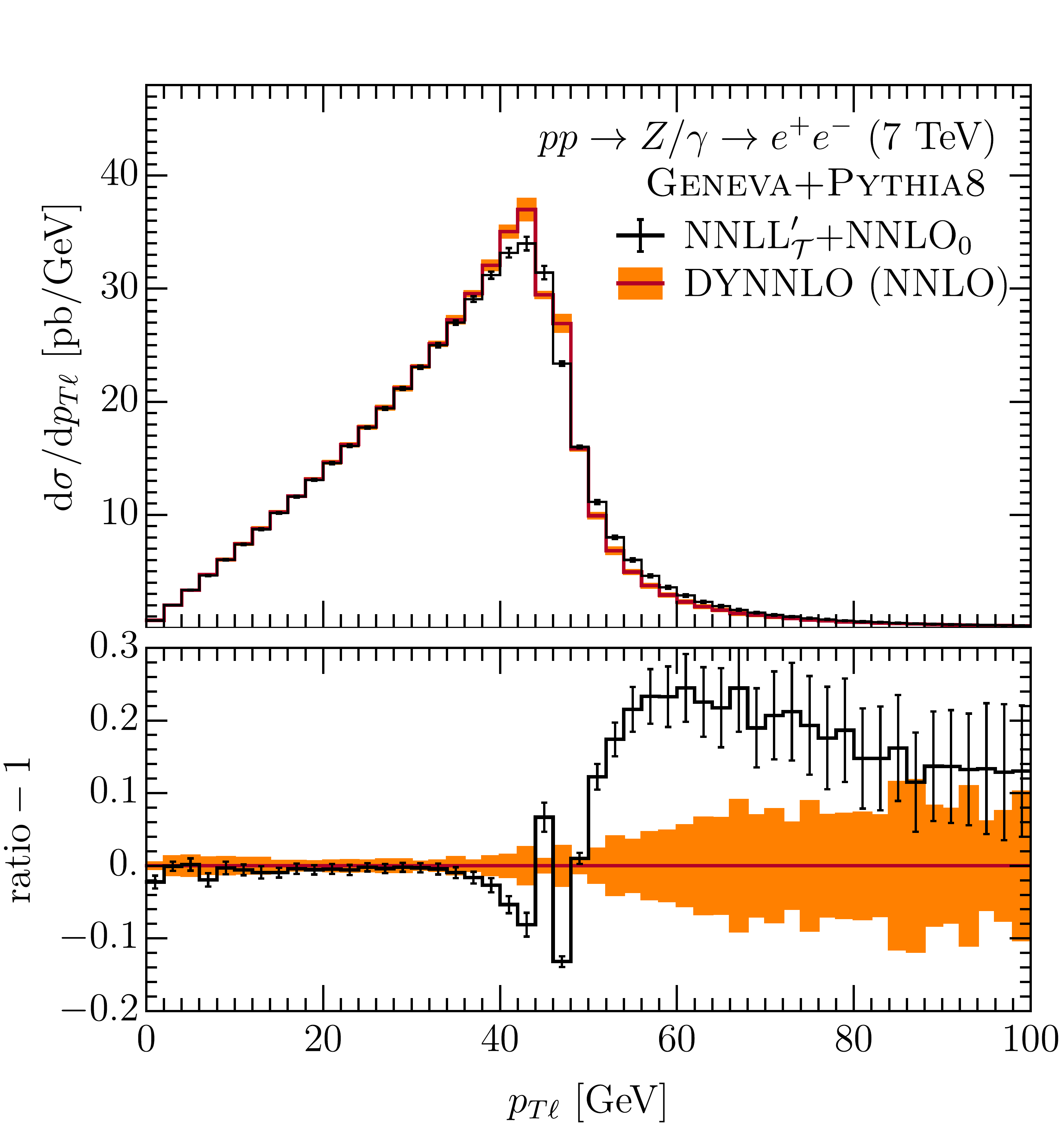}%
\vspace{-1ex}
\caption{Comparison of \geneva with NNLO transverse-momentum distribution of the negatively charged lepton. The orange curve shows the results from \dynnlo, while the black histogram shows the \geneva result.  For \geneva, the uncertainties shown are the FO uncertainties as described in the text.}
\label{fig:pTlepNNLO}
\end{figure}

The 0-jet resolution variable chosen in \geneva is $0$-jettiness, such that partonic configurations with $\Tau_0 < \Tau_0^\cut$ are part of the exclusive 0-jet MC cross section, while partonic configurations with $\Tau_0 > \Tau_0^\cut$ contribute to the inclusive 1-jet cross section. Since the dependence on the 0-jet resolution variable at NNLL$'$ accuracy is used as input in \geneva, it must reproduce the $\Tau_0$ spectrum at NNLL$'$ accuracy  both for the central value and for the event-by-event scale uncertainties.

In \fig{Tau0validation} we show the $\Tau_0$ spectrum in the peak, transition, and tail regions. All the curves are cut off at the $\Tau_0^\cut = 1$ GeV value used in the simulation. We show the analytic expressions with a red band, where the uncertainties are obtained using the same profile scale variations discussed in \subsec{SCETResummation}. The green histogram with error bars shows the results of \geneva before attaching the parton shower. One can see that \geneva is reproducing the analytic results exactly, as it should, for both the central value and the scale variations. Finally, the blue histogram shows the results of \geneva after the parton shower. As required, the parton shower changes the $\Tau_0$ spectrum only within the perturbative uncertainties. In fact, over most of the $\Tau_0$ range the spectrum is essentially unchanged.

The nonperturbative effects are shown in \fig{Tau0hadronized}, where we compare the predictions for the  $\Tau_0$  spectrum after the shower (again in blue) and after the addition of proton remnants, intrinsic $\kt$ smearing, and hadronization by \pythia8 in orange. These effects change the $\Tau_0$ distribution significantly in the peak region, while they become power corrections in the transition and tail regions. This behaviour is precisely as dictated by factorization, from which one expects that these effects should behave for $\Tau_0$ analogous to thrust in $e^+e^-$. Comparing to our $e^+e^-$ results~\cite{Alioli:2012fc}, this is precisely what we observe. A benefit of the \geneva approach is that it allows to directly combine the higher-order analytic resummation with these nonperturbative corrections provided by the hadronization model in \pythia8.

\subsection{Partonic NNLO$_0$ Observables}
\label{subsec:NNLO0}

We now show that \geneva reproduces fully inclusive observables at
NNLO accuracy, by comparing to dedicated NNLO
calculations. In this section, we only consider the profile scale variations
that reproduce the FO scale variations, as described in \subsec{SCETResummation}.

In \fig{YVNNLO} and \fig{YlepNNLO} we show the result for the rapidity
distribution of the vector boson and the negatively charged lepton
from its decay, respectively. The  orange band shows
the NNLO result from \dynnlo~\cite{Catani:2009sm}. We
show the results of \geneva as a black histogram, with the error bars
representing FO uncertainties as described above. In the
lower part of each plot, we show the ratio to
the \dynnlo central value.

The central value of \geneva agrees very well with the
fixed NNLO prediction.  The perturbative uncertainties from \geneva are also in
reasonable agreement with those from \dynnlo. The few fluctuations
observed in the plot are of statistical nature, as evidenced by
the fact that they grow larger toward larger rapidities, where the
statistics is poorer.  The rapidity distribution of the vector boson has also been
validated against the independent NNLO calculation provided by
\vrap~\cite{Anastasiou:2003yy}.

In \fig{pTlepNNLO} we show the result of the transverse momentum distribution of the negatively charged lepton. For $p_{T\ell}  < m_{\ell^+\ell^-}/2$ this observable is a true NNLO distribution and \geneva agrees very well with the NNLO prediction. The region above and below $m_{\ell^+\ell^-} / 2$ is very sensitive to Sudakov shoulder logarithms~\cite{Catani:1997xc}. It is known that the FO calculations perform very poorly in this region and fail to provide an accurate description of the physics. On the other hand, \geneva will have some of these logarithms resummed and should therefore provide a more accurate prediction. Finally, the region above $p_{T\ell} > m_{\ell^+\ell^-}/2$ can only be populated if the transverse momentum of the vector boson is nonzero, making this region only NLO accurate. Close to $p_{T\ell} \gtrsim m_{\ell^+\ell^-}/2$ \geneva lies above the FO prediction and converges back to the FO result for large values of $p_{T\ell}$. This is likely still the effect from the Sudakov shoulder at $m_{\ell^+\ell^-}/2$. Since the cross section above is much lower, any spillover from below $m_{\ell^+\ell^-}/2$ caused by the resummation can have a large relative effect.

\subsection{0-Jet Observables with Resummation}
\label{subsec:0jetresummed}

Since we resum our 0-jet resolution variable $\Tau_0$ to NNLL$'$, it is interesting to study how accurately \geneva is able to predict other observables sensitive to the $0/1-$jet separation. Since the distribution for other $0-$jet resolution variables is only obtained indirectly, via the $\Tau_0$ spectrum made fully differential by using splitting functions and a parton shower that are only correct to LL, we cannot formally claim the same NNLL$'$ accuracy for these observables. However, since the overall distribution of events in the $0/1$-jet region of phase space is clearly improved, we expect some of this accuracy to carry over to other observables as well. Hence, we expect to get predictions that are numerically much closer to NNLL resummed results also for other 0-jet observables. The same behaviour was already observed for $e^+e^-$~\cite{Alioli:2012fc}. 

\begin{figure}[t!]
\includegraphics[width=\columnwidth]{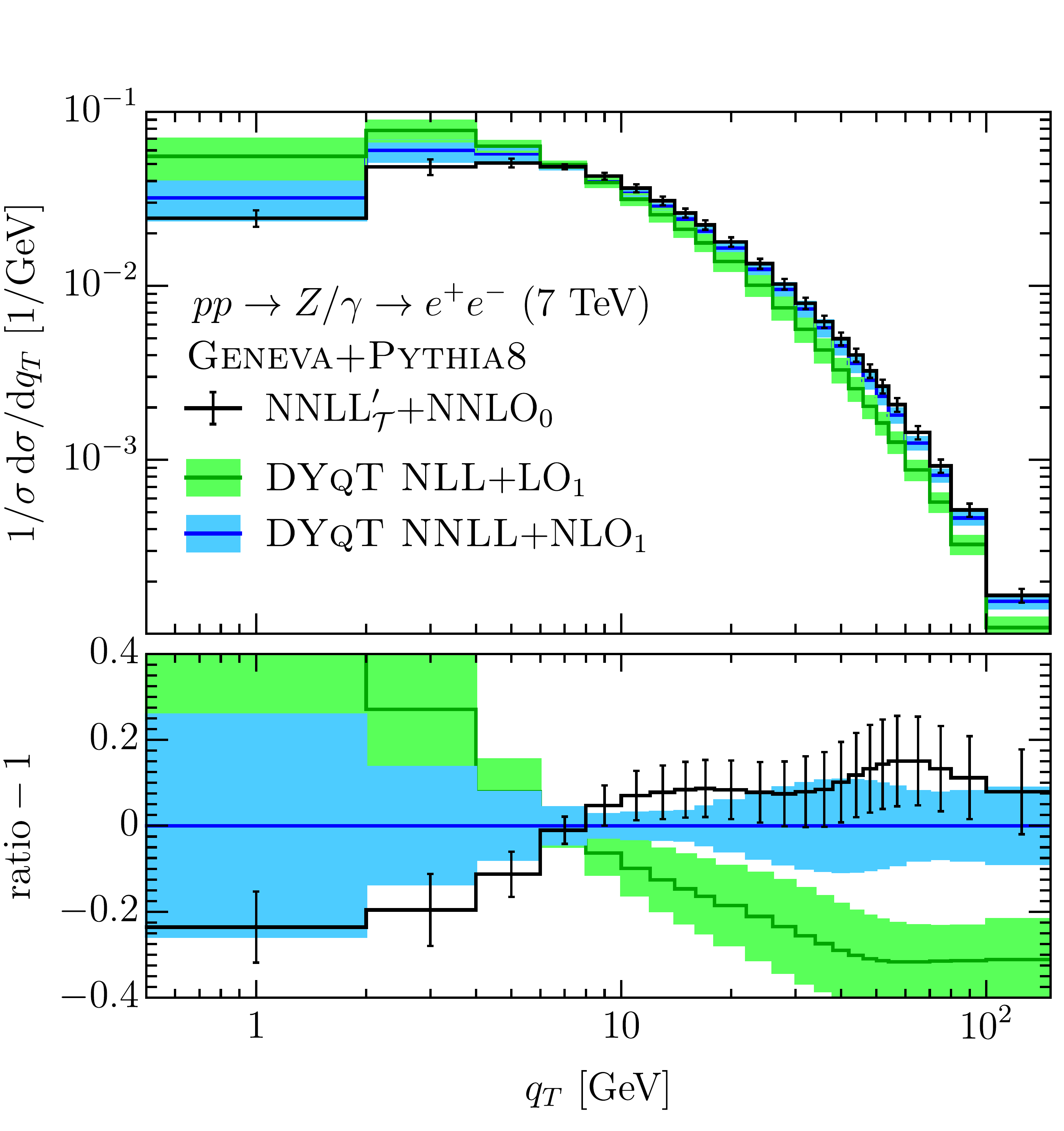}%
\caption{Comparison of \geneva with the $q_T$ distribution from \dyqt.  The NNLL$+$NLO$_1$ (NLL$+$LO$_1$) results from \dyqt are shown in blue (green), and the \geneva results are shown in black. The \geneva results are much closer to \dyqt's NNLL results than NLL results.}
\label{fig:DYqT}
\end{figure}

\begin{figure}[t!]
\includegraphics[width=\columnwidth]{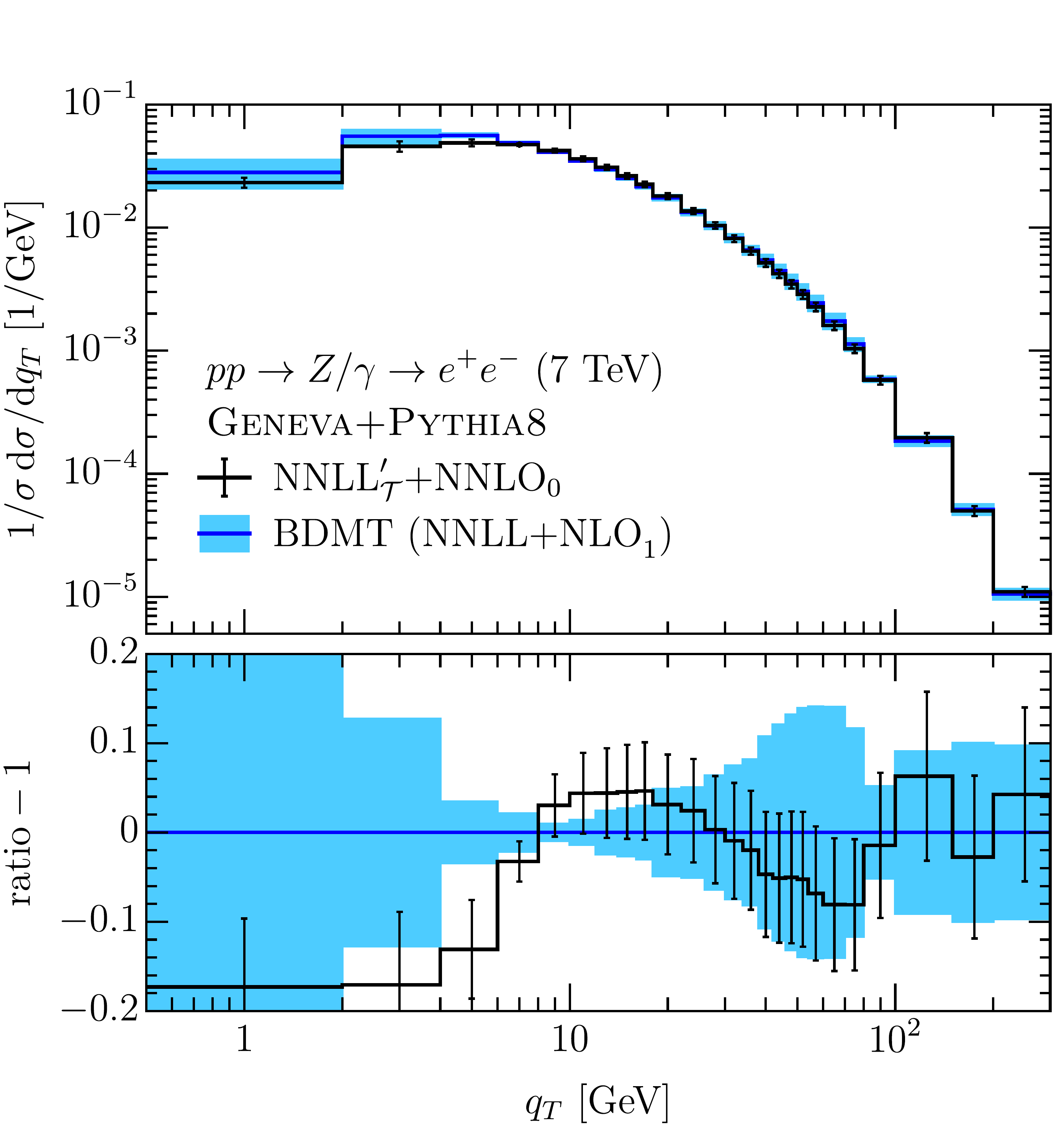}%
\caption{Comparison of \geneva with the analytic $q_T$ distribution at NNLL$+$NLO$_1$ from Ref.~\cite{Banfi:2012du}.  The analytic results are shown in blue, and the \geneva results are shown in black.}
\label{fig:qT_analytic}
\end{figure}

\begin{figure}[t!]
\includegraphics[width=\columnwidth]{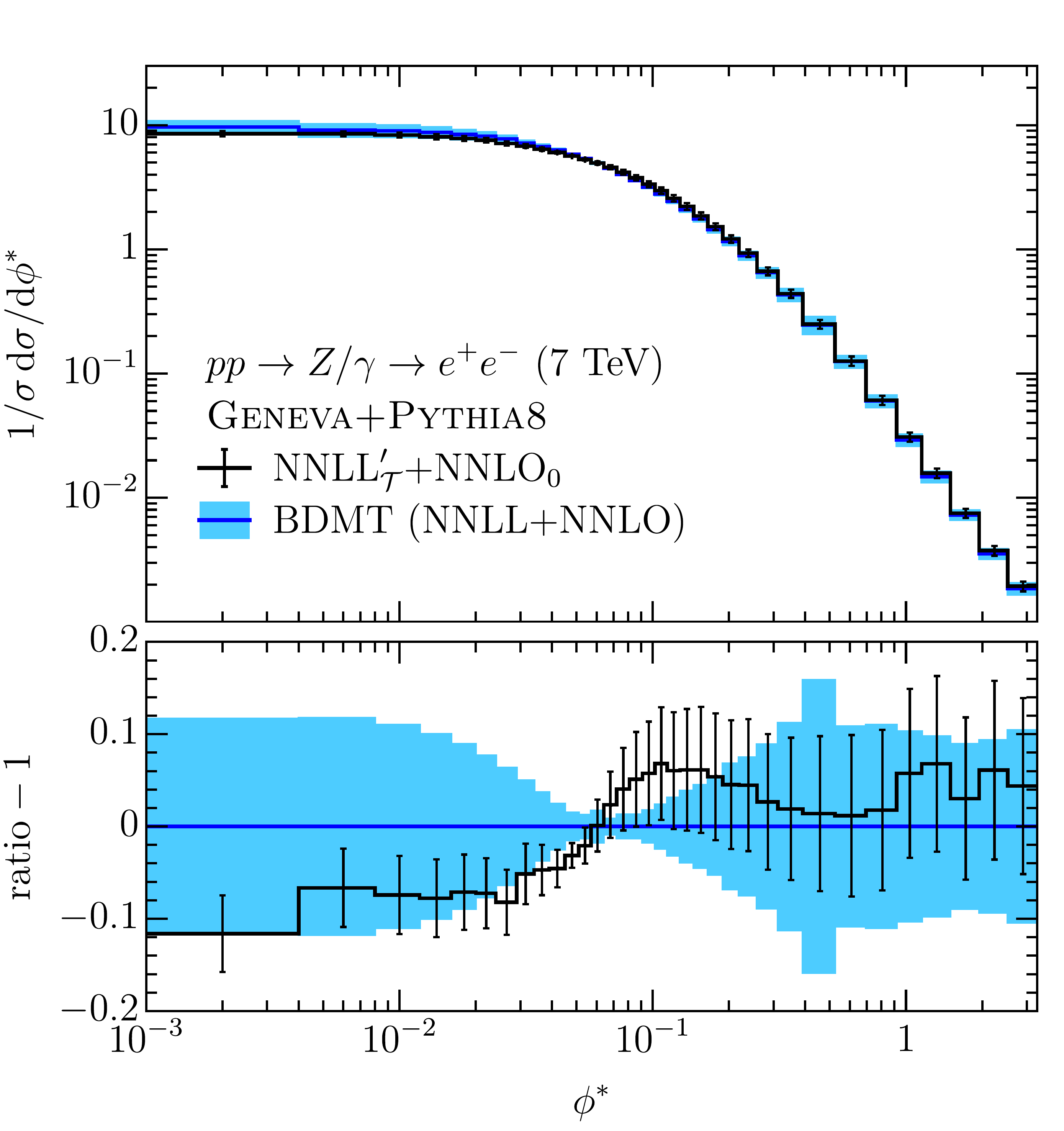}
\caption{Comparison of \geneva with the analytic $\phi^*$ distribution at NNLL$+$NLO$_1$ from Ref.~\cite{Banfi:2012du}.  The analytic results are shown in blue, and the \geneva results are shown in black.}
\label{fig:phiStar_analytic}
\end{figure}

\begin{figure}[t!]
\begin{center}
\includegraphics[width=\columnwidth]{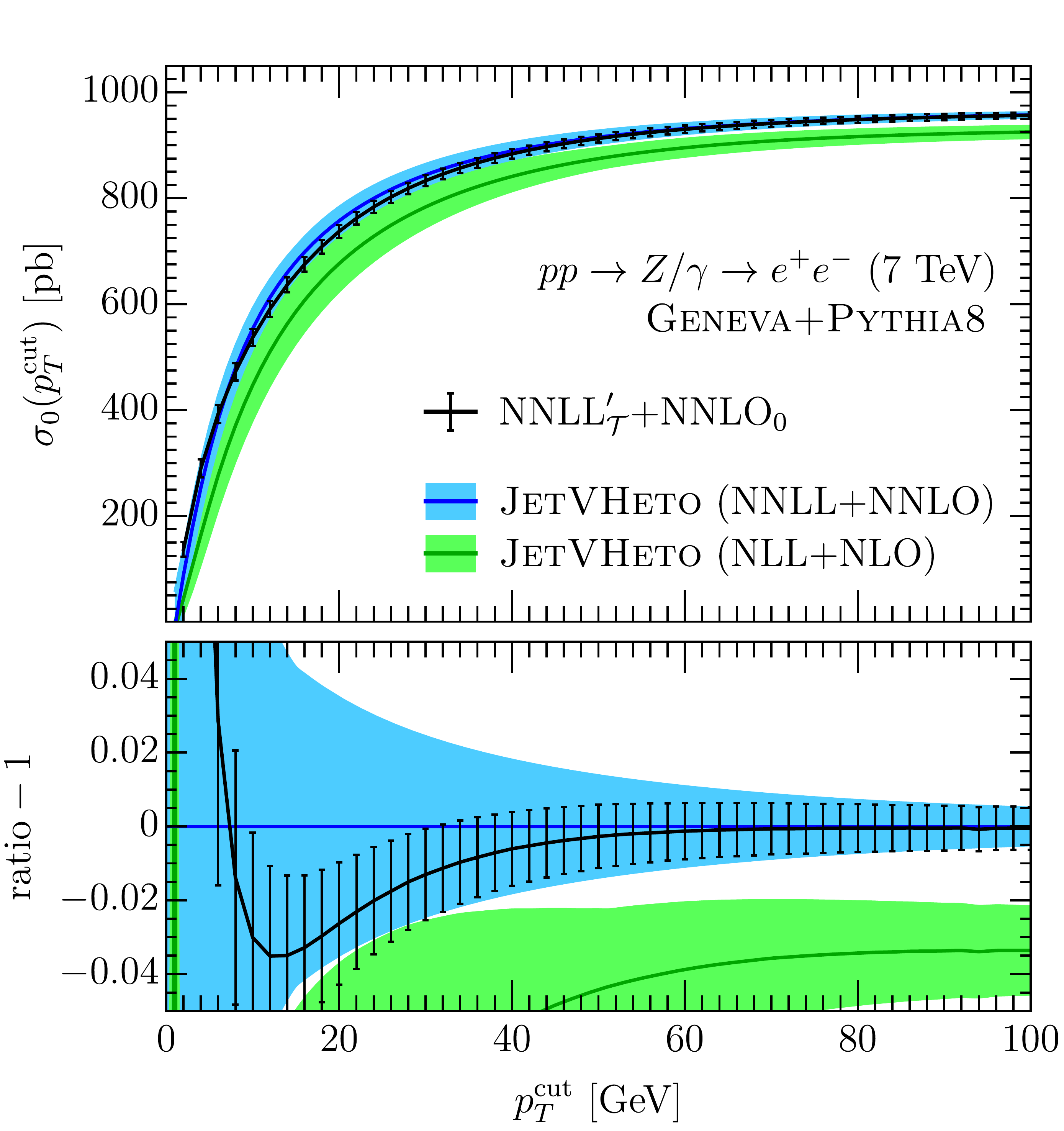}
\caption{Comparison of \geneva with the 0-jet cross section as a function of $p_T^\cut$ from \jetvheto~\cite{Banfi:2012jm} at NNLL$+$NNLO.  The analytic results are shown in blue, and the \geneva results are shown in black.  For \geneva, the uncertainties are the FO uncertainties only, see text for details.}
\label{fig:sigma0JetVHeto}
\end{center}
\end{figure}

In \fig{DYqT} we show the transverse momentum distribution of the vector boson compared to its analytic NNLL resummation from \dyqt. The predictions of \dyqt have been manually switched to agree with the FO results in the tail, according to the procedure outlined in Ref.~\cite{Catani:2015vma}.
We see that \geneva agrees reasonably well within the perturbative uncertainties with the NNLL resummed result.%
\footnote{Although \geneva's perturbative uncertainties appear smaller than \dyqt's at very low $q_T$, this should not be misinterpreted as \geneva being more accurate here. This is mostly an artifact of smearing out the uncertainties from a range of $\Tau_0$ values, which tends to reduce their numerical impact. In addition, we have not included here any uncertainties associated with the interface to the parton shower and the showering itself. Such uncertainties should eventually also be included as an additional part of the (resummation) uncertainties. As this will require a dedicated study, we leave this for future work.}
The NLL result from \dyqt has a significantly different shape, and the \geneva prediction is certainly in much better agreement with the NNLL predictions than the NLL ones. 

A similar prediction for the  transverse momentum distribution, but fully differential on the momenta of  vector boson decay products has been presented in Ref.~\cite{Banfi:2012du}. This allows the direct  comparison including the acceptance cuts  used by experimental analyses. Using the same lepton cuts as in that study, we show our comparison in  \fig{qT_analytic}. Again, we observe a fairly good agreement with the analytic NNLL prediction matched to NLO$_1$.
Another variable, quite similar to the transverse momentum of the vector boson, is the $\phi^*$ between the two leptons, with the precise definition of $\phi^*$ given in~\cite{Banfi:2011dm}. The comparison of \geneva to the NNLL$+$NLO$_1$ calculation of Ref.~\cite{Banfi:2012du} is shown in \fig{phiStar_analytic}, and we again observe good agreement.

Finally, we show the result for the exclusive 0-jet cross section as a function of $p_T^\cut$ in \fig{sigma0JetVHeto}, where the 0-jet sample is defined as all events containing no jets with $p_T > p_T^\cut$. The jets are reconstructed with the anti$-k_T$ algorithm~\cite{Cacciari:2008gp} as implemented in {\sc Fastjet}~\cite{Cacciari:2006sm,Cacciari:2011ma}, within a radius $R=0.4$.  We find good agreement between \geneva and the dedicated NNLL$+$NNLO calculation given by \jetvheto~\cite{Banfi:2012jm} within the perturbative uncertainties.  For this plot, we use the FO scale uncertainties discussed in \subsec{SCETResummation}, such that the uncertainties at large $p_T^\cut$ are estimated correctly and thus precisely reproduce those of \jetvheto. At small $p_T^\cut$ they are now underestimated and here the resummation uncertainties should be added. The better agreement with the NNLL$+$NNLO prediction compared to the lower order  NLL$+$NLO one,  especially in the large  $p_T^\cut$ region, is of course driven by the correct inclusion of the NNLO corrections in \geneva.

\section{Comparison to LHC Data}
\label{sec:ComparisonWithData}
\begin{figure}[t!]
\begin{center}
\includegraphics[width=\columnwidth]{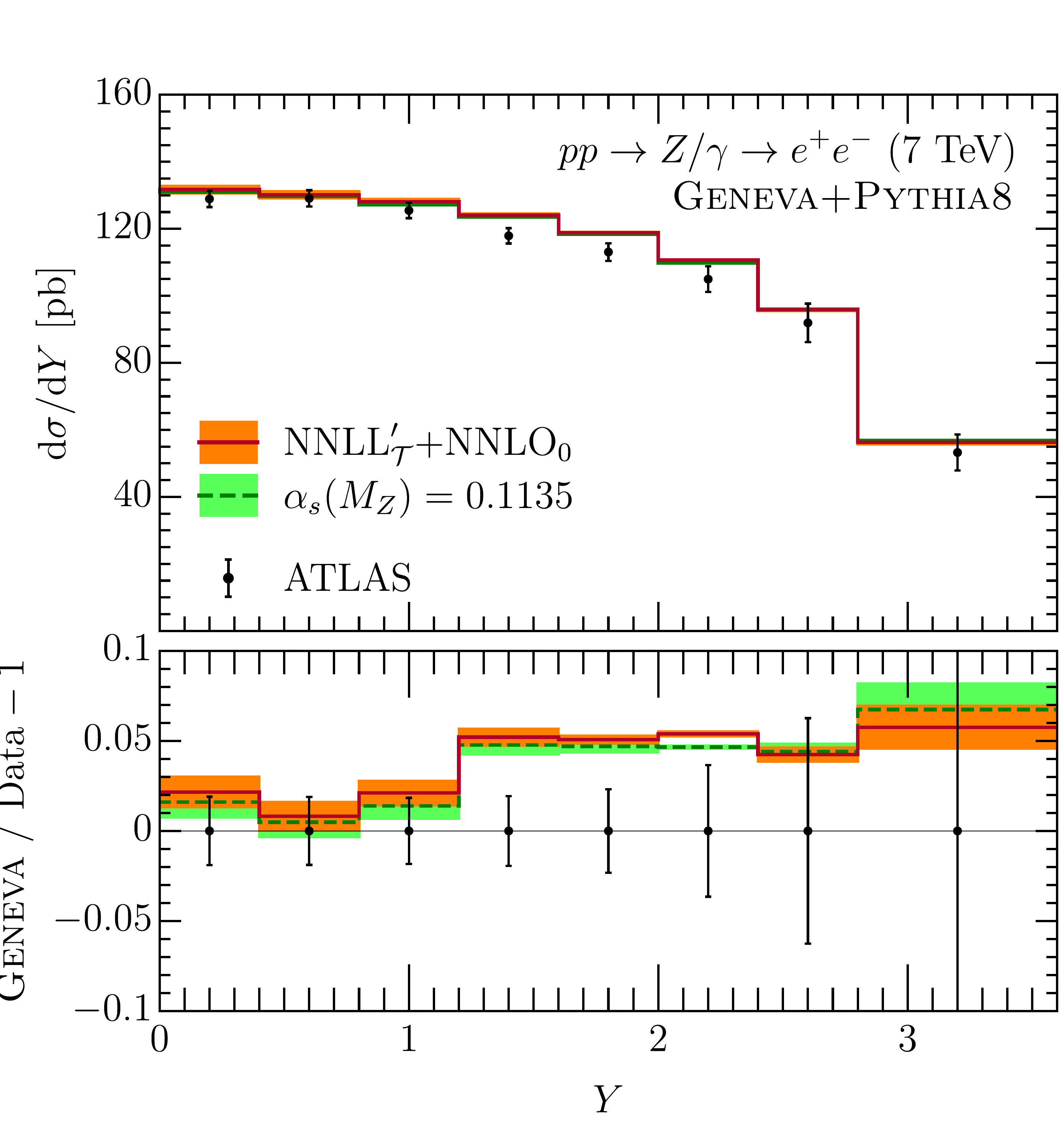}%
\hfill%
\caption{Comparison of \geneva with the rapidity of the vector boson from the \atlas study in \cite{Aad:2011dm}. The \geneva results with default values of $\as(M_Z)$  and with $\as(M_Z)=0.1135$ are shown in orange and green, respectively, while the \atlas points are shown in black.}
\label{fig:YVdataATLAS}
\end{center}
\end{figure}

In this section, we compare the results from \geneva with measurements using data collected during the $7\TeV$ LHC Run I. We use \geneva$+$\pythia8 to produce fully showered and hadronized events in \hepmc format, and then use the \rivet routines provided by the experimental collaborations to produce the histograms that can be compared with the experimental measurements. By using \rivet~\cite{Buckley:2010ar}, we ensure that event selection cuts and object definitions agree with what was used in the experimental analyses. We use the same run that was used for the comparison to the perturbative calculations, such that all parameters are identical to those given in \sec{ComparisonWithPerturbative}. 

In addition, we also include in our data comparisons \geneva results with a lower value of $\as(M_Z) = 0.1135$, since as we will see, the value of $\as(M_Z)$ can have a noticeable effect on resummation-sensitive observables. In the following, the results with $\alpha_s(M_Z) = 0.118$ will be shown by the orange bands, while those with $\alpha_s(M_Z) = 0.1135$ are shown by a green band. For the latter results, we only change the $\alpha_s(M_Z)$ value, holding all other parameters including the PDF set fixed.%
\footnote{We hasten to add that using a different $\alpha_s(M_Z)$ than the default value provided by the PDF set is not problematic as it may seem at first. In the resummation-dominated region, this only leads to a small inconsistency inside the beam functions, where the terms that are supposed to cancel the PDF's $\mu_F$ dependence will now use a different $\alpha_s$ value leading to a small noncancellation in the coefficients of some single-logarithmic terms. It is safe to assume that this will not be numerically relevant compared to the dominant $\alpha_s$ dependence coming from the double-logarithmic beam function terms, as well as the hard and soft function, and all the resummation kernels. All of these pieces are completely unrelated to the PDFs, and are in fact the same as for the corresponding $\Tau_2$ resummation in $e^+e^-$ collisions.}
This value is motivated by the fact, that all $\alpha_s(M_Z)$ determinations from LEP data that include higher-order resummation together with some form of analytic treatment of nonperturbative hadronization effects, find considerably lower values of $\alpha_s(M_Z)$ than $0.118$~\cite{Davison:2008vx, Gehrmann:2009eh, Abbate:2010xh, Abbate:2012jh, Gehrmann:2012sc, Hoang:2014wka, Hoang:2015hka}, see also Ref.~\cite{Alekhin:2012ig} for determinations from DIS.
Consistent with these findings, in Ref.~\cite{Alioli:2012fc} it was observed that the higher-order resummation using $\as(M_Z) = 0.1135$ coupled with \pythia8's hadronization model gave excellent agreement with data for $e^+e^-\to2$-jet resummation observables, namely thrust, $C$-parameter, heavy-jet mass, and jet broadening.
On the other hand, the predictions with $\alpha_s(M_Z) = 0.118$ gave consistently spectra that were somewhat too hard, i.e., shifted from lower to higher values compared to data. We will observe precisely the same qualitative trend here.

\begin{figure}[t!]
\begin{center}
\includegraphics[width=\columnwidth]{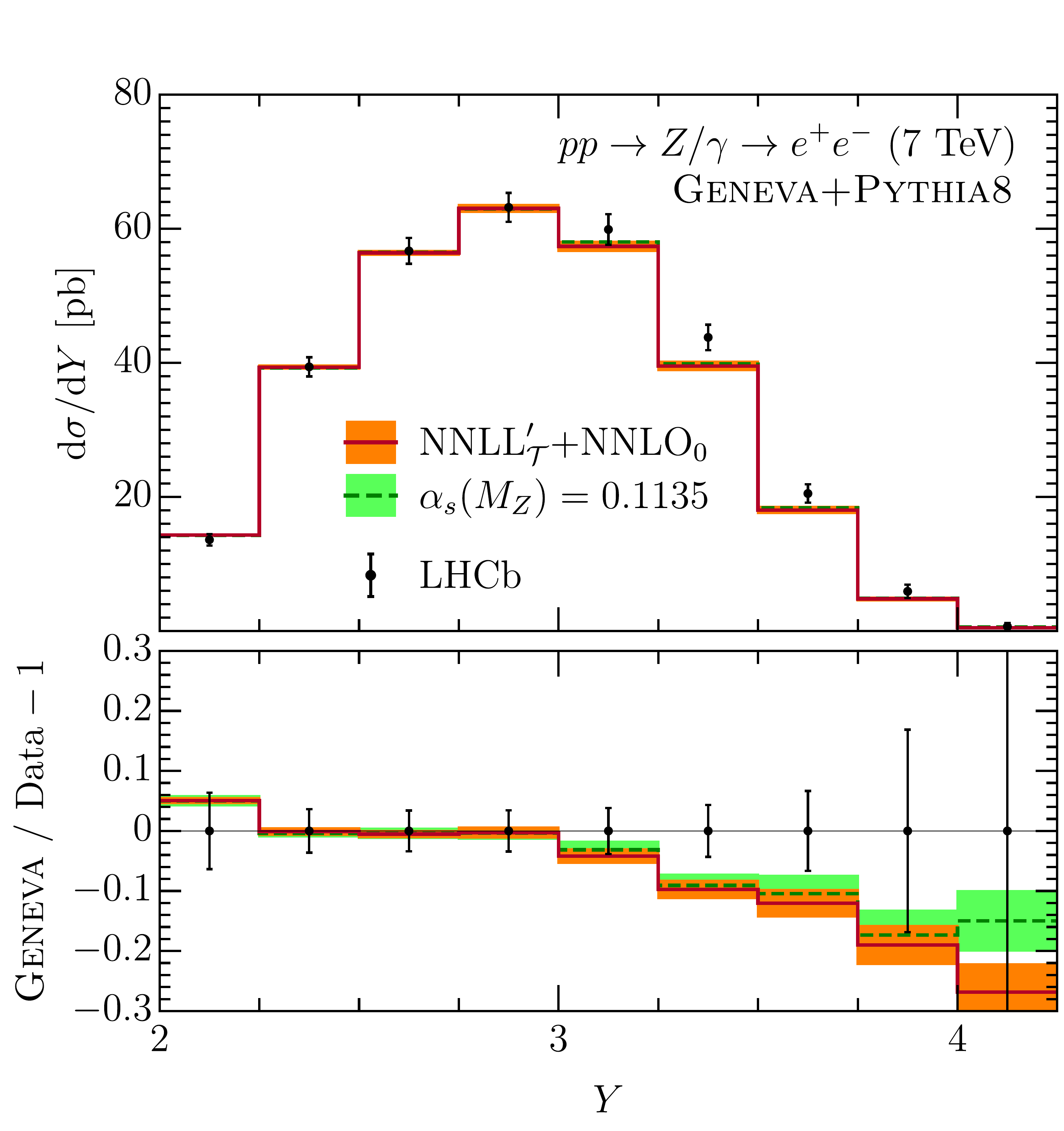}%
\caption{Comparison of \geneva with the rapidity of the vector boson from the \lhcb study in \cite{Aaij:2012mda}. The \geneva results with default values of $\as(M_Z)$  and with $\as(M_Z)=0.1135$ are shown in orange and green, respectively, while the \lhcb points are shown in black.}
\label{fig:YVdataLHCB}
\end{center}
\end{figure}

The rapidity distribution of the vector boson was measured by \atlas~\cite{Aad:2011dm}, \cms~\cite{Chatrchyan:2011wt}, and \lhcb~\cite{Aaij:2012mda}. Since we already saw that \geneva agrees very well with the NNLO predictions, these comparisons are mostly driven by the PDFs we use.  We show the comparison to the rapidity spectrum from \atlas and \cms in \figs{YVdataATLAS}{YVdataCMS}. The normalized results from \cms are reproduced very well for the entire $Y$ range. The ATLAS results are not normalized, and while \geneva agrees very well at low $Y$ values, it slightly overshoots the data (by about 2 standard deviations) for larger $Y$. A similar discrepancy was observed in Ref.~\cite{Aad:2011dm} for the MSTW08, HERAPDF1.5, and ABKM09 PDF sets. The comparison with \lhcb is shown in \fig{YVdataLHCB}. The agreement at moderate values of $Y$ is good here too. For forward events with $Y > 3.25$ our predictions show a trend to be lower than the data. This discrepancy was already noted in the \lhcb paper~\cite{Aaij:2012mda} when comparing with the NNLO calculation provided by \fewz~\cite{Gavin:2010az}. As we would expect, the lower $\as(M_Z)$ value makes almost no difference for the inclusive predictions.

\begin{figure}[t!]
\begin{center}
\includegraphics[width=\columnwidth]{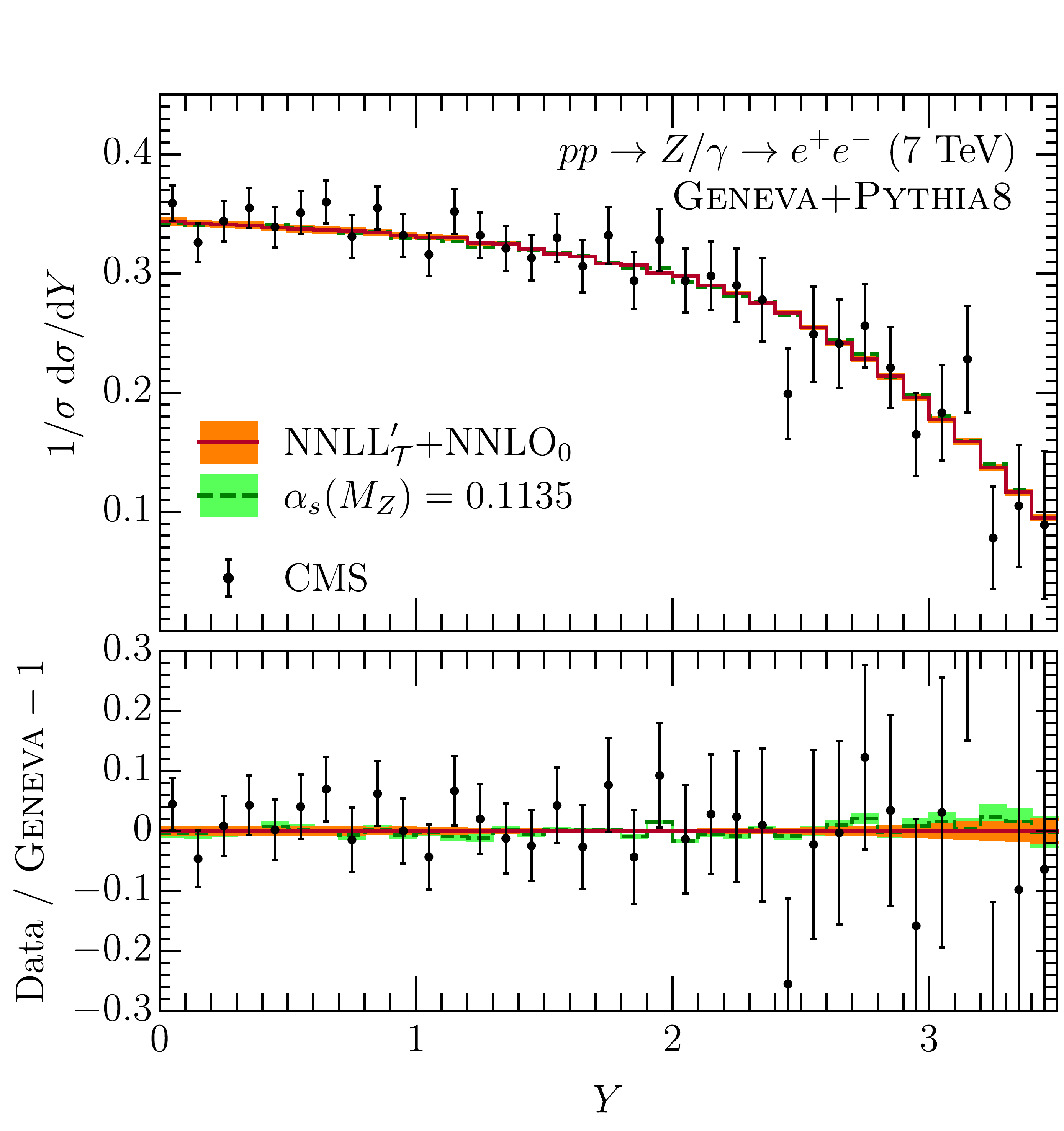}%
\caption{Comparison of \geneva with the rapidity of the vector boson from the \cms study in \cite{Chatrchyan:2011wt}. The \geneva results with default values of $\as(M_Z)$  and with $\as(M_Z)=0.1135$ are shown in orange and green, respectively, while the \cms points are shown in black.}
\label{fig:YVdataCMS}
\end{center}
\end{figure}

\begin{figure}[t!]
\begin{center}
\includegraphics[width=0.95\columnwidth]{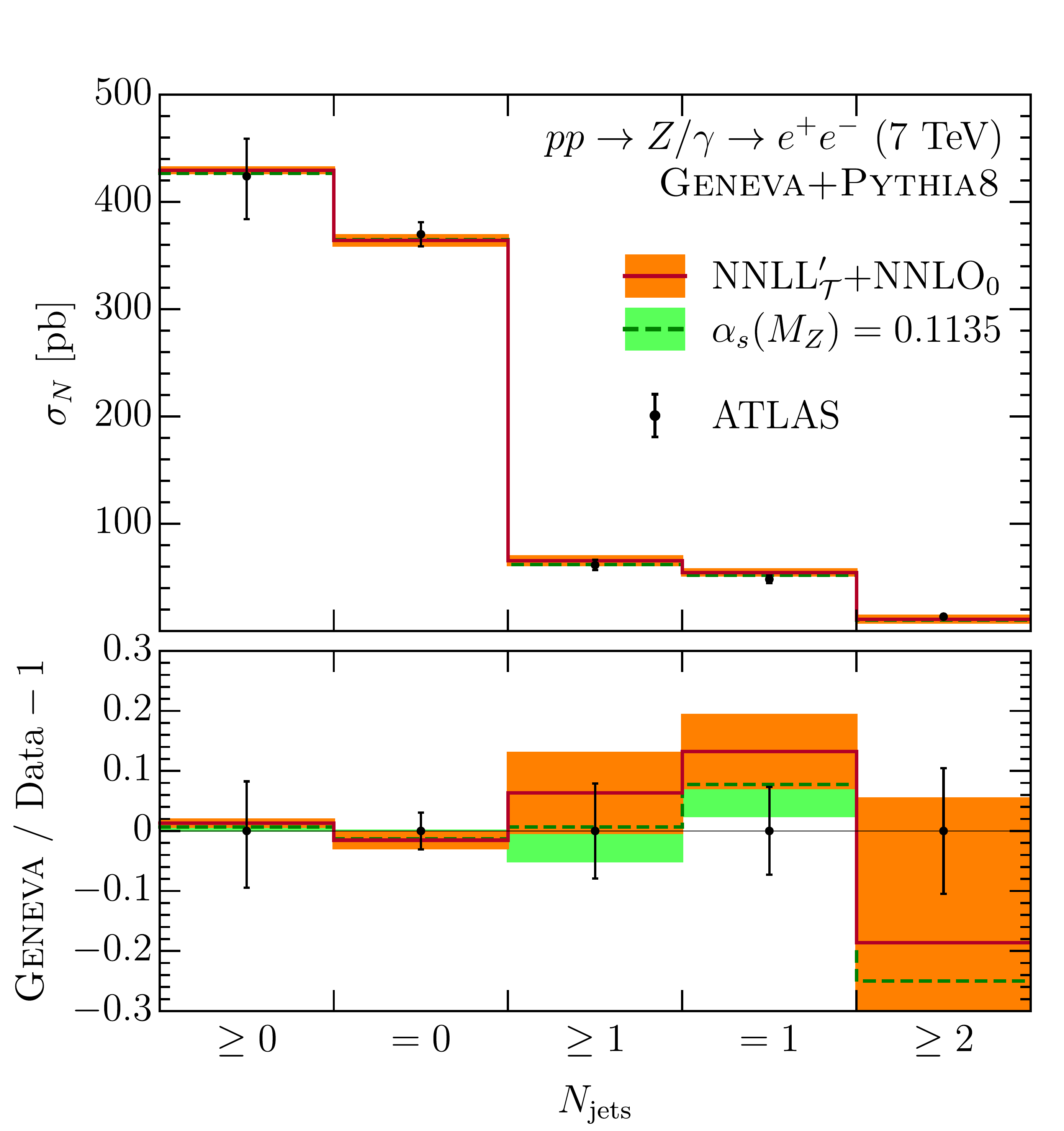}
\caption{Comparison of \geneva with data on inclusive and exclusive jet cross sections from the \atlas study in \cite{Aad:2013ysa}.  The \geneva results with default values of $\as(M_Z)$  and with $\as(M_Z)=0.1135$ are shown in orange and green, respectively, while the \atlas points are shown in black.}
\label{fig:NjetdataATLAS}
\end{center}
\end{figure}

A measurement of the production cross section of a $Z$ boson in association with jets was presented by \atlas in Ref.~\cite{Aad:2013ysa}, for both inclusive and exclusive jet cross sections. We compare our results with the measurements having up to 2 jets in \fig{NjetdataATLAS}. We choose to limit ourselves to up to 2 jets because any additional jet would only be provided at LL accuracy by \pythia8.
Our predictions for the 0-jet cross section agrees well with data, not only for the inclusive cross section, but also for the exclusive one, where resummation plays a role. The uncertainties for our predictions of the inclusive $Z+1$-jet cross section are larger, since they are only predicted to NLO$_1$, though they can still benefit from the resummation of the $0/1$ jet boundary. The separation into exclusive $Z+1$-jet and inclusive $Z+2$-jet is only at LL accuracy, and the FO accuracy of the inclusive $Z+2$-jet cross section is only LO$_2$, with correspondingly larger perturbative uncertainties. \geneva agrees well with the data, somewhat better for lower $\alpha_s(M_Z)$.

\begin{figure}[H]
\begin{center}
\includegraphics[width=0.95\columnwidth]{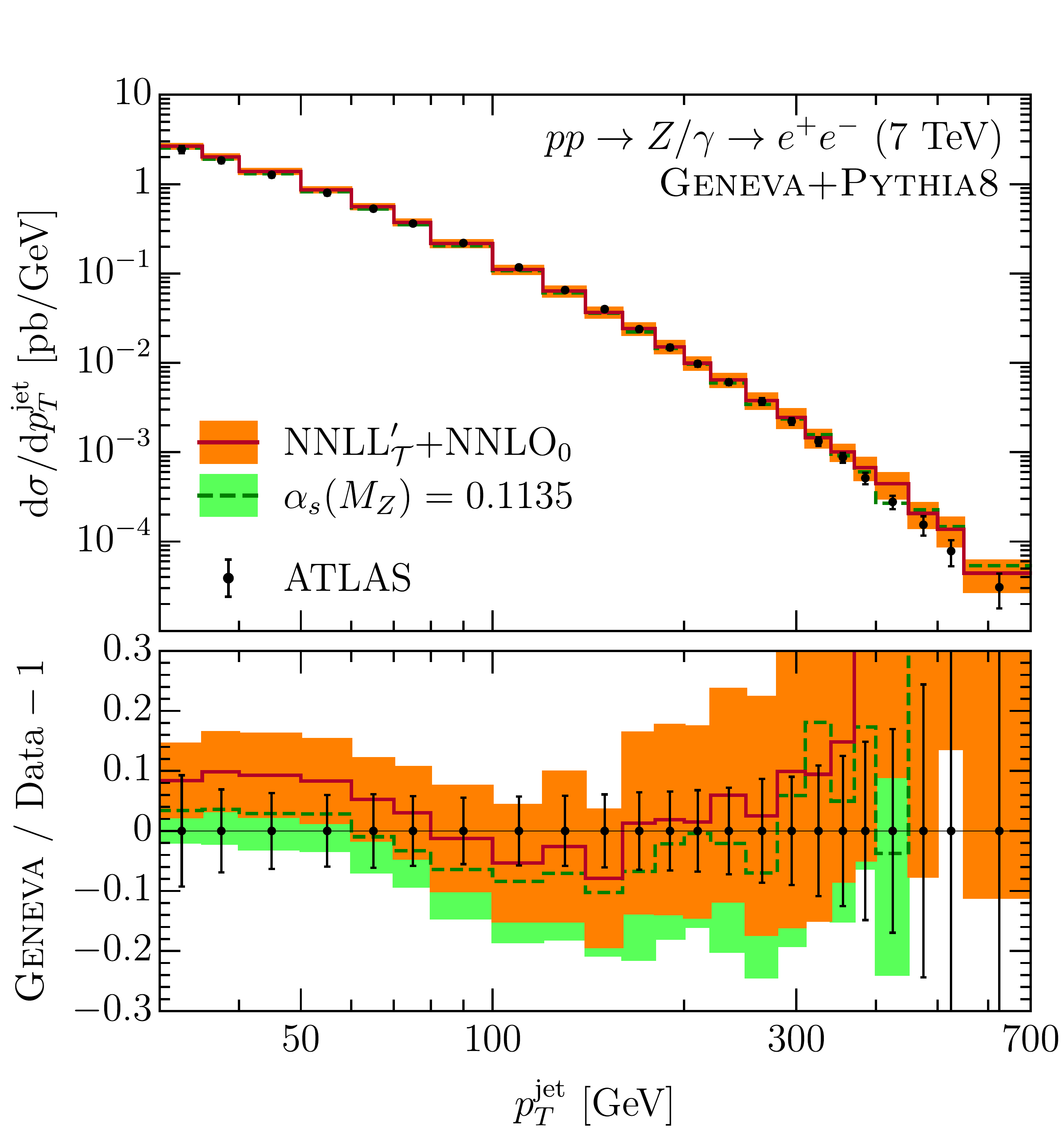}%
\caption{Comparison of \geneva with the transverse momentum of the leading jet from the \atlas study in \cite{Aad:2013ysa}. The \geneva results with default values of $\as(M_Z)$  and with $\as(M_Z)=0.1135$ are shown in orange and green, respectively,  while the \atlas points are shown in black.}
\label{fig:pTjetdata}
\end{center}
\end{figure}

In Ref.~\cite{Aad:2013ysa}, \atlas also presented a measurement of the transverse momentum distribution of the hardest jet, and the comparison to these results is shown in \fig{pTjetdata}.
The predictions from \geneva are in good agreement with the data. Below $p_T^{\rm jet} \lesssim m_Z$ the agreement is noticeably better for lower $\alpha_s(M_Z)$.
Above $p_T^{\rm jet} \gtrsim 300$ GeV the predictions tend to be higher than the data, but still consistent within the larger uncertainties. This could be due to the fact that we use a renormalization scale of $\mu = m_{\ell^+\ell^-}$, while at such large transverse momenta a better choice might be $\mu = p_T^{\rm jet}$.

Next, we compare our predictions to the $\phi^*$ between the leptons, and the transverse momentum of the vector boson. Both of these observables are 0-jet resummation variables, since at low values they are dominated by events without any hard emissions.
Both of these  distribution are quite sensitive to the choice of parameters of the parton shower and the nonperturbative model used in \pythia8. Without higher-order perturbative corrections included, the MC tune will partially adjust the available parameters to mimic missing higher-order perturbative effects. 
This implies that when using \pythia8 in conjunction with \geneva, which includes much more higher-order perturbative information, a retuning of the parameters becomes necessary. We stress that no attempt at a systematic retuning of \pythia8 has been done for this work, and we expect that a dedicated tune of \pythia8 together with \geneva, will improve the data agreement.

\begin{figure}[H]
\begin{center}
\includegraphics[width=0.95\columnwidth]{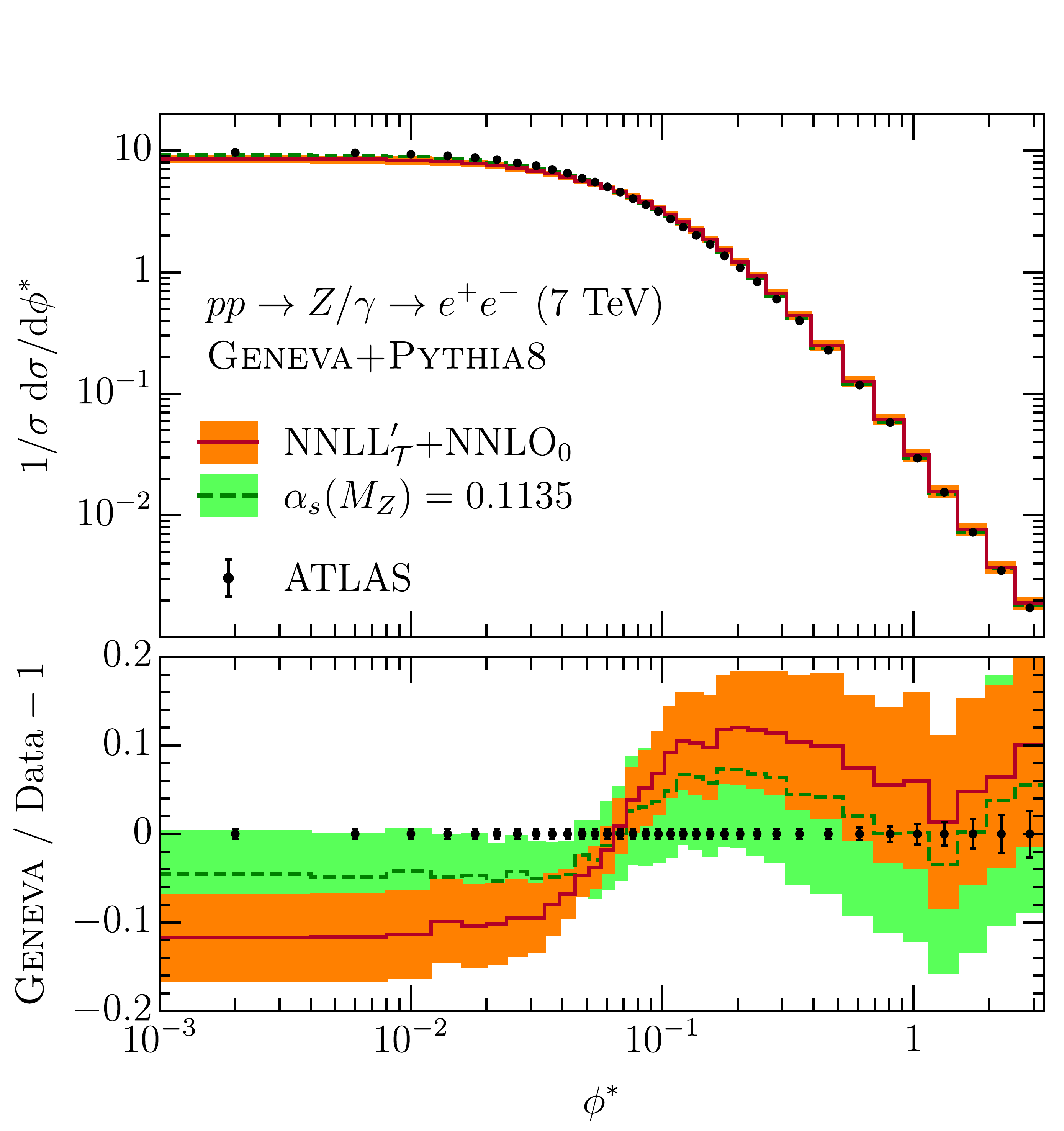}%
\caption{Comparison of \geneva with the $\phi^*$ distribution from the \atlas study in \cite{Aad:2012wfa}. The \geneva results with default values of $\as(M_Z)$  and with $\as(M_Z)=0.1135$ are shown in orange and green, respectively,  while the \atlas points are shown in black.}
\label{fig:phiStardataATLAS}
\end{center}
\end{figure}

The $\phi^*$ distribution has been measured by \atlas in Ref.~\cite{Aad:2012wfa} and \lhcb in Ref.~\cite{Aaij:2012mda}. The measurement from \atlas is a normalized spectrum, while \lhcb quotes the unnormalized distribution. The comparisons with these two measurements are shown in \figs{phiStardataATLAS}{phiStardataLHCB}.
\geneva agrees well with the results from \lhcb, though the measurement has relatively large uncertainties. Comparing with the much more precise results from \atlas, \geneva predicts a wider distribution, such that our predictions are below the data in the peak region for $\phi^* \lesssim 0.1$, and above the data for larger values of $\phi^*$.


\begin{figure}[H]
\begin{center}
\includegraphics[width=0.95\columnwidth]{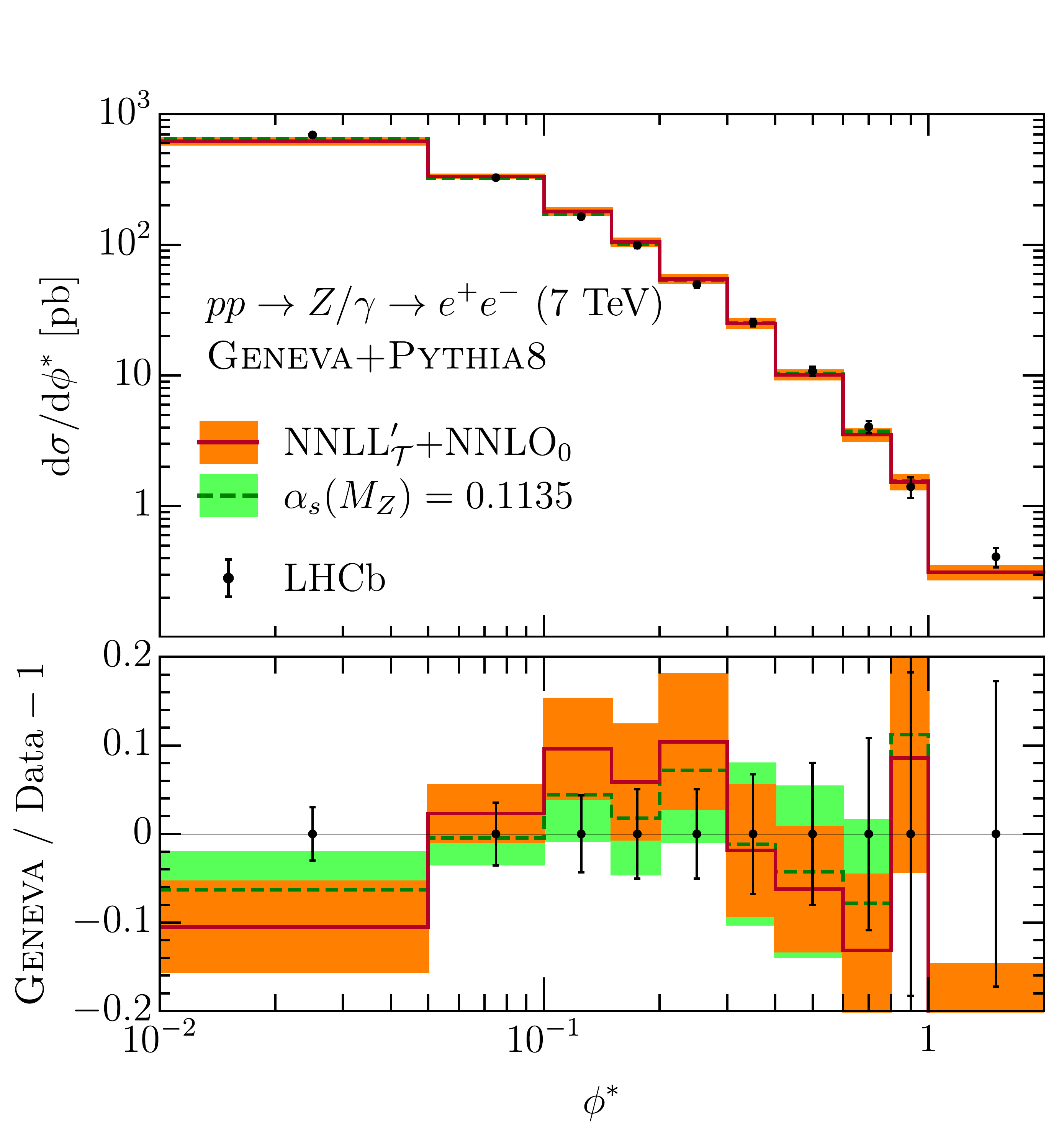}%
\caption{Comparison of \geneva with the $\phi^*$ distribution from the \lhcb study in \cite{Aaij:2012mda}. The \geneva results with default values of $\as(M_Z)$  and with $\as(M_Z)=0.1135$ are shown in orange and green, respectively,  while the \lhcb points are shown in black.}
\label{fig:phiStardataLHCB}
\end{center}
\end{figure}

\begin{figure}[H]
\begin{center}
\includegraphics[width=\columnwidth]{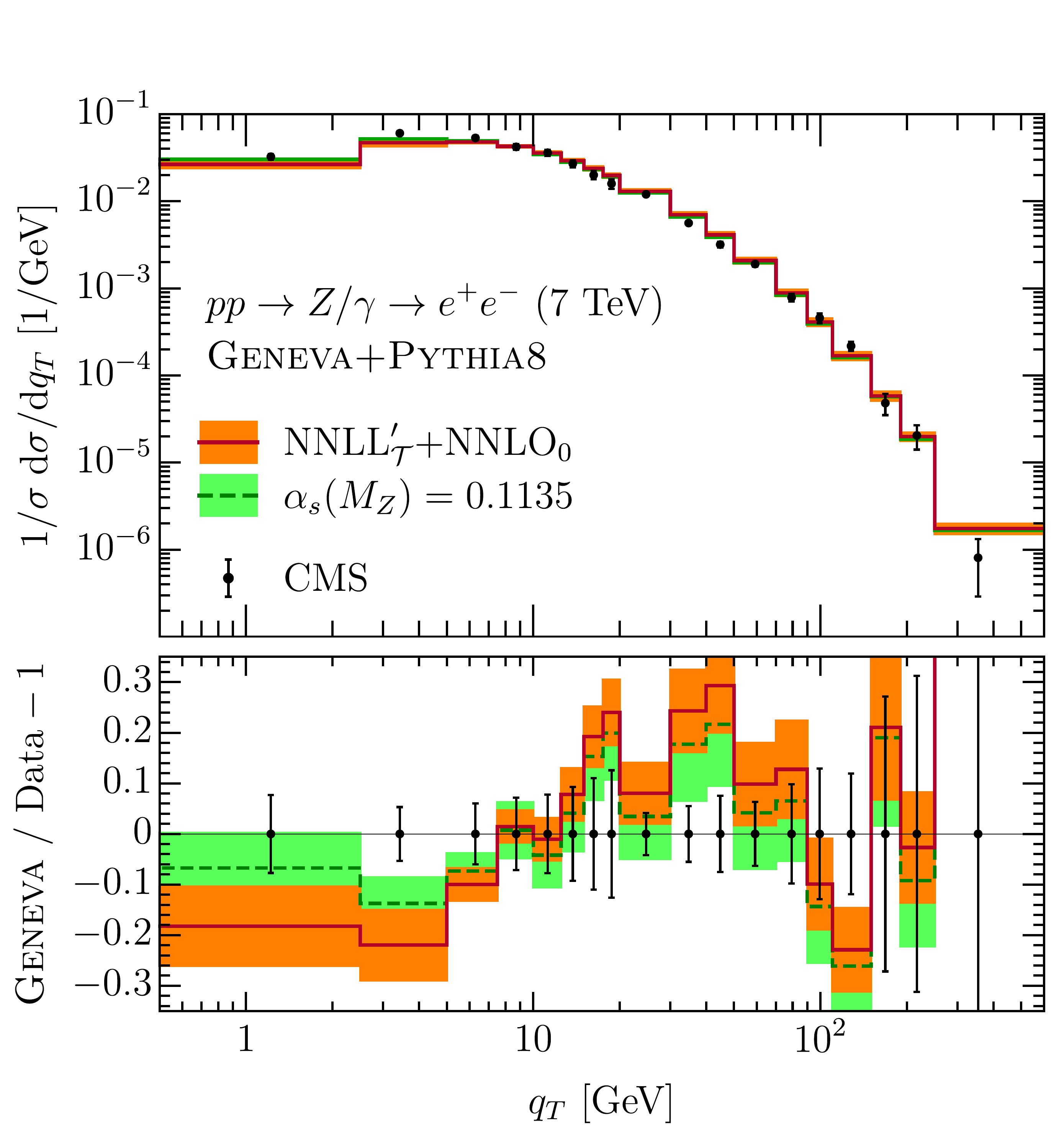}%
\caption{Comparison of \geneva with the transverse momentum of the vector boson from the \cms study in \cite{Chatrchyan:2011wt}. The \geneva results with default values of $\as(M_Z)$  and with $\as(M_Z)=0.1135$ are shown in orange and green, respectively,  while the \cms points are shown in black.}
\label{fig:qTdataCMS}
\end{center}
\end{figure}

\begin{figure*}[t!]
\includegraphics[width=\columnwidth]{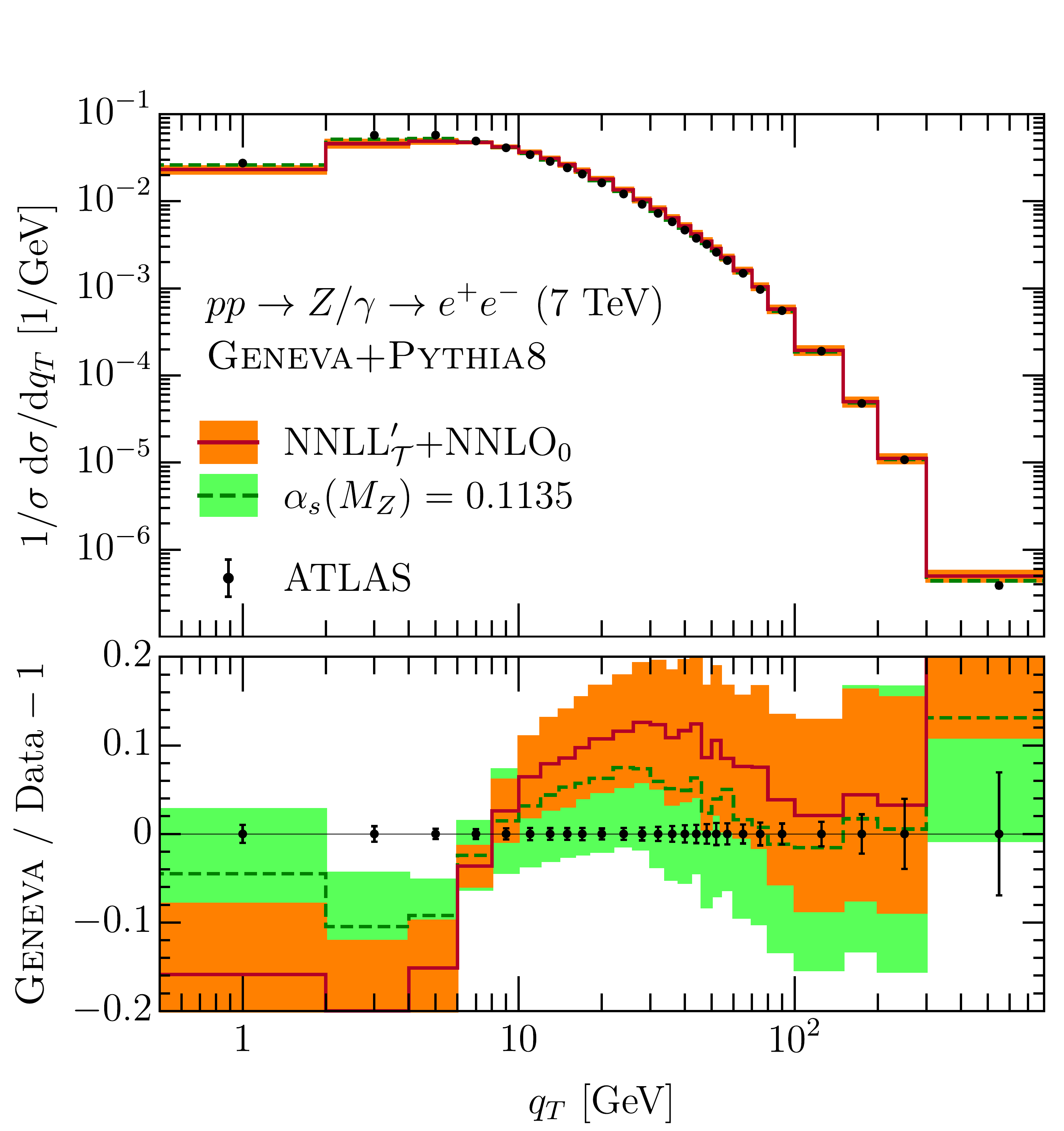}%
\hfill%
\includegraphics[width=\columnwidth]{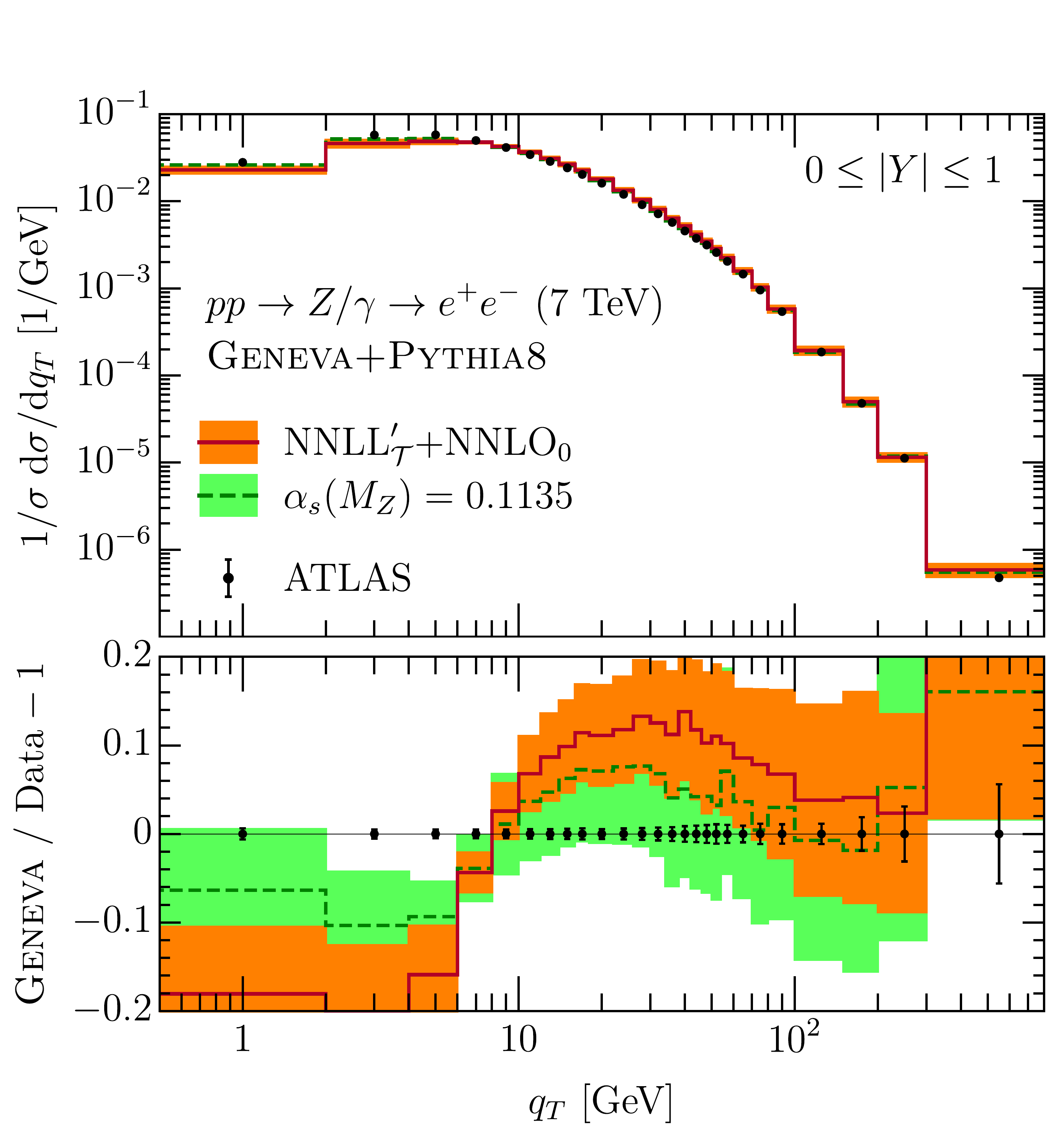}
\\[-2ex]
\includegraphics[width=\columnwidth]{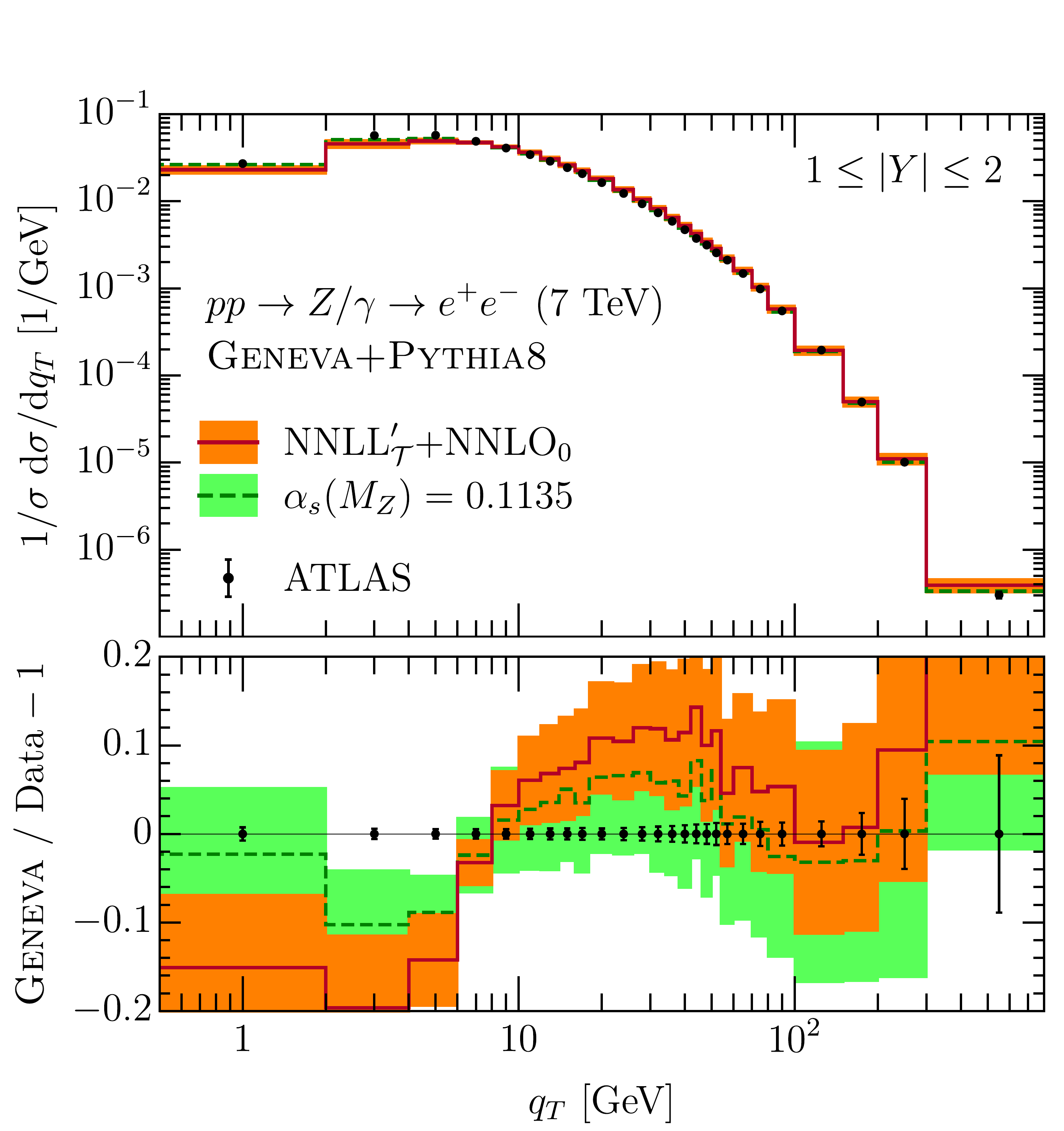}%
\hfill%
\includegraphics[width=\columnwidth]{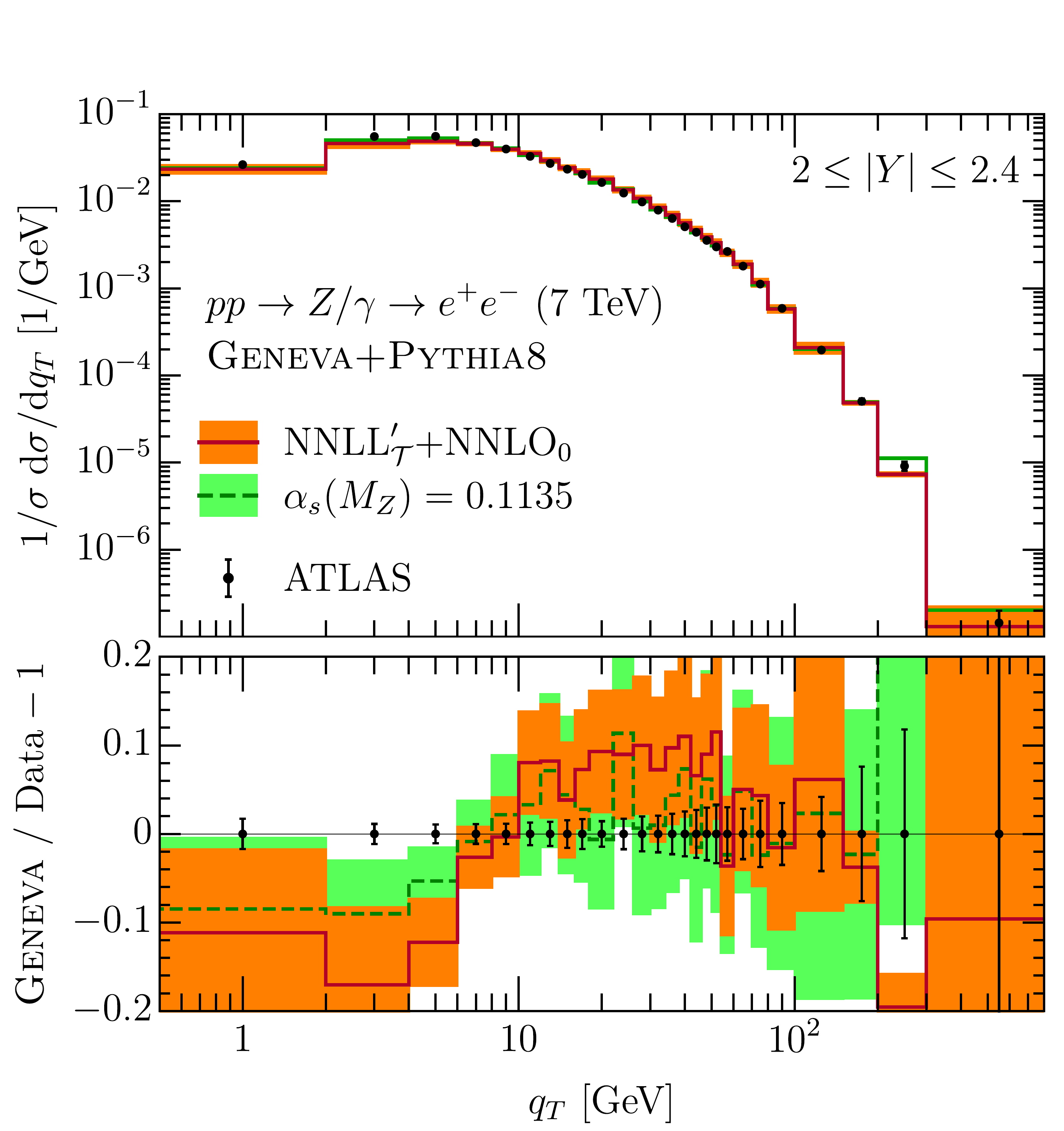}
\caption{Comparison of \geneva with the transverse momentum of the vector boson  both inclusive (top-left) and in different bins of vector boson rapidity $Y$ from the \atlas study in \cite{Aad:2014xaa}. The \geneva results with default values of $\as(M_Z)$  and with $\as(M_Z)=0.1135$ are shown in orange and green, respectively,  while the \atlas points are shown in black.}
\hspace{-2ex}
\label{fig:qTdataATLAS}
\end{figure*}

Discrepancies with this same trend and at a similar level were observed for other MC predictions in Ref.~\cite{Aad:2012wfa}. The agreement with data is considerable better when choosing a lower value of $\alpha_s(M_Z) = 0.1135$.

As a final comparison we consider the transverse momentum ($q_T$) distribution of the $Z$ boson. The $q_T$ distribution was measured by \cms in Ref.~\cite{Chatrchyan:2011wt} and by \atlas in Ref.~\cite{Aad:2014xaa}, where the latter also measured the $q_T$ spectrum in different bins of $Y$.  The comparison of our results with these measurements is shown in \figs{qTdataCMS}{qTdataATLAS}. The situation is similar to the $\phi^*$ distribution discussed above. For $\alpha_s(M_Z) = 0.118$, \geneva is below the data in the peak region $q_T \lesssim 10$ GeV, while it is above the data in the transition and FO regions for $q_T \gtrsim 10$ GeV. This is of course not unexpected, since the observables $q_T$ and $\phi^*$ are highly correlated. The same level of disagreement was already observed in Ref.~\cite{Aad:2014xaa}, where the data was compared with the NNLL$+$NLO$_1$ results of Ref.~\cite{Banfi:2012jm}. As for the $\phi^*$ distribution, the agreement between \geneva and the data is noticeably improved for a lower $\alpha_s(M_Z) = 0.1135$.

\section{Conclusions}
\label{sec:conclusions}

We have presented a combination of the fully-differential NNLO calculation for Drell-Yan production $pp \to \gamma / Z \to \ell^+ \ell^-$ combined with the NNLL$'$ resummation of $0$-jettiness and interfaced with the parton shower provided by \pythia8 within the \geneva Monte-Carlo framework. The starting point of \geneva is the formulation of Monte-Carlo cross sections, which are IR safe partonic jet cross sections, defined using a jet resolution variable $\Tau_N$, which is chosen to be $N$-jettiness in our implementation. We include $0$-jet, $1$-jet, and $2$-jet cross sections at NNLO$_0$, NLO$_1$ and LO$_2$. Furthermore, the dependence on the 0-jet resolution variable $\Tau_0$ is resummed to NNLL$'$ accuracy, while that of the 1-jet resolution variable $\Tau_1$ is presently resummed to LL.

Interfacing these partonic results with a parton shower such as \pythia8 requires a careful treatment. First, one needs to deal with the fact that the parton shower evolution variable is different from $N$-jettiness. Second, care has to be taken that the parton shower's lower resummation accuracy and different phase-space map does not destroy the higher logarithmic accuracy of the calculated Monte-Carlo cross sections. We discussed in detail how to solve these issues by performing the first two shower emissions by hand using the jet-resolution parameter as ordering variable. We then showed that any subsequent showering by \pythia8 does not affect the formal perturbative accuracy included in \geneva.

We have validated the FO perturbative accuracy by comparing our results for inclusive $0$-jet observables to \dynnlo. We agree with this dedicated NNLO calculation within the small perturbative (and statistical) uncertainties. We also studied how \geneva's improved perturbative accuracy in the resummation region from resumming $0$-jettiness to NNLL$'$, translates to other $0$-jet resummation variables, such as the transverse momentum of the vector boson $q_T$, the $\phi^*$ between the two leptons, and the exclusive $0$-jet cross section as a function of the jet $p_T^\cut$. Since \geneva partially relies on the parton shower for these observables, they formally do not have NNLL$'$ accuracy. Nevertheless, we find that \geneva reproduces dedicated NNLL resummations for these three observables rather well, and in particular the \geneva predictions are much closer to the exact NNLL$+$NLO$_1$ results than to the NLL$+$LO$_1$ results. This is a clear indication that the gain in resummation accuracy for $\Tau_0$, when implemented into a fully exclusive prediction, also translates into more accurate predictions for other $0$-jet resolution variables.

Finally, we presented a comparison of \geneva with measurements by \atlas, \cms, and \lhcb using the $7\TeV$ LHC data. Since \geneva agrees with the NNLO results for the rapidity distribution of the vector boson, the agreement between \geneva and the LHC data is very similar to that observed when comparing to other NNLO calculations. Here, we note that experimental measurements of the lepton $p_{T\ell}$ spectrum would be very valuable.

We find good agreement between \geneva and the \atlas measurement of exclusive jet cross sections and of the transverse momentum distribution of the hardest jet. For the transverse momentum of the vector boson or the $\phi^*$ between the leptons our predictions when using the default value of $\alpha_s(M_Z) = 0.118$ are lower in the peak region and higher in the transition region behind the peak compared to the measurements. The same trend has been observed before by other MC predictions and also dedicated resummed calculations for these observables. When using a lower value $\alpha_s(M_Z) = 0.1135$, we observe a noticeable improvement in agreement with data for essentially all resummation-sensitive observables. A similar effect was observed previously for a variety of $e^+e^-$ $2$-jet resummation observables. We believe this deserves further attention.

We encourage the experiments to also measure $\Tau_0$-like jet-based variables (see Ref.~\cite{Gangal:2014qda}) and to also perform differential measurements of resummation observables at much higher $m_{\ell^+\ell^-}$ values. Since the resummation type and regime for these cases are very different, they would add valuable complementary information for theory comparisons. In general, continued precise measurements of resummation observables at the LHC are essential for our understanding of higher-order QCD effects in exclusive and differential observables, which ultimately will also provide important inputs for the theoretical interpretation of Higgs measurements and new-physics searches.

\begin{acknowledgments}

We are grateful to
Giancarlo Ferrera for providing us with a preliminary version
of DYqT and to Andrea Banfi and Lee Tomlinson for help in comparing with their resummed  $q_T$ and $\phi^*$ predictions.
We also thank Z.~Marshall for comments to the manuscript.
SA thanks E.~Re, P.~Monni, and  G.~Salam for useful discussions.
CWB thanks the CERN theory group for its hospitality and partial support during the completion of this work.
This work was supported by the Department of Energy Early Career Award
with Funding Opportunity No. DE-PS02-09ER09-26 (SA, CWB, CB, JW), the
DFG Emmy-Noether Grant No. TA 867/1-1 (FT), the COFUND Fellowship
under grant agreement PCOFUND-GA-2012-600377 (SA), and the Director,
Office of Science, Office of High Energy Physics of the
U.S. Department of Energy under the Contract No. DE-AC02-05CH11231
(CWB, CB, JW).
This research used resources of the National Energy Research Scientific Computing Center, which is supported by the Office of Science of the U.S. Department of Energy under Contract No. DE-AC02-05CH11231.

\end{acknowledgments}


\bibliography{../geneva}

\end{document}